\documentclass[a4paper,fleqn]{cas-sc}

\usepackage[numbers,sort&compress]{natbib}

\usepackage{graphicx}%
\usepackage{mathrsfs}%
\usepackage{textcomp}%
\usepackage{booktabs}%
\usepackage{multirow}%

\newtheorem{theorem}{Theorem}
\newtheorem{proposition}{Proposition}
\newtheorem{lemma}{Lemma}
\newdefinition{example}{Example}
\newdefinition{remark}{Remark}
\newdefinition{definition}{Definition}
\newproof{proof}{Proof}

\newcommand{\Z}{\mathbb{Z}_n}
\newcommand{\CL}{\mathbf{CL}}

\begin{document}

\shorttitle{The Schur products of some nonbinary narrow-sense antiprimitive BCH codes}
\shortauthors{H. Xu et~al.}

\title[mode = title]{The Schur products of some nonbinary narrow-sense antiprimitive BCH codes}
\tnotemark[1]

\tnotetext[1]{The work of Haojie~Xu was supported by Postgraduate Research $\&$ Practice Innovation Program of Jiangsu Province under Grant KYCX25\_0417. The work of Xia Wu and Wei Lu were supported by the National Natural Science Foundation of China under Grant 12371035, and Jiangsu Provincial Scientific Research Center of Applied Mathematics under Grant BK20233002. The work of Xiwang Cao was supported by the National Natural Science Foundation of China under Grants 12171241 and 12571575.}

\author[1]{Haojie Xu}
\ead{xuhaojiechn@163.com}

\author[1]{Xia Wu}
\fnmark[1]
\ead{wuxia80@seu.edu.cn}

\author[1]{Wei Lu}
\cormark[1]
\fnmark[1]
\ead{luwei1010@seu.edu.cn}

\author[2]{Xiwang Cao}
\fnmark[1]
\ead{xwcao@nuaa.edu.cn}

\affiliation[1]{organization={School of Mathematics, Southeast University},
                city={Nanjing},
                postcode={210096},
                country={China}}

\affiliation[2]{organization={Department of Math, Nanjing University of Aeronautics and Astronautics},
                city={Nanjing},
                postcode={210096},
                country={China}}

\cortext[cor1]{Corresponding author}
\fntext[fn1]{These authors contributed equally to this work.}

\begin{abstract}
In recent years, the Schur products of linear codes have been extensively studied due to their favorable applications in cryptography. In this paper, we aim to study the parameters of the Schur products of some nonbinary narrow-sense antiprimitive BCH codes. Let $\mathcal{C}_1$ and $\mathcal{C}_2$ be two narrow-sense antiprimitive BCH codes with designed distances $\delta_{1}$ and $\delta_{2}$, respectively, where $2\le \delta_{1},\delta_{2}\le n$. Let $\mathcal{C}_1\star\mathcal{C}_2$ be the Schur product of $\mathcal{C}_1$ and $\mathcal{C}_2$. We present some necessary and sufficient conditions for $\mathcal{C}_1\star\mathcal{C}_2\ne\mathbb{F}_q^n$ by applying some constraints to $\delta_{1}$ and $\delta_{2}$. Furthermore, we determine the dimensions of these products and lower bounds on their minimum distance. Several linear codes with best known parameters are found.
\end{abstract}

\begin{keywords}
Schur product \sep coset leader \sep BCH code \sep cyclic code
\end{keywords}


\maketitle

\section{Introduction}\label{sec1}

Throughout this paper, let $q$ be the power of a prime $p$, $\mathbb{F}_q$ be the finite field with $q$ elements, and $\mathbb{F}_q^\ast=\mathbb{F}_q\setminus\{0\}$. An $[n,k]$ linear code $\mathcal{C}$ over $\mathbb{F}_q$ is a vector subspace of $\mathbb{F}_q^n$ with dimension $k$. The \textit{minimum distance} $d$ of $\mathcal{C}$ is represented by $d=\min\{\mathrm{wt}(\mathbf{c}):\mathbf{c}\in\mathcal{C}\setminus \{\mathbf{0}\}\}$, where $\mathrm{wt}(\mathbf{c})$ is the number of nonzero coordinates of $\mathbf{c}$ and called the \textit{Hamming weight} of $\mathbf{c}$. Let $\mathbf{x}=(x_1, x_2,\dots,x_n)$ and $\mathbf{y} = (y_1, y_2,\dots, y_n)\in\mathbb{F}_q^n$. The \textit{Schur product} (or \textit{componentwise product}) of $\mathbf{x}$ and $\mathbf{y}$ is denoted by $\mathbf{x}\ast\mathbf{y} = (x_1y_1, x_2y_2,\dots, x_ny_n)$. Let $\mathcal{C}_1$ and $\mathcal{C}_2$ be two linear codes of length $n$ over $\mathbb{F}_q$. Then the \textit{Schur product} (for short, \textit{product}) $\mathcal{C}_1\star\mathcal{C}_2$ of $\mathcal{C}_1$ and $\mathcal{C}_2$ is the linear code over $\mathbb{F}_q$ spanned by the set $\{\mathbf{c}_1\ast\mathbf{c}_2\ \mid\ \mathbf{c}_1\in\mathcal{C}_1,\mathbf{c}_2\in\mathcal{C}_2\}$. If $\mathcal{C}_1=\mathcal{C}_2$, $\mathcal{C}_1\star\mathcal{C}_2$ is referred to as the \textit{Schur square} (for short, \textit{square}) of $\mathcal{C}_1$.

The products of codes were first proposed to enable the development of algebraic decoding algorithms for linear codes \cite{Kotter1992Unified,Pellikaan1992decoding}. Recently, they have been used to construct structural attacks on the McEliece public key cryptosystem based on GRS subcodes \cite{Wieschebrink2010Cryptanalysis} or on subcodes of algebraic geometry codes \cite{Couvreur2015Cryptanalysis}. The products of codes have also been investigated for use in secret sharing schemes \cite{Cascudo2009Asymptotically} and secure multiparty computation \cite{Cramer2000General}. Moreover, further applications are discussed in \cite{Cascudo2015Squares}. Owing to its good performance in cryptographic applications, investigating the parameters of products of linear codes has turned into a key research focus recently. It is easily seen that $$\dim(\mathcal{C}_1\star\mathcal{C}_2)\le\dim(\mathcal{C}_1)\dim(\mathcal{C}_2),$$ where $\mathcal{C}_1$ and $\mathcal{C}_2$ are two linear codes of length $n$ over $\mathbb{F}_q$. 

As a special case of the Schur products, the Schur squares of linear codes have been studied and some progress has been made. Cascudo et al. investigated the Schur squares of random linear codes \cite{Cascudo2015Squares} and matrix product codes \cite{Cascudo2020Squares}, respectively. Moreover, Cascudo systematically studied the Schur squares of cyclic codes and constructed some cyclic codes with large dimensions and large minimum distances \cite{Cascudo2019Squares}. Inspired by Cascudo's work, Dong et al. studied the Schur squares of narrow-sense primitive BCH codes and their complementary codes \cite{Dong2023Parameters}. Shortly afterward, the parameters of the Schur squares of LCD primitive BCH codes $C_{(q, q^m-1, 2\delta, -\delta+1)}$ and their complementary codes were investigated in \cite{Dong2024Squares,Zhang2026Constructions}. Zheng and Zhu analyzed the Schur squares of four classes of BCH codes of length $\frac{(q^m-1)}{2}$ and their complementary codes \cite{Zheng2026Four}.

Currently, literature on Schur products is relatively limited. Randriambololona provided a Schur product Singleton bound, linking the dimensions of some given number of linear codes to the minimum distance of their Schur product \cite{Randriambololona2013Upper}. Mirandola and Z\'{e}mor characterized the features of linear code pairs that achieve the Schur product Singleton bound \cite{Mirandola2015Critical}. Falk et al. presented some properties of the Schur product of constacyclic codes \cite{Falk2020Properties}. Xu and Li gave some conditions for the product of certain primitive BCH codes and their complementary codes not being equal to $\mathbb{F}_q^n$, and studied the parameters of some of these products \cite{Xu2024Products}. Hu and Heng also provided some conditions under which the products of some primitive narrow-sense BCH codes are not equal to $\mathbb{F}_q^n$ \cite{Hu2025Schur}.

As far as we know, there is currently no research on the Schur products of antiprimitive BCH codes (length $q^m+1$). Therefore, the objective of this paper is to investigate the parameters of the product for some nonbinary narrow-sense antiprimitive BCH codes. Let $\Delta_1$ and $\Delta_{2}$ be the first and second largest $q$-cyclotomic coset leader modulo $q^m+1$, respectively. Let $\mathcal{C}_1$ and $\mathcal{C}_2$ be two nonbinary narrow-sense antiprimitive BCH codes with designed distances $\delta_{1}$ and $\delta_{2}$, respectively, where $2\le \delta_{1},\delta_{2}\le n$. Let $\Omega_{(q,m)}$ and $\Lambda_{(q,m)}$ be as in Theorems~\ref{sec3 thm1}, \ref{sec3.2 thm1}, \ref{sec3.2 thm2} and \ref{sec3.2 thm3}. Taking $\mathcal{C}_1\star\mathcal{C}_2=\mathcal{C}(\delta_{1},\delta_{2})$, our main contributions are as follows:
\begin{itemize}
	\item \textbf{$\boldsymbol{q}$ odd and $\boldsymbol{m\ge2}$, or $\boldsymbol{q}$ even, $\boldsymbol{q>2}$ and $\boldsymbol{2\le m\le11}$}. For $\delta_{1}\le\Omega_{(q,m)}$, we prove that $\mathcal{C}(\delta_{1},\delta_{2})\ne \mathbb{F}_q^n$ if and only if $\delta_{2}\ge\Delta_1+1$. 
	\item \textbf{$\boldsymbol{q}$ even, $\boldsymbol{q>2}$ and $\boldsymbol{m\not\equiv0\bmod 4}$}. For $\delta_{1}\le\Lambda_{(q,m)}$, we have $\mathcal{C}(\delta_{1},\delta_{2})\ne \mathbb{F}_q^n$ if and only if $\delta_{2}\ge\Delta_1+1$. 
	\item \textbf{$\boldsymbol{q}$ odd and $\boldsymbol{m\ge2}$, or $\boldsymbol{q}$ even, $\boldsymbol{q>2}$ and $\boldsymbol{2\le m\le5}$}. For $\delta_{1}\ge\Omega_{(q,m)}+1$ and $\Delta_2+1\le\delta_{2}\le\Delta_{1}$, we determine the dimension of $\mathcal{C}(\delta_{1},\delta_{2})$ and a lower bound on its minimum distance.
\end{itemize}

\section{Preliminaries}

In this section, we present some preliminaries on BCH codes, products of cyclic codes and largest coset leaders.

\subsection{Cyclic codes and BCH codes}\label{sec2.1}

An $[n,k]$ code $\mathcal{C}$ over $\mathbb{F}_q$ is called \textit{cyclic} if $\mathbf{c}=(c_0,c_1,\dots,c_{n-2},c_{n-1})\in\mathcal{C}$ implies $(c_{n-1},c_0,c_1,\dots,c_{n-2})\in\mathcal{C}$. Throughout this subsection, we assume that $\gcd(n,q)=1$. The residue class ring $\mathbb{F}_q[x]/(x^n-1)$ is isomorphic to $\mathbb{F}_q^n$ as a vector space over $\mathbb{F}_q$, where the isomorphism is given by $(c_0,c_1,\cdots,c_{n-1})\leftrightarrow c_0+c_1x+\cdots+c_{n-1}x^{n-1}$. Because of this isomorphism, any linear code $\mathcal{C}$ of length $n$ over $\mathbb{F}_q$ corresponds to a subset of $\mathbb{F}_q[x]/(x^n-1)$. Then we can interpret $\mathcal{C}$ as a subset of $\mathbb{F}_q[x]/(x^n-1)$. It is easily checked that the linear code $\mathcal{C}$ is cyclic if and only if $\mathcal{C}$ is an ideal of $\mathbb{F}_q[x]/(x^n-1)$. Since every ideal of $\mathbb{F}_q[x]/(x^n-1)$ is principal, every cyclic code $\mathcal{C}$ is generated by the unique monic polynomial $g(x)\in\mathbb{F}_q[x]$ of the lowest degree, i.e., $\mathcal{C}=\left<g(x)\right>$. $g(x)$ is called the \textit{generator polynomial} of $\mathcal{C}$ and $h(x)=(x^n-1)/g(x)$ is called the \textit{parity-check polynomial} of $\mathcal{C}$. 

We always assume that $\gcd(q,n)=1$ in this paper. Let $m$ be the multiplicative order of $q$ modulo $n$. Let $\alpha$ be a generator of $\mathbb{F}^\ast_{q^m}$ and $\beta=\alpha^{(q^m-1)/n}$. Then $\beta$ is a primitive $n$-th root of unity in $\mathbb{F}_{q^m}$. The set $T = \{0 \le i \le n-1\ \mid\ g(\beta^i)=0\}$ is called the \textit{defining set} of $\mathcal{C}$ with respect to $\beta$. It should be noticed that the generator polynomial $g(x)$ is a factor of $x^n-1$, thus we have to study the canonical factorization of $x^n-1$ over $\mathbb{F}_q$ to handle $g(x)$. With this intention, we need to introduce $q$-cyclotomic cosets modulo $n$ \cite{Ding2014CodesDifferenceSetsa}. 

Let $\mathbb{Z}_n$ denote the set $\{0,1,2,\dots,n-1\}$ and let $s<n$ be a nonnegative integer. The \textit{$q$-cyclotomic coset of $s$ modulo $n$} is given by 
\begin{equation*}
	C_s=\{s,sq,sq^2,\ldots,sq^{\ell_s-1}\}\bmod n\subseteq\mathbb{Z}_n,
\end{equation*}
where $\ell_s$ is the smallest positive integer such that $s \equiv sq^{\ell_s} \pmod n$, and is the size of the $q$-cyclotomic coset. The smallest integer $\CL(s)$ in $C_s$ is called the \textit{coset leader} of $C_s$. Let $\Gamma_{(n,q)}$ be the set of all the coset leaders. We have then $C_s \cap C_t = \emptyset$ for any two distinct elements $s$ and $t$ in $\Gamma_{(n,q)}$. Denote the minimal polynomial over $\mathbb{F}_q$ of $\beta^s$ by $g_s(x)$. And this polynomial is given by
\begin{equation}\label{sec2.1 equ3}
	g_s(x)=\prod\limits_{i\in C_s}(x-\beta^i)\in\mathbb{F}_q[x],
\end{equation}
which is irreducible over $\mathbb{F}_q$. 

BCH codes are a crucial class of cyclic codes due to their exceptional error-correcting capabilities and simple encoding and decoding algorithms. Let $h$ be a nonnegative integer and $m$ be the multiplicative order of $q$ modulo $n$. Suppose that $\beta\in\mathbb{F}_{q^m}$ is a primitive $n$-th root of unity and $\delta$ is an integer with $2\le\delta\le n$. Then a \textit{BCH code} over $\mathbb{F}_q$ of length $n$ and \textit{designed distance} $\delta$, denoted by $\mathcal{C}_{(q,n,\delta,h)}$, is a cyclic code generated by 
\begin{equation*}
	g_{(q,n,\delta,h)}=\mathrm{lcm}\left(g_h(x),g_{h+1}(x),\dots,g_{h+\delta-2}(x)\right),
\end{equation*}
where the least common multiple is computed over $\mathbb{F}_q$ and $g_s(x)$ is represented in (\ref{sec2.1 equ3}). If $h=1$, the BCH code $\mathcal{C}_{(q,n,\delta,h)}$ is called \textit{narrow-sense BCH code}. If $n=q^m-1$, the corresponding BCH codes are said to be \textit{primitive}. If $n=q^m+1$, then $\mathcal{C}_{(q,n,\delta,h)}$ is referred to as the \textit{antiprimitive} BCH code. According to the \textit{BCH bound}, the minimum distance of the BCH code $\mathcal{C}_{(q,n,\delta,h)}$ is at least $\delta$.

Although BCH codes have been widely investigated for decades, determining their parameters is generally challenging \cite{Ding2024BCH}. Until now, some results regarding the dimensions and minimum distances of certain BCH codes have been established. Here we mainly list relevant references for antiprimitive BCH codes, e.g.,  \cite{Geng2022Class,Liu2019Some,Liu2023DiemnsionNonbinary,Xu2025Sufficient,Xu2026Dual,Yan2018Parameters,Zhu2022AryAntiprimitive}.

\subsection{Products of cyclic codes}
For two subsets $A,B\subseteq\Z$, we define $A+B:=\{(a+b) \bmod n \ \Big|\  a \in A, b \in B\}$. The following lemma concerns the product of two cyclic codes.

\begin{lemma}\cite{Cascudo2020Squares}\label{sec2.2 lem1}
	Let $\mathcal{C}_1$ and $\mathcal{C}_2$ be two cyclic codes over $\mathbb{F}_q$ with generator polynomials
	\begin{equation*}
		g_{1}(x)=\frac{x^{n}-1}{\prod\limits_{i\in I_{1}}(x-\beta^{i})}\ and\ g_{2}(x)=\frac{x^{n}-1}{\prod\limits_{i\in I_{2}}(x-\beta^{i})},
	\end{equation*}
	respectively, where $I_{1},I_{2}\subseteq\mathbb{Z}_{n}$. Then the product $\mathcal{C}=\mathcal{C}_{1}\star\mathcal{C}_{2}$ is a cyclic code over $\mathbb{F}_q$ with generator polynomial
	\begin{equation*}
		g(x)=\frac{x^{n}-1}{\prod\limits_{i\in I_{1}+I_{2}}(x-\beta^{i})}.
	\end{equation*}
\end{lemma}

Throughout this paper, let $\mathcal{C}_1=\mathcal{C}_{(q,n,\delta_1,1)}$ and $\mathcal{C}_2=\mathcal{C}_{(q,n,\delta_{2},1)}$ be two narrow-sense BCH codes, where
$2\le\delta_1,\delta_{2}\le n$. Then the defining sets of
$\mathcal{C}_1$ and $\mathcal{C}_2$ are equal to $T_{1} = C_{1}\cup C_{2}\cup\cdots\cup C_{\delta_{1}-1}$ and $T_{2} = C_{1}\cup C_{2}\cup\cdots\cup C_{\delta_{2}-1}$, respectively. Denote $I_1 = \mathbb{Z}_n\setminus T_1$, $I_2 = \mathbb{Z}_n\setminus T_2$ and $\mathcal C(\delta_1,\delta_{2})=\mathcal C_1\star\mathcal C_2$. It follows from Lemma~\ref{sec2.2 lem1} that the defining set of the	product $\mathcal C(\delta_1,\delta_{2})$ is equal to $T(\delta_{1},\delta_{2})=\mathbb{Z}_{n}\backslash(I_{1}+I_{2})$. Then
\begin{equation}\label{sec2.2 equ1}
	\mathcal{C}(\delta_1,\delta_{2})\neq\mathbb{F}_q^n \Leftrightarrow T(\delta_1,\delta_{2})\neq\emptyset \Leftrightarrow I_1+I_2\neq\mathbb{Z}_n.
\end{equation}
Note that 
\begin{equation}\label{sec2.2 equ2}
	T(\delta_1',\delta_{2})\subseteq T(\delta_1'',\delta_{2})\quad\mathrm{if}\quad\delta_1'\leq\delta_1''
\end{equation}
and
\begin{equation}\label{sec2.2 equ3}
	T(\delta_1,\delta_{2}')\subseteq T(\delta_1,\delta_{2}'')\quad\mathrm{if}\quad\delta_{2}'\leq\delta_{2}''.
\end{equation}

When $\delta_{1}$ and $\delta_{2}$ are both sufficiently small, $T(\delta_{1},\delta_{2})$ may be empty. Hence it is necessary to give some conditions for $T(\delta_{1},\delta_{2})\ne\emptyset$. According to the proof of \cite[Theorem 3.1]{Xu2024Products}, we have Lemma~\ref{sec2.2 lem2}, which shows that $T(\delta_{1},\delta_{2})\ne\emptyset$ provided that either $\delta_{1}$ or $\delta_{2}$ is sufficiently large.

\begin{lemma}\label{sec2.2 lem2}
	Let $m\ge 2$ be an integer and $\Delta_1$ be the largest coset leader modulo $n$. If $\delta_{1}\ge2$ and $\delta_{2}\ge\Delta_{1}+1$, then $\mathcal{C}(\delta_{1},\delta_{2})\neq\mathbb{F}_{q}^{n}$. Specifically, $\mathcal{C}(\delta_{1},\delta_{2}) = \mathcal{C}_{1}$ in this case.
\end{lemma}
\begin{proof}
	Combining (\ref{sec2.2 equ1}), (\ref{sec2.2 equ2}) and (\ref{sec2.2 equ3}), it is enough to prove that
	\begin{equation*}
		T(2,\Delta_1+1)\neq\emptyset.
	\end{equation*}
	Suppose that $\delta_1=2$ and $\delta_2=\Delta_1+1$. Since $T_1=C_1$ and $T_2= C_{1}\cup C_{2}\cup\cdots\cup C_{\Delta_1}$, then $I_1=\mathbb{Z}_n\setminus C_1$ and $I_2=\mathbb{Z}_n\setminus(C_{1}\cup C_{2}\cup\cdots\cup C_{\Delta_1})=C_0=\{0\}$ as $\Delta_{1}$ is the largest coset leader modulo $n$. Hence, we have $I_1+I_2=I_1$, which implies $T(\delta_{1}, \delta_{2}) = T_1\ne\emptyset$ and $C(\delta_{1}, \delta_{2}) = C_1$. This completes the proof.
\end{proof}

\subsection{Largest coset leaders}

In this subsection, we introduce the first several largest coset leaders modulo $q^m+1$. Throughout this paper, we denote $\Delta_i$ the $i$-th largest coset leader modulo $q^m+1$. The results in the subsection are mainly from \cite{Liu2023DiemnsionNonbinary}.

We first provide the first and second largest coset leaders for $q$ odd.
\begin{lemma}\cite{Liu2023DiemnsionNonbinary}
	\label{sec2.3.1 lem1}
	Let $n=q^m+1$, where $q$ is an odd prime power and $m\ge2$ is an integer. Then $\Delta_{1}=\frac{q^m+1}{2}$ and
	\begin{equation*}
		\Delta_2=\left\{
		\begin{array}{ll}
			\frac{(q-1)(q^m+1)}{2(q+1)} & \mathrm{if}\ m\ \mathrm{is\ odd}, \\
			\frac{\left(q-1\right)\left(q^m+1\right)}{2\left(q^{2^{v_2\left(m\right)}}+1\right)}\prod_{j=0}^{v_2\left(m\right)-1}\left(q^{2^j}-1\right) & \mathrm{if}\ m\ \mathrm{is\ even}, 
		\end{array}
		\right.
	\end{equation*}
	where $v_2\left(m\right)$ is the largest integer such that $2^{v_2\left(m\right)}\ \Big|\  m$ but $2^{v_2\left(m\right)+1}\nmid m$. Moreover, $\left|C_{\Delta_1}\right|=1$, $\left|C_{\Delta_2}\right|=2$ if $m$ is odd, and
	\begin{equation*}
		\left|C_{\Delta_2}\right|=\left\{
		\begin{array}{ll}
			2m & \mathrm{if}\ m=2^{v_2(m)}, \\
			2^{v_2(m)+1} & \mathrm{otherwise}.
		\end{array}
		\right.
	\end{equation*}
\end{lemma}

Next, we consider the case of even $q$. Since the case of $m=3$ is not covered in \cite{Liu2023DiemnsionNonbinary}, we need the following proposition to provide it.

\begin{proposition}\cite[Proposition III.1]{Zhu2022AryAntiprimitive}
	\label{sec2.3.2 pro1}
	Let $m \ge 2$ and $q$ be an even prime power. Let $1\le i < m$, $l$ and $h$ be integers satisfying
	\begin{equation*}
		1\leq l\leq\left\lfloor\frac{(q^i-1)q}{2(q+1)}\right\rfloor 
	\end{equation*}
	and 
	\begin{equation*}
		-\frac{l(q^{m-i}-1)}{q^i+1}<h<\frac{l(q^{m-i}+1)}{q^i-1}.
	\end{equation*}
	Then $0\le a \le q^m$ is a coset leader modulo $q^m+1$ if and only if $0\leq a\leq\left\lfloor{\frac{q^{m+1}+q}{2(q+1)}}\right\rfloor$ and $a\neq lq^{m-i}+h.$
\end{proposition}

The following lemma provides the first largest coset leader for even $q$ and $m=3$.
\begin{lemma}\label{sec2.3.2 lem1}
	Let $m =3$ and $q>2$ be an even prime power. Then $\Delta_1=\frac{q(q^3+1)}{2(q+1)}$ is the largest $q$-cyclotomic coset leader modulo $q^m+1$. Further, $|C_{\Delta_{1}}|=2$.
\end{lemma}
\begin{proof}
	According to Proposition~\ref{sec2.3.2 pro1}, $0\le a \le q^3$ is a coset leader if and only if $0\leq a\leq\left\lfloor{\frac{q^{4}+q}{2(q+1)}}\right\rfloor$ and $a\neq lq^{3-i}+h,$ where $1\le i \le 2$, $1\leq l\leq\left\lfloor\frac{(q^i-1)q}{2(q+1)}\right\rfloor$ and $-\frac{l(q^{3-i}-1)}{q^i+1}<h<\frac{l(q^{3-i}+1)}{q^i-1}$.
	\begin{itemize}
		\item If $i=1$, we have $1\le l \le \frac{q}{2}-1$ and $-l(q-1)<h<\frac{l(q^{2}+1)}{q-1}$. Since
		\begin{equation*}
			\frac{q}{2}-1\cdot\frac{q^{2}+1}{q-1}=\frac{(q-2)(q+1)}{2}-\frac{1}{q-1}+1,
		\end{equation*}
		then $-\frac{(q-2)(q-1)}{2}+1\le h\le \frac{(q-2)(q+1)}{2}$. Thus the largest value of $a=lq^2+h$ is
		\begin{equation*}
			\frac{q}{2}-1\cdot q^2+\frac{(q-2)(q+1)}{2}=\frac{q^4-2q^2-3q-2}{2(q+1)}<\frac{q(q^3+1)}{2(q+1)}.
		\end{equation*}
		\item If $i=2$, we have $1\le l \le \frac{q(q-1)}{2}$ and $-\frac{l(q-1)}{q^2+1}<h<\frac{l}{q-1}$. Since
		\begin{equation*}
			\frac{q(q-1)}{2}\cdot\frac{(q-1)}{(q^{2}+1)}=\frac{q}{2}+\frac{1}{q^2+1}-1,
		\end{equation*}
		then $-\frac{q}{2}-1\le h\le \frac{q}{2}-1$. Thus the largest value of $a=lq+h$ is
		\begin{equation*}
			\frac{q(q-1)}{2}\cdot q+\frac{q}{2}-1=\frac{q(q^3+q)}{2(q+1)}-1<\frac{q(q^3+1)}{2(q+1)}.
		\end{equation*}
	\end{itemize}
	Therefore, $\frac{q(q^3+1)}{2(q+1)}$ is the largest $q$-cyclotomic coset leader modulo $q^3+1$. It is easily checked that $C_{\Delta_1}=\big\{\frac{q(q^3+1)}{2(q+1)},\frac{(q+2)(q^3+1)}{2(q+1)}\big\}$. Thus, $|C_{\Delta_{1}}|=2$.
\end{proof}

Combining Lemma~\ref{sec2.3.2 lem1} and \cite[Theorems~16 and 24]{Liu2023DiemnsionNonbinary}, we have the following result.
\begin{lemma}\label{sec2 lem2}
	Let $n=q^m+1$, where $q>2$ is an even prime power and $m\ge2$ is an integer. Then 
	\begin{equation*}
		\Delta_1=\left\{
		\begin{array}{ll}
			\frac{q(q^m+1)}{2(q+1)} & \mathrm{if}\ m\ \mathrm{is\ odd}, \\
			\frac{q\left(q^m+1\right)}{2\left(q^{2^{v_2\left(m\right)}}+1\right)}\prod_{j=0}^{v_2\left(m\right)-1}\left(q^{2^j}-1\right) & \mathrm{if}\ m\ \mathrm{is\ even},
		\end{array}
		\right.
	\end{equation*}
	where $v_2\left(m\right)$ is the largest integer such that $2^{v_2\left(m\right)}\ \Big|\  m$ but $2^{v_2\left(m\right)+1}\nmid m$. If $m\ge5$ is odd, then
	\begin{equation*}
		\Delta_2=\Delta_1-\frac{2\Delta_1+\left(q-1\right)q}{q^2}.
	\end{equation*}
\end{lemma}

\begin{remark}
	When $q$ and $m$ are even, the second largest coset leader modulo $q^m+1$ is not determined. However, our subsequent proofs do not require its specific value. Therefore, this will not affect the readability of our work.
\end{remark}

\section{Minimum coset leader in a set}

Throughout this section, let $n=q^m+1$, where $q$ is a prime power and $m\ge 2$ is an integer. In this section, we introduce a method for obtaining the minimum coset leader modulo $n$ in a set. Let $x$ be an integer with $0\leq x\leq q^{m}-1$ and 
\begin{equation*}
	x=x_{m-1}q^{m-1}+x_{m-2}q^{m-2}+\cdots+x_1q+x_0=(x_{m-1},x_{m-2},\ldots,x_{1},x_{0})_{q}
\end{equation*}
be the \textit{formal $q$-adic expansion} of $x$, where $0\leq x_{j}\leq q-1$ for $0\le j\le m-1$. Let $a,b$ be two integers with $q$-adic expansion $a=(a_{m-1},\dots,a_0)_q$ and $b=(b_{m-1},\dots,b_0)_q$. Note that $a\ge b$ if and only if $(a_{m-1},\dots,a_0)_q\ge (b_{m-1},\dots,b_0)_q$ if and only if there exists an integer $0\le k\le m-1$ such that $a_k\ge b_k$ and $a_j=b_j$ for $k+1\le j\le m-1$. 

If there exists an integer  $0\le j\le m-1$ such that $-(q-1)\le x_j<0$ or $q\le x_j\le 2q-1$, we denote its $q$-adic expansion as
\begin{equation*}
	(x_{m-1},x_{m-2},\ldots,x_{1},x_{0}),
\end{equation*}
and refer it to as an \textit{informal $q$-adic expansion} of $x$. Unlike its formal $q$-adic expansion, this expansion is not unique. For convenience, we denote $q^m=(q-1,\ldots,q-1,q)$ and $q^m+1=(q-1,\dots,q-1,q+1)$ henceforth. Suppose that $x\in \mathbb{Z}_n$, we next consider the coset leader $\CL(i)$ of $x$. It is easily seen that $\CL(0)=0$ and $\CL(q^m)=1$. 

Now suppose that $x=(x_{m-1},x_{m-2},\dots,x_{1},x_{0})_{q}\in\mathbb{Z}_n\setminus\{0,q^m\}$, where $0\le x_j\le q-1$ for $0\le j\le m-1$. Then
\begin{equation*}
	xq=x_{m-1}(q^m+1)+(x_{m-2},x_{m-3},\dots,x_0,-x_{m-1})\equiv (x_{m-2},x_{m-3},\ldots,x_0,-x_{m-1})\bmod n.
\end{equation*}
Assume that $0\le j\le m-1$ is the smallest integer such that $x_j\ge1$. 
\begin{itemize}
	\item If $x_{m-1}=0$, then
	\begin{equation*}
		xq\equiv (x_{m-2},x_{m-3},\ldots,x_0,0)_q\bmod n.
	\end{equation*}
	\item If $x_{m-1}>0$ and $j\le m-2$,
	\begin{equation*}
		xq\equiv (x_{m-2},x_{m-3},\dots,x_j-1,q-1,\dots,q-1,q-x_{m-1})_q\bmod n.
	\end{equation*}
	\item If $x_{m-1}>1$ and $j=m-1$,
	\begin{equation*}
		xq\equiv (0,\dots,0,-x_{m-1})+(q-1,\dots,q-1,q+1)\equiv (q-1,q-1,\dots,q-1,q-x_{m-1}+1)_q\bmod n.
	\end{equation*}
	\item If $x_{m-1}=1$ and $j=m-1$, then $x=q^{m-1}$, and so $xq=q^m$.
\end{itemize}

Let $n=q^m+1$ and $X$ be a subset of $\mathbb{Z}_n$, where $q>2$ is a prime power. Suppose that $x=(x_{m-1},\dots,x_1,x_0)_q\in X$ and $x_{m-1}\notin\{0,q-1\}$. According to the lengths of consecutive $0$'s and consecutive $(q-1)$'s in the $q$-ary expansion of $x$, we introduce the following definition.
\begin{itemize}
	\item Define $l_1(x):=\max\{0\le k\le m-1 \mid x_{k-1}=x_{k-2}=\dots=x_0=0\}$. Then let 
	$$X_1 := \{ x\in X \mid l_1(x) > 0  \},	\ L_1 := \max\limits_{x\in X_1} l_1(x), \text{\ and\ } S_1 := \{ x\in X_1 \mid l_1(x) = L_1 \}.$$
	Multiplying each $x \in S_1$ by $q^{2m-L_1}$ modulo $n$, we obtain
	$$R_1:=\{(\underbrace{0,\dots,0}_{L_1},x_{m-1},\dots,x_{L_1})_q \mid  x \in S_1\}.$$
	
	\item Define $l_2(x):=1+\max\{0\le k\le m-2 \mid x_{k}=x_{k-1}=\dots=x_1=0,x_0=1\}$ and set $l_{2}(x):=0$ if $x_0\ne1$. Then let 
	$$X_2 := \{ x\in X \ |\ l_2(x)>0 \},\ L_2 := \max\limits_{x\in X_2} l_2(x), \text{\ and\ } S_2 := \{ x\in X_2 \mid l_2(x) = L_2 \}.$$
	Multiplying each $x \in S_2$ by $q^{m-L_2}$ modulo $n$, we get
	$$R_2:=\{(\underbrace{0,\dots,0}_{L_2},q-1-x_{m-1},\dots,q-1-x_{L_2+1},q-x_{L_2})_q \mid  x \in S_2\}.$$
		
	\item For each $x\in X_1$ with $l_1(x)\le m-2$, define
	$l_{3}(x):=1+\max\{0\le k\le m-2-l_1(x) \mid x_{l_1(x)+k} = \cdots = x_{l_1(x)+1} = 0,x_{l_1(x)}=1\},$
	and set $l_{3}(x):=0$ if $x_{l_1(x)}\ne1$. Let
	\begin{equation*}
		X_3:=\{ x\in X_1 \ \mid\ l_3(x)>0 \},L_3 := \max\limits_{x\in X_3} l_3(x), \text{\ and\ } S_3 := \{ x\in X_3 \mid l_3(x) = L_3 \}.
	\end{equation*}
	Multiplying each $x \in S_3$ by $q^{m-l_1(x)-L_3}$ modulo $n$, we have
	$$R_3=\{(\underbrace{0,\dots,0}_{L_3},\underbrace{q-1,\dots,q-1}_{l_1(x)},q-1-x_{m-1},\dots,q-1-x_{l_1(x)+L_3+1},q-x_{l_1(x)+L_3})_q \mid x \in S_{3}\}.$$

	\item Define $l_4(x):=\max\{k\ge0 \mid \text{there exists an integer } 1\le e\le m-1-k \text{ such that }
	x_{e+k-1}=\cdots=x_{e}=0 \text{ and } x_{e-1}\neq0\}$, and for $l_4(x)>0$ let
	$$E(x):=\{1\le e\le m-1-l_4(x) \mid x_{e+l_4(x)-1}=\cdots=x_{e}=0,x_{e-1}\neq0\}.$$
	Then let
	\begin{align*}
		X_4 &:= \{ x\in X \mid l_4(x) > 0,\text{there exists } e\in E(x) \text{ such that } l_1(x)\ne e-1 \text{ or } x_{e-1}\ne1 \},\\
		&\quad L_4 := \max_{x\in X_4} l_4(x),\text{ and } S_4 := \{ x\in X_4 \mid l_4(x) = L_4 \}.
	\end{align*}
	Multiplying each $x \in S_4$ by $q^{m-e-L_4}$ modulo $n$ for every $e\in E(x)$, we get
	$$R_4:=\{(\underbrace{0,\dots,0}_{L_4},x_{e-1},\dots,x_0-1,
	q-1-x_{m-1},\dots,q-1-x_{e+L_4+1},q-x_{e+L_4}) \mid x \in S_4, e\in E(x) \}.$$

	\item Define $l_5(x):=\max\{0\le k \mid \text{there exists an integer } 1\le f\le m-1-k \text{ such that }
	x_{f+k-1}=\cdots=x_{f}=q-1 \text{ and an integer } 0\le t\le f-1 \text{ such that } x_t\neq0\}$, and for $l_5(x)>0$ let
	\begin{align*}
		F(x)&:=\{1\le f\le m-1-l_5(x) \mid x_{f+l_5(x)-1}=\cdots=x_{f}=q-1\\
		&\quad \text{ and there exists an integer } 0\le t\le f-1 \text{ such that } x_t\neq0\}.
	\end{align*}
	Then let
	\begin{equation*}
		X_5 := \{x\in X \mid l_5(x)>0 \}, L_5 := \max\limits_{x\in X_5} l_5(x)\text{ and }S_5 := \{ x\in X_5 \mid l_5(x)=L_5 \}.
	\end{equation*}
	Multiplying each $x \in S_5$ by $q^{2m-f-L_5}$ modulo $n$ for every $f\in F(x)$, we get
	$$R_5:=\Big\{(\underbrace{0,\dots,0}_{L_5},q-1-x_{f-1},\dots,q-1-x_1,q-x_0, x_{m-1},\dots,x_{f+L_5+1},x_{f+L_5}+1) \mid x \in S_5,\ f\in F(x)\Big\} $$
\end{itemize}
For $1\le j\le 5$, take $L_j := 0$ provided that $X_j=\emptyset$. 

According to the above preparations, we have the following lemma, which is important for the proof of the subsequent theorems.
\begin{lemma}\label{sec3 lem0}
	Let $n=q^m+1$ and $X$ be a subset of $\mathbb{Z}_n$, where $q>2$ is a prime power. Suppose that $x=(x_{m-1},\dots,x_1,x_0)_q\in X$ and $x_{m-1}\notin\{0,q-1\}$. Then $\min\limits_{x\in X}\CL(x)\ge q^{m-1-\max\limits_{1\le j\le 5}L_j}$. If $\max\limits_{1\le j\le 5}L_j>0$, then $\min\limits_{x\in X}\CL(x)=\min \bigcup\limits_{1\le j\le 5}R_j$.
\end{lemma}

\begin{proof}
	Suppose that $\min\limits_{x\in X}\CL(x)=y=(\underbrace{0,\dots,0}_L,y_{m-L-1},\dots,y_0)_q$, where $y_{m-L-1}y_0\ne0$. Since $\bigcup\limits_{1\le j\le 5}R_j$ is a subset of $\bigcup\limits_{x\in X}C_x$, $y=\min\limits_{x\in X}\CL(x)\le\min \bigcup\limits_{1\le j\le 5}R_j$. Since $y_{m-L-1}\ne0$, we have $y\ge q^{m-1-L}$. For each $j$ with $R_j\ne\emptyset$, every element $r\in R_j$ belongs to some $C_x$ with $x\in X$ and possesses $L_j$ leading zeros. Hence, $q^{m-1-L}\le y\le r<q^{m-L_j}$, and so $L_j\le L$. Therefore, we have $L\ge\max\limits_{1\le j\le 5}L_j$. If $L=0$, then $\max\limits_{1\le j\le 5}L_j=0$ and $y\ge q^{m-1}$, that is, $\min\limits_{x\in X}\CL(x)\ge q^{m-1-\max\limits_{1\le j\le 5}L_j}$. 
	
	 Next we consider the case of $L>0$. Assume that $l$ is the smallest integer such that $y_{m-L-1}=y_{m-L-2}=\dots=y_{l}=q-1$. We take $l:=m-L$ if $y_{m-L-1}\ne q-1$. Since $y_{m-L-1}\ne0$ and $y_0\ne0$, we have $0\le l\le m-L$. In particular, $l=0$ holds if and only if $y_{m-L-1}=y_{m-L-2}=\dots=y_0=q-1$, that is,
	 $y=q^{m-L}-1$. The displayed forms of $y\cdot q^k\bmod n$ below involve $y_{l-1}$, which is undefined when $l=0$; therefore the case $l=0$ is listed separately in the decomposition and treated separately in the case analysis below.
	 
	 The $q$-cyclotomic coset of $y$ modulo $n$ can be represented as
	\begin{align*}
		C_y = &\{y\cdot q^k \bmod n\ \Big|\  0\le k\le 2m-1\}\\
		=&\{(\underbrace{0,\dots,0}_{L-k},y_{m-L-1},\dots,y_0,\underbrace{0,\dots,0}_k)_q\ \Big|\  0\le k\le L\}\\
		&\cup\{(y_{m-1-k},\dots,y_1,y_0-1,\underbrace{q-1,\dots,q-1}_L,q-1-y_{m-L-1},\dots,q-1-y_{m-k+1},q-y_{m-k}) \\ 
		&\quad \Big|\  L+1\le k\le m-1\} \\
		&\cup\{(\underbrace{q-1,\dots,q-1}_L,q-1-y_{m-L-1},\dots,q-1-y_{1},q-y_{0}+1)\}\cup\\
		&\cup\{(\underbrace{q-1,\dots,q-1}_{L-(k-m)},q-1-y_{m-L-1},\dots,q-1-y_{1},q-y_{0},\underbrace{0,\dots,0}_{k-m-1},1)_q \Big|\  m+1\le k\le m+L\} \\
		&\cup\{(\underbrace{0,\dots,0}_{2m-k-1},1,\underbrace{0,\dots,0}_{L-1},1,\underbrace{0,\dots,0}_{k-m-L})_q \Big|\  l=0,\ m+L+1\le k\le 2m-1\} \\
		&\cup\{(\underbrace{0,\dots,0}_{2m-l-k},q-1-y_{l-1},\dots,q-1-y_{1},q-y_{0},\underbrace{0,\dots,0}_{L-1},1,\underbrace{0,\dots,0}_{k-m-L})_q \Big|\  l\ge1,\ m+L+1\le k\le 2m-l\} \\
		&\cup\{(q-1-y_{2m-k-1},\dots,q-1-y_1,q-y_0,\underbrace{0,\dots,0}_L,\underbrace{q-1,\dots,q-1}_{m-l-L},y_{l-1},\dots,y_{2m-k+1},y_{2m-k}+1) \\
		&\quad \Big|\  l\ge1,\ 2m-l+1\le k\le 2m-1\}.
	\end{align*}
	Since $y=\min\limits_{x\in X}\CL(x)$, there exists $z=(z_{m-1},\dots,z_0)_q\in X$ such that $y=\CL(z)$. Thus, $z\in C_y\cap X$. Note that $z_{m-1}\notin\{0,q-1\}$. 
	
	According to the forms of elements in $C_y$, we have:
	\begin{itemize}
		\item if $z\in\{y\cdot q^k \bmod n \mid k=L\}$, then $l_1(z)=k=L>0$. Hence, $z\in X_1$, which gives $L\le L_1$.
		\item if $z\in\{y\cdot q^k \bmod n \mid k= m+L\}$, then $l_2(z)=k-m=L>0$, and so $z\in X_2$ and $L\le L_2$.
		\item if $l=0$ and $z\in\{y\cdot q^k \bmod n \mid k= 2m-1\}$, then $l_1(z)=k-m-L>0$ and $l_3(z)=L>0$. Hence, $z\in X_3$ and $L\le L_3$.
		\item if $l\ge1$ and $z\in\{y\cdot q^k \bmod n \mid k= 2m-l\}$, then $l_1(z)=k-m-L>0$ and $l_3(z)=L>0$. Thus, $z\in X_3$ and $L\le L_3$.
		\item if $l\ge1$ and $z\in\{y\cdot q^k \bmod n \mid 2m-l+1\le k\le 2m-1\}$, then $e=k-m-L$ and $l_4(z)\ge L>0$. If $l_1(z)=e-1=k-m-L-1$, then $y_{m-L-2}=\dots=y_{2m-k}=q-1$, which implies that $z_{e-1}=y_{m-L-1}+1\ne1$. Hence, $z\in X_4$ and $L\le L_4$.
		\item if $z\in\{y\cdot q^k \bmod n \mid L+1\le k\le m-1\}$, then $l_5(z)\ge L>0$, and hence $z\in X_5$ and $L\le L_5$. 
	\end{itemize}   
	Therefore, we have $z\in \bigcup\limits_{1\le j\le 5}X_j$ and $L\le L_j\le\max\limits_{1\le j\le 5}L_j$ for some $j$. Hence, $L=\max\limits_{1\le j\le 5}L_j$, which indicates that $\min\limits_{x\in X}\CL(x)=y\ge q^{m-1-L}=q^{m-1-\max\limits_{1\le j\le 5}L_j}$.
	
	Note that there exists an index $j$ such that $z\in X_j$ and $L\le l_j(z)$; since $z\in X_j$ we also have $l_j(z)\le L_j$. Thus, if $L=\max\limits_{1\le j\le 5}L_j>0$, then $L_j\ge l_j(z)\ge L=\max\limits_{1\le j\le 5}L_j\ge L_j$, which forces $l_j(z)=L_j$, i.e., $z\in S_j\subseteq \bigcup\limits_{1\le j\le 5}S_j$. For $1\le j\le 5$, we get $y\in R_j$ by the construction of $R_j$. In summary, $y\in \bigcup\limits_{1\le j\le 5}R_j$, and so $y\ge\min \bigcup\limits_{1\le j\le 5}R_j$. Therefore, we have $\min\limits_{x\in X}\CL(x)=y=\min \bigcup\limits_{1\le j\le 5}R_j$. The proof is completed.
\end{proof}

According to Lemmas~\ref{sec2.2 lem2} and \ref{sec3 lem0}, we can investigate the products of two narrow-sense antiprimitive BCH codes.

\section{Products of narrow-sense antiprimitive BCH codes}
In this section, we give some conditions for $\mathcal{C}(\delta_{1},\delta_{2})\neq\mathbb{F}_{q}^{n}$, where $\mathcal{C}(\delta_{1},\delta_{2})$ is the product of two narrow-sense antiprimitive BCH codes $\mathcal{C}_1$ and $\mathcal{C}_2$. Moreover, we study the parameters of some products.

\subsection{The existence of $\Omega_{(q,m)}$}
Recall from Lemmas~\ref{sec2.3.1 lem1} and \ref{sec2 lem2} that $\Delta_{1}$ and $\Delta_{2}$ are the first and second largest coset leaders modulo $n$, respectively. We prove the following theorem on the existence of $\Omega_{(q,m)}$ for any $q>2$.

\begin{theorem}\label{sec3.2 thm0}
	Let $n=q^m+1$, where $q>2$ is a prime power and $m\ge 2$ is an integer. Then there exists an integer $\Omega_{(q,m)}$ such that 
	
	i) when $2\le\delta_{1}\le\Omega_{(q,m)}$, $\mathcal{C}(\delta_{1},\delta_{2})\neq\mathbb{F}_{q}^{n}$ if and only if $\delta_{2}\ge\Delta_{1}+1$. 
	
	ii) when $\delta_{1}>\Omega_{(q,m)}$, if $\delta_{2}\ge\Delta_{2}+1$, then $\mathcal{C}(\delta_{1},\delta_{2})\neq\mathbb{F}_{q}^{n}$. 
\end{theorem}
\begin{proof}
	Define $S :=\{\Omega\in\{2,\ldots,n\}\mid T(\Omega,\Delta_{1})=\emptyset\}$. First, we show that $S\ne\emptyset$. Actually, $2\in S$, i.e., $T(2,\Delta_{1})=\emptyset$. Suppose that $\delta_{1}=2,\delta_{2}=\Delta_1$. Then $T_1=C_{1}$ and $T_2=C_{1}\cup C_{2}\cup\dots\cup C_{\Delta_1-1}$, and so $I_1=\mathbb{Z}_n\setminus C_1$ and $I_2=C_0\cup C_{\Delta_1}$. Then we have $I_1+I_2=(\mathbb{Z}_n\setminus C_1)\cup(\mathbb{Z}_n\setminus C_1+C_{\Delta_1})$. We only need to prove that $C_1\subseteq(\mathbb{Z}_n\setminus C_1+C_{\Delta_1})$, which indicates that $I_1+I_2=\mathbb{Z}_n$ and $T(2,\Delta_{1})=\emptyset$. Note that
	\begin{equation*}
		C_{\Delta_1}=\{\Delta_{1},\Delta_1\cdot q,\dots,\Delta_1\cdot q^{m-1}, -\Delta_{1},-\Delta_1\cdot q,\dots,-\Delta_1\cdot q^{m-1}\}
	\end{equation*}
	and
	\begin{equation*}
		C_{\Delta_1+1}=\{-\Delta_1-1,-\Delta_1\cdot q-q,\dots,-\Delta_1\cdot q^{m-1}-q^{m-1},\Delta_{1}+1,\Delta_1\cdot q+q,\dots,\Delta_1\cdot q^{m-1}+q^{m-1}\}.
	\end{equation*}
	It is easily verified that $C_1\subseteq(C_{\Delta_1+1}+C_{\Delta_1})$. We next prove that $C_{\Delta_1+1}\subseteq\mathbb{Z}_n\setminus C_1$, which is equivalent to $C_{\Delta_1+1}\ne C_1$. Otherwise, there exists an integer $0\le j\le m-1$ such that $\Delta_{1}=q^j-1$ or $q^m-q^j$. By the expression of $\Delta_{1}$ (see Lemmas~\ref{sec2.3.1 lem1} and \ref{sec2 lem2}), $\Delta_{1}$ cannot be of the form $q^j-1$ or $q^m-q^j$ since $q>2$. Thus, $C_1\ne C_{\Delta_{1}+1}$, and so $C_1\subseteq(\mathbb{Z}_n\setminus C_1+C_{\Delta_1})$. Hence, $S\ne\emptyset$. 
	
	Let $\Omega_{(q,m)}$ be the largest integer of $S$. Now we show that $\Omega_{(q,m)}$ satisfies i) and ii).
	
	i) The sufficiency is a direct consequence of Lemma~\ref{sec2.2 lem2}. For necessity, we have $T\left(\Omega_{(q,m)},\Delta_1\right)=\emptyset$ by the definition of $\Omega_{(q,m)}$. Then for $2\le\delta_{1}\le\Omega_{(q,m)}$ and $2\le\delta_2\le\Delta_{1}$, $\mathcal{C}(\delta_{1},\delta_{2})=\mathbb{F}_{q}^{n}$. Equivalently, if $2\le\delta_{1}\le\Omega_{(q,m)}$ and $\mathcal{C}(\delta_{1},\delta_{2})\ne\mathbb{F}_{q}^{n}$, then $\delta_{2}\ge\Delta_{1}+1$. 
	
	ii) By the definition of $\Omega_{(q,m)}$, we have $T(\delta_{1},\Delta_1)=T(\delta_{1},\Delta_2+1)\ne\emptyset$ for $\delta_{1}>\Omega_{(q,m)}$. By (\ref{sec2.2 equ1}), (\ref{sec2.2 equ2}) and (\ref{sec2.2 equ3}), if $\delta_{2}\ge\Delta_{2}+1$, then $\mathcal{C}(\delta_{1},\delta_{2})\ne\mathbb{F}_{q}^{n}$. The proof is completed.
\end{proof}

\begin{remark}
	For $q$ odd, we give the value of $\Omega_{(q,m)}$ (see Theorem~\ref{sec3 thm1}). For $q$ even, we have not found a unified expression for $\Omega_{(q,m)}$ through our investigation. When $2\le m\le 11$, we present the specific value of $\Omega_{(q,m)}$ (see Table~\ref{sec3.2 thm1 table1}).
\end{remark}

\subsection{$q$ odd}\label{sec3.1}

Recall from Lemma~\ref{sec2.3.1 lem1} that $\Delta_{1}$ and $\Delta_{2}$ are the first and second largest coset leaders modulo $n$, respectively. Then we have $\Delta_{1}=\frac{n}{2}$ and the following theorem for $q$ odd.
\begin{theorem}\label{sec3 thm1}
	Let $n=q^m+1$, where $q$ is an odd prime power and $m\ge 2$ is an integer. Suppose that $\Omega_{(q,m)}=\frac{q^{\lfloor\frac{m+1}{2}\rfloor}-q^{\mu(m)}}{2}+1$, where $\mu(m)$ equals $1$ if $m$ is odd, and $0$ if $m$ is even. Then 
	
	i) when $2\le\delta_{1}\le\Omega_{(q,m)}$, $\mathcal{C}(\delta_{1},\delta_{2})\neq\mathbb{F}_{q}^{n}$ if and only if $\delta_{2}\ge\Delta_{1}+1$. 
	
	ii) when $\delta_{1}>\Omega_{(q,m)}$, if $\delta_{2}\ge\Delta_{2}+1$, then $\mathcal{C}(\delta_{1},\delta_{2})\neq\mathbb{F}_{q}^{n}$. 
\end{theorem}
\begin{proof}
	By Theorem~\ref{sec3.2 thm0}, we need to prove that 
	\begin{equation}\label{sec3 thm1 equ4}
		T\left(\Omega_{(q,m)},\Delta_1\right)=\emptyset.
	\end{equation}
	and
	\begin{equation}\label{sec3 thm1 equ5}
		T\left(\Omega_{(q,m)}+1,\Delta_1\right)=T\left(\Omega_{(q,m)}+1,\Delta_2+1\right)\ne\emptyset.
	\end{equation}
	
	We first prove (\ref{sec3 thm1 equ4}). Now suppose that $\delta_{1}=\Omega_{(q,m)}$ and $\delta_{2}=\Delta_1$, then $T_1=C_{1}\cup C_{2}\cup\cdots\cup C_{\Omega_{(q,m)}-1}$ and $T_2=C_{1}\cup C_{2}\cup\cdots\cup C_{\Delta_1-1}$, and so $I_1=\mathbb{Z}_n\setminus T_1$ and $I_2=\mathbb{Z}_n\setminus T_2=C_0\cup C_{\Delta_1}=\{0,\frac{n}{2}\}$. Note that~(\ref{sec3 thm1 equ4}) is equivalent to 
	\begin{equation*}
		I_1+I_2=\mathbb{Z}_n.
	\end{equation*}
	It is obvious that $I_{1}+I_{2}\subseteq\mathbb{Z}_{n}$. Next we prove that $\mathbb{Z}_{n}\subseteq I_{1}+I_{2}$, it is sufficient to show that $i\in I_{1}+I_{2}$ for any coset leader $i\in\mathbb{Z}_n$.
	\begin{itemize}
		\item If $i \in I_1$, then $i = i+0\in I_1+I_2$.
		\item If $i\in\mathbb{Z}_n\setminus I_1=T_1$, we have
		\begin{equation*}
			i=(\underbrace{0,0,\ldots,0}_{m-r-1},i_r\ne0,i_{r-1},\ldots,i_1,i_0\ne0)_q.
		\end{equation*}
		Note that $i\equiv(\frac{n}{2}+i)+\frac{n}{2}\bmod n$. Since $\frac{n}{2}\in I_2$, we next show that $\frac{n}{2}+i \in I_1$ for any coset leader $i \in T_1$. With the same notation as in Lemma~\ref{sec3 lem0}, take $a:=\frac{n}{2}$. Then 
		\begin{equation*}
			a+i=\frac{n}{2}+i=\Big(\underbrace{\frac{q-1}{2},\dots,\frac{q-1}{2}}_{m-r-1},\frac{q-1}{2}+i_r,\dots,\frac{q-1}{2}+i_1,\frac{q+1}{2}+i_0\Big).
		\end{equation*}
		Actually, we only need to prove that $\min\limits_{i\in T_1}\CL(a+i)\ge\delta_{1}$.
		\begin{itemize}
			\item If $m$ is even, we have \begin{equation*}
				\delta_1=\frac{q^{\frac{m}{2}}+1}{2}=\Big(\underbrace{0,\ldots,0}_{\frac{m}{2}},\frac{q-1}{2},\dots,\frac{q-1}{2},\frac{q+1}{2}\Big)_q
			\end{equation*}
			and $r\le \frac{m}{2}-1$ as $i<\delta_1$. Through calculation, $\delta_1-1$ is a coset leader, and so the largest coset leader in $T_1$ is 
			\begin{equation*}
				\delta_1-1=\Big(\underbrace{0,\ldots,0}_{\frac{m}{2}},\frac{q-1}{2},\dots,\frac{q-1}{2}\Big)_q.
			\end{equation*}
			\begin{itemize}
				\item If $m=2$, then $a+i=(\frac{q-1}{2},\frac{q+1}{2}+i_0)$. Since $i\le (0,\frac{q-1}{2})_q$ and $i_0\ne0$, it is easily checked that  \begin{equation*}
					\min\limits_{i\in T_1}\CL(a+i)=\CL\Big(\Big(\frac{q-1}{2},\frac{q+1}{2}+\frac{q-1}{2}\Big)\Big)=\Big(0,\frac{q+1}{2}\Big)_q=\frac{q^{\frac{m}{2}}+1}{2}=\delta_{1}.
				\end{equation*}
				
				\item If $m\ge4$, take $X := a + T_1$. By Lemma~\ref{sec3 lem0}, $L_1=\frac{m}{2},L_2=L_5=\frac{m}{2}-1$, and $L_3,L_4\le \frac{m}{2}-2$. Then $\max\limits_{1\le j\le 5}L_j=L_1=\frac{m}{2}$, and so $\min\limits_{i\in T_1}\CL(a+i)=\min\limits_{x\in X}\CL(x)=\min R_1$. Note that 
				\begin{align*}
					S_1=&\Big\{\Big(\underbrace{\frac{q-1}{2},\dots,\frac{q-1}{2}}_{\frac{m}{2}},\frac{q-1}{2}+i_{\frac{m}{2}-1},\dots,\frac{q-1}{2}+i_1,\frac{q+1}{2}+i_0\Big)\ \Big|\\
					&\quad i_{\frac{m}{2}-1}=\dots=i_0=\frac{q-1}{2}\Big\}\\
					=&\Big\{\Big(\underbrace{\frac{q-1}{2},\dots,\frac{q-1}{2}}_{\frac{m}{2}-1},\frac{q+1}{2},\underbrace{0,\ldots,0}_{\frac{m}{2}}\Big)\Big\}.
				\end{align*} 
				Hence, we have 
				\begin{align*}
					&\min\limits_{i\in T_1}\CL(a+i)=\min R_1=\min\{x\cdot q^{2m-\frac{m}{2}}\mid x\in S_1\}\\
					=&\Big(\underbrace{\vphantom{\frac{q-1}{2}}0,\ldots,0}_{\frac{m}{2}},\underbrace{\frac{q-1}{2},\dots,\frac{q-1}{2}}_{\frac{m}{2}-1},\frac{q+1}{2}\Big)_q=\frac{q^{\frac{m}{2}}+1}{2}=\delta_{1}.
				\end{align*}
				\end{itemize}
				\item If $m$ is odd, we get 
				\begin{equation*}
				\delta_1=\frac{q^{\frac{m+1}{2}}-q}{2}+1=\Big(\underbrace{\vphantom{\frac{q-1}{2}}0,\ldots,0}_{\frac{m-1}{2}},\underbrace{\frac{q-1}{2},\dots,\frac{q-1}{2}}_{\frac{m-1}{2}},1\Big)_q
				\end{equation*}
				and $r\le \frac{m-1}{2}$ as $i<\delta_1$. By calculation, $\delta_1-1$ is not a coset leader, whereas $\delta_1-2$ is. Hence, the largest coset leader in $T_1$ is
				\begin{equation*}
				\delta_{1}-2=\Big(\underbrace{\vphantom{\frac{q-1}{2}}0,\ldots,0}_{\frac{m-1}{2}},\underbrace{\frac{q-1}{2},\dots,\frac{q-1}{2}}_{\frac{m-3}{2}},\frac{q-3}{2},q-1\Big)_q.
				\end{equation*}
				Take $X := a + T_1$. By Lemma~\ref{sec3 lem0}, we have $L_1=L_2=L_5=\frac{m-1}{2}$ and $L_3,L_4\le \frac{m-3}{2}$. Then $\max\limits_{1\le j\le 5}L_j=\frac{m-1}{2}$, and so $\min\limits_{i\in T_1}\CL(a+i)=\min\limits_{x\in X}\CL(x)=\min (R_1\cup R_2\cup R_5)$. Note that
				\begin{align*}
				S_1=&\Big\{\Big(\underbrace{\frac{q-1}{2},\dots,\frac{q-1}{2}}_{\frac{m-1}{2}},\frac{q-1}{2}+i_{\frac{m-1}{2}},\dots,\frac{q-1}{2}+i_1,\frac{q+1}{2}+i_0\Big)\ \Big|\\
				&\quad i_{\frac{m-3}{2}}=\dots=i_0=\frac{q-1}{2},i_{\frac{m-1}{2}}\le \frac{q-3}{2}\Big\}\\
				=&\Big\{\Big(\underbrace{\frac{q-1}{2},\dots,\frac{q-1}{2}}_{\frac{m-1}{2}},\frac{q+1}{2}+i_{\frac{m-1}{2}},\underbrace{0,\ldots,0}_{\frac{m-1}{2}}\Big)\ \Big|\ i_{\frac{m-1}{2}}\le \frac{q-3}{2}\Big\},\\
				S_2=&\Big\{\Big(\underbrace{\frac{q-1}{2},\dots,\frac{q-1}{2}}_{\frac{m-1}{2}},\frac{q-1}{2}+i_{\frac{m-1}{2}},\dots,\frac{q-1}{2}+i_1,\frac{q+1}{2}+i_0\Big)\ \Big|\\
				&\quad i_{\frac{m-3}{2}}=\dots=i_1=\frac{q-1}{2},i_0=\frac{q+1}{2},i_{\frac{m-1}{2}}\le \frac{q-3}{2}\Big\}\\
				=&\Big\{\Big(\underbrace{\frac{q-1}{2},\dots,\frac{q-1}{2}}_{\frac{m-1}{2}},\frac{q+1}{2}+i_{\frac{m-1}{2}},\underbrace{0,\ldots,0}_{\frac{m-3}{2}},1\Big)\ \Big|\ i_{\frac{m-1}{2}}\le \frac{q-3}{2}\Big\},\ \mathrm{and}\\
				S_5=&\Big\{\Big(\underbrace{\frac{q-1}{2},\dots,\frac{q-1}{2}}_{\frac{m-1}{2}},\frac{q-1}{2}+i_{\frac{m-1}{2}},\dots,\frac{q-1}{2}+i_1,\frac{q+1}{2}+i_0\Big)\ \Big|\\
				&\quad i_{\frac{m-1}{2}}=\dots=i_2=\frac{q-1}{2},i_1=\frac{q-3}{2},\frac{q+1}{2}\le i_0\le q-1\Big\}\\
				=&\Big\{\Big(\underbrace{\frac{q-1}{2},\dots,\frac{q-1}{2}}_{\frac{m-1}{2}},\underbrace{q-1,\dots,q-1}_{\frac{m-1}{2}},i_0-\frac{q-1}{2}\Big)\ \Big|\ \frac{q+1}{2}\le i_0\le q-1\Big\}.
				\end{align*}
				Hence, we have
				\begin{align*}
				\min R_1=&\min\{x\cdot q^{2m-\frac{m-1}{2}}\mid x\in S_1\}\\
				=&\min\Big\{\Big(\underbrace{\vphantom{\frac{q-1}{2}}0,\ldots,0}_{\frac{m-1}{2}},\underbrace{\frac{q-1}{2},\dots,\frac{q-1}{2}}_{\frac{m-1}{2}},\frac{q+1}{2}+i_{\frac{m-1}{2}}\Big)\ \Big|\ i_{\frac{m-1}{2}}\le \frac{q-3}{2}\Big\}\\
				=&\Big(\underbrace{\vphantom{\frac{q-1}{2}}0,\ldots,0}_{\frac{m-1}{2}},\underbrace{\frac{q-1}{2},\dots,\frac{q-1}{2}}_{\frac{m-1}{2}},\frac{q+1}{2}\Big)_q>\frac{q^{\frac{m+1}{2}}-q}{2}+1=\delta_{1},\\
				\min R_2=&\min\{x\cdot q^{m-\frac{m-1}{2}}\mid x\in S_2\}\\
				=&\min\Big\{\Big(\underbrace{\vphantom{\frac{q-1}{2}}0,\ldots,0}_{\frac{m-1}{2}},\underbrace{\frac{q-1}{2},\dots,\frac{q-1}{2}}_{\frac{m-1}{2}},\frac{q-1}{2}-i_{\frac{m-1}{2}}\Big)\ \Big|\ i_{\frac{m-1}{2}}\le \frac{q-3}{2}\Big\}\\
				=&\Big(\underbrace{\vphantom{\frac{q-1}{2}}0,\ldots,0}_{\frac{m-1}{2}},\underbrace{\frac{q-1}{2},\dots,\frac{q-1}{2}}_{\frac{m-1}{2}},\frac{q-1}{2}-\frac{q-3}{2}\Big)_q=\frac{q^{\frac{m+1}{2}}-q}{2}+1=\delta_{1},\ \mathrm{and}\ \\
				\min R_5=&\min\{x\cdot q^{2m-f-\frac{m-1}{2}}\mid x\in S_5,f=1\}\\
				=&\min\Big\{\Big(\underbrace{\vphantom{\frac{q-1}{2}}0,\ldots,0}_{\frac{m-1}{2}},\frac{3q-1}{2}-i_0,\underbrace{\frac{q-1}{2},\dots,\frac{q-1}{2}}_{\frac{m-3}{2}},\frac{q+1}{2}\Big)\ \Big|\ \frac{q+1}{2}\le i_0\le q-1\Big\}\\
				=&\Big(\underbrace{\vphantom{\frac{q-1}{2}}0,\ldots,0}_{\frac{m-1}{2}},\frac{3q-1}{2}-(q-1),\underbrace{\frac{q-1}{2},\dots,\frac{q-1}{2}}_{\frac{m-3}{2}},\frac{q+1}{2}\Big)_q>\frac{q^{\frac{m+1}{2}}-q}{2}+1=\delta_{1}.
			\end{align*}
			That is, $\min\limits_{i\in T_1}\CL(a+i)=\frac{q^{\frac{m+1}{2}}-q}{2}+1=\delta_{1}$.
		\end{itemize}
		Therefore, we have $a+i\in I_1$, and so $n+i=(a+i)+(n-a) \in I_1+I_2$ for any $i\in T_1$.
	\end{itemize}
	Thus, $\Z \subseteq I_1 + I_2$.
	
	We then prove (\ref{sec3 thm1 equ5}). Suppose that $\delta_{1}=\Omega_{(q,m)}+1$ and $\delta_{2}=\Delta_2+1$, then $I_2=C_0\cup C_{\Delta_1}$. If $m$ is even, then $\delta_1=\frac{q^{\frac{m}{2}}+3}{2}$ and $I_1=\Z\setminus T_1=\Z\setminus\left(C_1\cup\dots\cup C_{\frac{q^{\frac{m}{2}}+1}{2}}\right)$. Then we will prove $\frac{q^{\frac{m}{2}}-1}{2}\in T\left(\delta_{1},\delta_{2}\right)$. Suppose on the contrary that $\frac{q^{\frac{m}{2}}-1}{2} \in I_1+I_2$, then there exist $x\in I_1$ and $y\in I_2$ such that
	\begin{equation*}
		x+y=\frac{q^{\frac{m}{2}}-1}{2}\ \text{or}\ x+y=\frac{q^{\frac{m}{2}}-1}{2}+n
	\end{equation*}
	since $0\leq x+y<2n$. If $y=0$, then $x=\frac{q^{\frac{m}{2}}-1}{2}\in T_1=\mathbb{Z}_n\setminus I_1$, a contradiction. Thus, we have $y=\Delta_1=\frac{n}{2}$ and $x=\frac{n}{2}+\frac{q^{\frac{m}{2}}-1}{2}$. Note that $\CL(x)=\frac{q^{\frac{m}{2}}+1}{2}<\frac{q^{\frac{m}{2}}+3}{2}$, which contradicts $x\in I_1$. Similarly, we have $\frac{q^{\frac{m+1}{2}}+1}{2}-q^{\frac{m-1}{2}}\in T\left(\Omega_{(q,m)}+1,\Delta_2+1\right)$ if $m$ is odd. Therefore, $T\left(\Omega_{(q,m)}+1,\Delta_2+1\right)\ne\emptyset$. The proof is completed.
\end{proof}

\begin{remark}
	When $\delta_{1}$ is small, Theorem~\ref{sec3 thm1} i) presents a necessary and sufficient condition for $\mathcal{C}(\delta_{1},\delta_{2})\ne\mathbb{F}_{q}^{n}$. When $\delta_{1}$ is larger, Theorem~\ref{sec3 thm1} ii) shows that $\mathcal{C}(\delta_{1},\delta_{2})\ne\mathbb{F}_{q}^{n}$ for some smaller $\delta_{2}$ (i.e., $\delta_{2}\ge\Delta_{2}+1$). By Lemma~\ref{sec2.3.1 lem1}, one can see that $\Delta_{2}$ is complicated. Therefore, we are not yet able to provide the necessary and sufficient conditions for $\mathcal{C}(\delta_{1},\delta_{2})\ne\mathbb{F}_{q}^{n}$ when $\delta_{1}>\Omega_{(q,m)}$.
\end{remark}

When $2\le\delta_{1}\le\Omega_{(q,m)}$ and $\delta_{2}\ge\Delta_{1}+1$, it follows from Lemma~\ref{sec2.2 lem2} that $\mathcal{C}(\delta_{1},\delta_{2})=\mathcal{C}_1$ is a narrow-sense antiprimitive BCH code over $\mathbb{F}_q$. It has been studied extensively for several years. 
\begin{example}
	When $2\le\delta_{1}\le\Omega_{(q,m)}$ and $\delta_{2}=\Delta_{1}+1$, we present the parameters of some products $\mathcal{C}(\delta_{1},\delta_{2})$ by Magma (see Table~\ref{sec3.1 thm1 table1}). Compared with the best known parameters in \cite{Grassl2026Bounds}, one can see that they have good parameters.
\end{example}

\begin{table}[!ht]
	\centering
	\caption{The parameters of $\mathcal{C}(\delta_{1},\delta_{2})$}\label{sec3.1 thm1 table1}
	\begin{tabular}{|c|c|c|c|c|c|}
		\hline
		$q$ & $m$ & $\delta_{1}$ & $\delta_{2}$ & Parameters of $\mathcal{C}(\delta_{1},\delta_{2})$ & Best known parameters \\ \hline
		$3$ & $2$ & $2$ & $6$ & $[10,6,2]$ & $[10,6,4]$\\ \hline
		$3$ & $3$ & $4$ & $15$ & $[28,16,4]$ & $[28,16,7]$\\ \hline
		$3$ & $4$ & $5$ & $42$ & $[82,58,9]$ & $[82,58,10]$\\ \hline
		$5$ & $2$ & $3$ & $14$ & $[26,18,5]$ & $[26,18,6]$ \\ \hline
		$5$ & $3$ & $4$ & $64$ & $[126,108,6]$ & $[126,108,8]$ \\ \hline
		$7$ & $2$ & $3$ & $26$ & $[50,42,5]$ & $[50,42,6]$ \\ \hline
		$9$ & $2$ & $3$ & $42$ & $[82,74,5]$ & $[82,74,6]$ \\ \hline
	\end{tabular}
\end{table}

When $\delta_{1}\ge\Omega_{(q,m)}+1$ and $\delta_{2}\ge\Delta_{1}+1$, $\mathcal{C}(\delta_{1},\delta_{2})=\mathcal{C}_1$ by Lemma~\ref{sec2.2 lem2}. When $\delta_{1}>\Omega_{(q,m)}+1$ and $\Delta_{2}+1\le\delta_{2}\le \Delta_{1}$, it is complicated to consider the parameters of $\mathcal{C}(\delta_{1},\delta_{2})$. We give the case of $\delta_{1}=\Omega_{(q,m)}+1$ and $\Delta_{2}+1\le\delta_{2}\le \Delta_{1}$ in the following theorem.

\begin{theorem}\label{sec3 thm3}
	Let $q$ be an odd prime power and $m\ge 2$ be an integer. Let $\Omega_{(q,m)}=\frac{q^{\lfloor\frac{m+1}{2}\rfloor}-q^{\mu(m)}}{2}+1$ be as in Theorem~\ref{sec3 thm1}. If $\delta_{1}=\Omega_{(q,m)}+1$ and $\Delta_{2}+1\le\delta_{2}\le \Delta_{1}$, then $\mathcal{C}(\delta_{1},\delta_{2})=\mathcal{C}(\delta_{1},\Delta_{1})$ has parameters $[q^m+1,q^m+1-4m,d]$, where $d\ge2$ if $m$ is odd, and $d\ge3$ if $m$ is even.
\end{theorem}
\begin{proof}		
	Suppose that $\delta_{1}=\Omega_{(q,m)}+1$ and $\delta_{2}=\Delta_{2}+1$. 
	\begin{itemize}
		\item If $m$ is odd, then $\delta_{1}=\frac{q^{\frac{m+1}{2}}-q}{2}+2$, and so $$I_1=\Z\setminus T_1=\Z\setminus\left(C_1\cup\dots\cup C_{\delta_{1}-1}\right) \mathrm{\ and\ }I_2=C_0\cup C_{\Delta_1}.$$ By the proof of Theorem~\ref{sec3 thm1}, $\Z=\left(I_1\cup C_{\delta_{1}-1}\right)+I_2$. Obviously,  $T(\delta_1,\delta_2)=\Z\setminus(I_1+I_2)\subseteq C_{\delta_{1}-1}+ I_2$. Next we will prove that $C_{\delta_{1}-1}+ I_2\subseteq T(\delta_1,\delta_2)$. Otherwise, there exist two integers $x\in I_1$ and $y\in I_2$ such that $i\equiv x+y \bmod n$ for some $i \in C_{\delta_{1}-1}+I_2$. Note that $C_{\delta_{1}-1}+I_2=(C_{\delta_{1}-1}+C_0)\cup(C_{\delta_{1}-1}+C_{\Delta_1})=C_{\delta_{1}-1}\cup C_{\delta_{1}-1+\frac{n}{2}}$. For $i \in C_{\delta_{1}-1}$, we have $\CL(i)=\delta_{1}-1=\frac{q^{\frac{m+1}{2}}-q}{2}+1\in T_1$. If $y=0\in C_0$, then $\CL(x)=\CL(i)\in T_1$, a contradiction. If $y=\frac{n}{2}\in C_{\Delta_1}$, then $\CL(x)=\CL(i+\frac{n}{2})=\frac{q^{\frac{m+1}{2}}+1}{2}-q^{\frac{m-1}{2}}\in T_1$, which is impossible. Similarly, we have $i \notin I_1+I_2$ for $i \in C_{\delta_{1}-1+\frac{n}{2}}$. Hence, 
		$$|T(\delta_1,\delta_2)|=|\Z\setminus(I_1+I_2)|=|C_{\delta_{1}-1}\cup C_{\delta_{1}-1+\frac{n}{2}}|.$$	
		It is easy to see that $\CL(\delta_{1}-1)=\frac{q^{\frac{m+1}{2}}-q}{2}+1\ne\frac{q^{\frac{m+1}{2}}+1}{2}-q^{\frac{m-1}{2}}=\CL(\delta_{1}-1+\frac{n}{2})$, and $|C_{\delta_{1}-1}|=|C_{\delta_{1}-1+\frac{n}{2}}|=2m$. The dimension of $\mathcal{C}(\delta_{1},\delta_{2})$ is $q^m+1-4m$. The longest	consecutive set in $T(\delta_1,\delta_2)$ has size 1, then we have $d\ge2$ by the BCH bound. 
		\item If $m$ is even, then $\delta_{1}=\frac{q^{\frac{m}{2}}+3}{2}$. Similar to the statement above, we only need to show that $\CL(i+\frac{n}{2})\in T_1$ for any $i \in C_{\delta_{1}-1+\frac{n}{2}}$. In fact, $\CL(i+\frac{n}{2})=\CL(\frac{q^{\frac{m}{2}}+1}{2}+\frac{n}{2})=\frac{q^{\frac{m}{2}}-1}{2}\in T_1$. Hence, 
		$$|T(\delta_1,\delta_2)|=|\Z\setminus(I_1+I_2)|=|C_{\delta_{1}-1}\cup C_{\delta_{1}-1+\frac{n}{2}}|.$$	
		The dimension of $\mathcal{C}(\delta_{1},\delta_{2})$ is $q^m+1-4m$. Note that $\{\frac{q^{\frac{m}{2}}-1}{2},\frac{q^{\frac{m}{2}}+1}{2}\}$ is one of the longest	consecutive set in $T(\delta_1,\delta_2)$, then we have $d\ge3$ by the BCH bound. 
	\end{itemize}
	This completes the proof.
\end{proof}

\begin{remark}
	For $(q,m)=(3,3)$, the minimum distance of $\mathcal{C}(\delta_{1},\delta_{2})$ in Theorem~\ref{sec3 thm3} is $2$, which implies that the lower bound on the minimum distance in Theorem~\ref{sec3 thm3} is tight when $m$ is odd. However, when $m$ is even, we did not find an example whose minimum distance is $3$.
\end{remark}

\begin{example}
	When $\delta_{1}=\Omega_{(q,m)}+1$ and $\delta_{2}=\Delta_{1}$, the parameters of $\mathcal{C}(\delta_{1},\delta_{2})$ obtained using Magma are listed in Table~\ref{sec3 thm3 table1}.
\end{example}

\begin{table}[!ht]
	\centering
	\caption{The parameters of $\mathcal{C}(\delta_{1},\delta_{2})$}\label{sec3 thm3 table1}
	\begin{tabular}{|c|c|c|c|c|c|}
		\hline
		$q$ & $m$ & $\delta_{1}$ & $\delta_{2}$ & Parameters of $\mathcal{C}(\delta_{1},\delta_{2})$ & Best known parameters \cite{Grassl2026Bounds} \\ \hline
		$3$ & $2$ & $3$ & $5$ & $[10,2,5]$ & $[10,2,7]$\\ \hline
		$3$ & $3$ & $5$ & $14$ & $[28,16,2]$ & $[28,16,7]$\\ \hline
		$3$ & $4$ & $6$ & $41$ & $[82,66,5]$ & $[82,66,8]$\\ \hline
		$5$ & $2$ & $4$ & $13$ & $[26,18,4]$ & $[26,18,6]$ \\ \hline
		$5$ & $3$ & $12$ & $63$ & $[126,114,3]$ & $[126,114,6]$ \\ \hline
		$7$ & $2$ & $5$ & $25$ & $[50,42,4]$ & $[50,42,6]$ \\ \hline
		$9$ & $2$ & $6$ & $41$ & $[82,74,4]$ & $[82,74,6]$ \\ \hline
	\end{tabular}
\end{table}

\subsection{$q$ even}\label{sec3.2}

For $2\le m\le 11$, the values of $\Omega_{(q,m)}$ satisfying the conditions of Theorem~\ref{sec3.2 thm0} are listed in Table~\ref{sec3.2 thm1 table1}.
\begin{theorem}\label{sec3.2 thm1}
	Let $n=q^m+1$, where $q>2$ is an even prime power and $2\le m\le 11$ is an integer. Suppose that $\Omega_{(q,m)}$ is as in Table~\ref{sec3.2 thm1 table1}.
	\begin{table}[!ht]
		\centering
		\caption{The value of $\Omega_{(q,m)}$}\label{sec3.2 thm1 table1}
		\begin{tabular}{|c|c|c|c|c|c|}
			\hline
			$m$   & 2 & 3 & 4 & 5 & 6 \\ \hline
			$\Omega_{(q,m)}$  & $\frac{q}{2}+1$ & $\frac{q^2-q}{2}$ & $\frac{q^3-q^2-q}{2}$ & $\frac{2q^3-3q^2+q}{2}-1$ & $\frac{2q^4-2q^3-q^2+q}{2}$  \\ \hline
			$m$   & 7 & 8 & 9 & 10 & 11 \\ \hline
			$\Omega_{(q,m)}$  & $\frac{q^5-q^4+2q}{2}-1$ & $\frac{2q^7-3q^3+q}{2(q+1)}$ & $\frac{q^7-2q^3+q}{2(q+1)}$ & $\frac{2q^7-2q^6+q^5+q^4+q^3-2q^2-2q}{2}+1$ & $\frac{2q^8-q^5-2q^4+q}{2(q+1)}-1$ \\ \hline
		\end{tabular}
	\end{table}
	Then 
	
	i) when $2\le\delta_{1}\le\Omega_{(q,m)}$, $\mathcal{C}(\delta_{1},\delta_{2})\neq\mathbb{F}_{q}^{n}$ if and only if $\delta_{2}\ge\Delta_{1}+1$. 
	
	ii) when $\delta_{1}>\Omega_{(q,m)}$, if $\delta_{2}\ge \Delta_{2}+1$, then $\mathcal{C}(\delta_{1},\delta_{2})\neq\mathbb{F}_{q}^{n}$. 
\end{theorem}
\begin{proof}[Proof of Theorem~\ref{sec3.2 thm1}]
	See Appendix~\ref{apdx a}.
\end{proof}

\begin{example}
	When $(q,m)=(4,2),(4,3)$ or $(8,2)$, we present the parameters of some products $\mathcal{C}(\delta_{1},\delta_{2})$ with $2\le\delta_{1}\le\Omega_{(q,m)}$ and $\delta_{2}=\Delta_{1}+1$ by Magma (see Table~\ref{sec3.2 thm1 table2}). They are either the best known or 1 away from the best known.
\end{example}

\begin{table}[!ht]
	\centering
	\caption{The parameters of $\mathcal{C}(\delta_{1},\delta_{2})$}\label{sec3.2 thm1 table2}
	\begin{tabular}{|c|c|c|c|c|c|}
		\hline
		$q$ & $m$ & $\delta_{1}$ & $\delta_{2}$ & Parameters of $\mathcal{C}(\delta_{1},\delta_{2})$ & Best known parameters \cite{Grassl2026Bounds} \\ \hline
		$4$ & $2$ & $2$ & $7$ & $[17,13,4]$ & $[17,13,4]$\\ \hline
		$4$ & $3$ & $2$ & $27$ & $[65,59,3]$ & $[65,59,4]$\\ \hline
		$4$ & $3$ & $3$ & $27$ & $[65,53,5]$ & $[65,53,6]$\\ \hline
		$4$ & $3$ & $6$ & $27$ & $[65,41,11]$ & $[65,41,12]$\\ \hline
		$8$ & $2$ & $2$ & $29$ & $[65,61,4]$ & $ [65,61,4]$\\ \hline
		$8$ & $2$ & $3$ & $29$ & $[65,57,5]$ & $[65,57,5]$ \\ \hline
	\end{tabular}
\end{table}

\begin{remark}
	According to Table~\ref{sec3.2 thm1 table1}, the values of $\Omega_{(q,m)}$ cannot be unified into a expression in terms of $q$ and $m$. Moreover, the proof becomes more complex as $m$ increases. Hence, when $q$ is even, the value of $\Omega_{(q,m)}$ is not provided for any $m$.
\end{remark}

For any odd $m\ge5$ and $2\le\delta_{1}\le\Lambda_{(q,m)}$, the following theorem gives a necessary and sufficient condition for $\mathcal{C}(\delta_{1},\delta_{2})\neq\mathbb{F}_{q}^{n}$.	

\begin{theorem}\label{sec3.2 thm2}
	Let $n=q^m+1$, where $q>2$ is an even prime power and $m\ge5$ is an odd integer. Recall from Lemma~\ref{sec2 lem2} that $\Delta_{1}$ is the largest coset leader modulo $n$. Suppose that $\Lambda_{(q,m)}$ is equal to $\frac{q^{\frac{m+3}{2}}+2q^{\frac{m+1}{2}}+2q^{\frac{m-1}{2}}-q}{2(q+1)}$ if $\frac{m-1}{2}$ is odd, and $\frac{q^{\frac{m+3}{2}}+2q^{\frac{m+1}{2}}+2q^{\frac{m-1}{2}}-2}{2(q+1)}$ if $\frac{m-1}{2}$ is even. Then when $2\le\delta_{1}\le\Lambda_{(q,m)}$, $\mathcal{C}(\delta_{1},\delta_{2})\neq\mathbb{F}_{q}^{n}$ if and only if $\delta_{2}\ge\Delta_{1}+1$. Specifically, if $2\le\delta_{1}\le\Lambda_{(q,m)}$ and $\mathcal{C}(\delta_{1},\delta_{2})\neq\mathbb{F}_{q}^{n}$, then $\mathcal{C}(\delta_{1},\delta_{2}) = \mathcal{C}_{1}$.
\end{theorem}
\begin{proof}
	See Appendix~\ref{apdx b}.
\end{proof}

For any $m\equiv2\bmod 4$, $m\ge6$ and $2\le\delta_{1}\le\Lambda_{(q,m)}$, we present a necessary and sufficient condition for $\mathcal{C}(\delta_{1},\delta_{2})\neq\mathbb{F}_{q}^{n}$ in the following theorem.	

\begin{theorem}\label{sec3.2 thm3}
	Let $n=q^m+1$, where $q>2$ is an even prime power, $m\ge6$ and $m\equiv2\bmod 4$. Recall from Lemma~\ref{sec2 lem2} that $\Delta_{1}$ is the largest coset leader modulo $n$. Suppose that $\Lambda_{(q,m)}$ is equal to $\frac{q^{\frac{m}{2}+3}-q^{\frac{m}{2}+2}+2q^{\frac{m}{2}+1}+2q^{\frac{m}{2}-1}+q^2-q}{2(q^2+1)}$ if $\frac{m}{2}\equiv1\bmod 4$, and $\frac{q^{\frac{m}{2}+3}-q^{\frac{m}{2}+2}+2q^{\frac{m}{2}+1}+2q^{\frac{m}{2}-1}+q^2+q+2}{2(q^2+1)}$ if $\frac{m}{2}\equiv3\bmod 4$. Then when $2\le\delta_{1}\le\Lambda_{(q,m)}$, $\mathcal{C}(\delta_{1},\delta_{2})\neq\mathbb{F}_{q}^{n}$ if and only if $\delta_{2}\ge\Delta_{1}+1$. Specifically, if $2\le\delta_{1}\le\Lambda_{(q,m)}$ and $\mathcal{C}(\delta_{1},\delta_{2})\neq\mathbb{F}_{q}^{n}$, then $\mathcal{C}(\delta_{1},\delta_{2}) = \mathcal{C}_{1}$.
\end{theorem}
\begin{proof}
	See Appendix~\ref{apdx c}.
\end{proof}

\begin{remark}
	Note that $\Omega_{(q,m)}\ge\Lambda_{(q,m)}$. That is, when $\delta_{1}>\Lambda_{(q,m)}$ and $\Delta_{2}+1\le\delta_{2}\le\Delta_{1}$, it cannot be concluded that $\mathcal{C}(\delta_{1},\delta_{2})\neq\mathbb{F}_{q}^{n}$. Hence, the conditions in Theorem~\ref{sec3.2 thm2} are weaker than those in Theorem~\ref{sec3.2 thm1}. The case of $m\equiv0\bmod4$ is more complex; we have not been able to derive a unified result.
\end{remark}

When $\delta_{1}\ge\Omega_{(q,m)}+1$ and $\delta_{2}\ge\Delta_{1}+1$, $\mathcal{C}(\delta_{1},\delta_{2})=\mathcal{C}_1$ by Lemma~\ref{sec2.2 lem2}. When $\delta_{1}>\Omega_{(q,m)}+1$, $\Delta_{2}+1\le\delta_{2}\le \Delta_{1}$ and $m\ge6$, it is no simple task to consider the parameters of $\mathcal{C}(\delta_{1},\delta_{2})$. When $q$ is even and $2\le m\le 5$, the cases where $\delta_{1}=\Omega_{(q,m)}+1$ and $\Delta_{2}+1\le\delta_{2}\le \Delta_{1}$ are given as follows.

\begin{theorem}\label{sec3.2 thm4}
	Let $q>2$ be an even prime power and $2\le m\le 5$ be an integer. Let $\Omega_{(q,m)}$ be as in Table~\ref{sec3.2 thm1 table1}. If $\delta_{1}=\Omega_{(q,m)}+1$ and $\Delta_{2}+1\le\delta_{2}\le\Delta_{1}$, then $\mathcal{C}(\delta_{1},\delta_{2})=\mathcal{C}(\delta_{1},\Delta_{1})$ has parameters $[q^m+1,q^m+1-2m,d]$, where $d\ge3$ if $m=2$, and $d\ge2$ if $3\le m\le5$.
\end{theorem}
\begin{proof}		
	Suppose that $\delta_{1}=\Omega_{(q,m)}+1$ and $\delta_{2}=\Delta_{2}+1$. 
	\begin{itemize}
		\item If $m=2$, then $\delta_{1}=\frac{q}{2}+2$, and so $$I_1=\Z\setminus T_1=\Z\setminus\left(C_1\cup\dots\cup C_{\delta_{1}-1}\right) \mathrm{\ and\ }I_2=C_0\cup C_{\Delta_1}.$$ By the proof of Theorem~\ref{sec3.2 thm1}, $\Z=\left(I_1\cup C_{\delta_{1}-1}\right)+I_2$. Obviously, $T(\delta_1,\delta_2)=\Z\setminus(I_1+I_2)\subseteq C_{\delta_{1}-1}+ I_2$. By computing,
		\begin{align*}
				&C_{\delta_{1}-1}+ I_2=C_{\delta_{1}-1}+ (C_0+C_{\Delta_1})=(C_{\delta_{1}-1}+C_0)\cup(C_{\delta_{1}-1}+C_{\Delta_1})\\
				=&C_{\frac{q}{2}-1}\cup C_{\frac{q}{2}}\cup C_{\frac{q}{2}+1}\cup C_{\frac{3q}{2}-1}\cup C_{\frac{3q}{2}}.
			\end{align*}
		Since $\frac{3q}{2}>\frac{3q}{2}-1>\delta_1$, $C_{\frac{3q}{2}-1}\cup C_{\frac{3q}{2}}=(C_{\frac{3q}{2}-1}\cup C_{\frac{3q}{2}})+C_0\in I_1+I_2$. For $C_{\frac{q}{2}-1}$ (or $C_{\frac{q}{2}+1}$), there exists an integer $a\in C_{\Delta_{1}}$ such that $\CL(a+\frac{q}{2}-1)$ (or $\CL(a+\frac{q}{2}+1$) $\ge\delta_1$, which implies that $C_{\frac{q}{2}-1}$ (or $C_{\frac{q}{2}+1}$) $\in I_1+I_2$. For $C_{\frac{q}{2}}$, we have $\max\limits_{a\in C_{\Delta_1}}\CL(a+\frac{q}{2})=\frac{q}{2}+1<\delta_1$. That is, $C_{\frac{q}{2}}\notin I_1+I_2$, and so $$T(\delta_1,\delta_2)=C_{\frac{q}{2}}=\Big\{\Big(0,\frac{q}{2}\Big), \Big(\frac{q}{2},0\Big), \Big(q-1,\frac{q}{2}+1\Big), \Big(\frac{q}{2},1\Big)\Big\}.$$
		Hence, $|T(\delta_1,\delta_2)|=4=2m$ and $d\ge3$ by the BCH bound.
		
		\item If $m=3$, then $\delta_{1}=\frac{q^2-q}{2}+1$, and so $$I_1=\Z\setminus T_1=\Z\setminus\left(C_1\cup\dots\cup C_{\delta_{1}-1}\right) \mathrm{\ and\ }I_2=C_0\cup C_{\Delta_1}.$$ By the proof of Theorem~\ref{sec3.2 thm1}, $\Z=\left(I_1\cup C_{\delta_{1}-1}\right)+I_2$. Obviously, $T(\delta_1,\delta_2)=\Z\setminus(I_1+I_2)\subseteq C_{\delta_{1}-1}+ I_2$. By computing,
		\begin{align*}
				C_{\delta_{1}-1}+ I_2=C_{\delta_{1}-1}+ (C_0+C_{\Delta_1})=(C_{\delta_{1}-1}+C_0)\cup(C_{\delta_{1}-1}+C_{\Delta_1})
				=C_{\frac{q}{2}}\cup C_{\frac{q^2}{2}-q+1}\cup C_{\frac{q^2-q}{2}}.
			\end{align*}
		For $C_{\frac{q}{2}}$ (or $C_{\frac{q^2}{2}-q+1}$), there exists an integer $a\in C_{\Delta_{1}}$ such that $\CL(a+\frac{q}{2})$ (or $\CL(a+\frac{q^2}{2}-q+1$) $\ge\delta_1$, which implies that $C_{\frac{q}{2}}$ (or $C_{\frac{q^2}{2}-q+1}$) $\in I_1+I_2$. For $C_{\frac{q^2-q}{2}}$, we have $\max\limits_{a\in C_{\Delta_1}}\CL(a+\frac{q^2-q}{2})=\frac{q^2}{2}-q+1<\delta_1$. That is, $C_{\frac{q^2-q}{2}}\notin I_1+I_2$, and so $$T(\delta_1,\delta_2)=C_{\frac{q^2-q}{2}}.$$ Hence, $|T(\delta_1,\delta_2)|=6=2m$ and $d\ge2$ by the BCH bound.
		
		\item If $m=4$, then $\delta_{1}=\frac{q^3-q^2-q}{2}+1$, and so $$I_1=\Z\setminus T_1=\Z\setminus\left(C_1\cup\dots\cup C_{\delta_{1}-1}\right) \mathrm{\ and\ }I_2=C_0\cup C_{\Delta_1}.$$ By the proof of Theorem~\ref{sec3.2 thm1}, $\Z=\left(I_1\cup C_{\delta_{1}-1}\right)+I_2$. Obviously, $T(\delta_1,\delta_2)=\Z\setminus(I_1+I_2)\subseteq C_{\delta_{1}-1}+ I_2$. By computing,
		\begin{align*}
				&C_{\delta_{1}-1}+ I_2=C_{\delta_{1}-1}+ (C_0+C_{\Delta_1})=(C_{\delta_{1}-1}+C_0)\cup(C_{\delta_{1}-1}+C_{\Delta_1})\\
				=&C_{\frac{q^3-q^2-q}{2}}\cup C_{\frac{q^2}{2}-1}\cup C_{\frac{2q^2-q}{2}}\cup C_{\frac{q^2-2q}{2}+1}\cup C_{\frac{q}{2}}\cup C_{\frac{2q^2-q}{2}+1}\cup C_{\frac{q^2+2q}{2}-1}\cup C_{\frac{2q^3+q}{2}}\cup C_{\frac{q^3-2q^2+2q}{2}+1}.
			\end{align*}
		Similar to the case where $m=3$, one can check that $$T(\delta_1,\delta_2)=C_{\frac{q^3-q^2-q}{2}}.$$ Hence, $|T(\delta_1,\delta_2)|=8=2m$ and $d\ge2$ by the BCH bound.
		
		\item If $m=5$, then $\delta_{1}=\frac{2q^3-3q^2+q}{2}$, and so $$I_1=\Z\setminus T_1=\Z\setminus\left(C_1\cup\dots\cup C_{\delta_{1}-1}\right) \mathrm{\ and\ }I_2=C_0\cup C_{\Delta_1}.$$ By the proof of Theorem~\ref{sec3.2 thm1}, $\Z=\left(I_1\cup C_{\delta_{1}-1}\right)+I_2$. Obviously, $T(\delta_1,\delta_2)=\Z\setminus(I_1+I_2)\subseteq C_{\delta_{1}-1}+ I_2$. By computing, $C_{\delta_{1}-1}+ I_2$ equals
		\begin{align*}
				C_{\frac{2q^3-3q^2+q}{2}}\cup C_{\frac{q^3+q^2+q}{2}-1}\cup C_{\frac{3q^3+q^2+q}{2}-1}
			\end{align*}
		if $q=4$, and
		\begin{align*}
				C_{\frac{2q^3-3q^2+q}{2}}\cup C_{\frac{q^3+q^2+q}{2}-1}\cup C_{\frac{3q^4-q^3+3q^2-4q}{2}+2}
			\end{align*}
		if $q>4$. Similar to the case where $m=3$, one can check that $$T(\delta_1,\delta_2)=C_{\frac{q^3+q^2+q}{2}-1}.$$ Hence, $|T(\delta_1,\delta_2)|=10=2m$ and $d\ge2$ by the BCH bound.
		
	\end{itemize}
	This completes the proof.
\end{proof}

\begin{example}
	For $(q,m)=(4,2),(4,3)$ or $(8,2)$, we use Magma to obtain the parameters of $\mathcal{C}(\delta_{1},\delta_{2})$ $\delta_{1}=\Omega_{(q,m)}+1$ and $\delta_{2}=\Delta_{1}$, as listed in Table~\ref{sec3.2 thm4 table1}. Note that these codes have the same parameters as those in Table~\ref{sec3.2 thm1 table2}. However, for $(q,m)=(4,3)$ or $(8,2)$, it has been verified by Magma that they are different codes.
\end{example}

\begin{table}[!ht]
	\centering
	\caption{The parameters of $\mathcal{C}(\delta_{1},\delta_{2})$}\label{sec3.2 thm4 table1}
	\begin{tabular}{|c|c|c|c|c|c|}
		\hline
		$q$ & $m$ & $\delta_{1}$ & $\delta_{2}$ & Parameters of $\mathcal{C}(\delta_{1},\delta_{2})$ & Best known parameters \cite{Grassl2026Bounds} \\ \hline
		$4$ & $2$ & $4$ & $6$ & $[17,13,4]$ & $[17,13,4]$\\ \hline
		$4$ & $3$ & $7$ & $26$ & $[65,59,3]$ & $[65,59,4]$\\ \hline
		$8$ & $2$ & $6$ & $28$ & $[65,61,4]$ & $[65,61,4]$\\ \hline
	\end{tabular}
\end{table}

\section{Conclusion}
In this paper, we investigated the Schur products $\mathcal{C}(\delta_{1},\delta_2)$ of two nonbinary narrow-sense antiprimitive BCH codes $\mathcal{C}_1$ and $\mathcal{C}_2$. Our main results are as follows:
\begin{itemize}
	\item For $q$ odd with $m\ge2$, or $q$ even and $q>2$ with $2\le m\le11$, a necessary and sufficient condition for $\mathcal{C}(\delta_{1},\delta_{2})\ne \mathbb{F}_q^n$ is established when $2\le\delta_{1}\le\Omega_{(q,m)}$ and $\delta_{2}\ge\Delta_{1}+1$ (see Theorems~\ref{sec3 thm1} and \ref{sec3.2 thm1}).
	\item For $q$ even, $q>2$ and $m\not\equiv0\bmod 4$, a necessary and sufficient condition for $\mathcal{C}(\delta_{1},\delta_{2})\ne \mathbb{F}_q^n$ is obtained when $2\le\delta_{1}\le\Lambda_{(q,m)}$ and $\delta_{2}\ge\Delta_{1}+1$ (see Theorems~\ref{sec3.2 thm2} and~\ref{sec3.2 thm3}).
	\item For $q$ odd with $m\ge2$, or $q$ even and $q>2$ with $2\le m\le5$, the dimension of $\mathcal{C}(\delta_{1},\delta_{2})$ and a lower bound on its minimum distance are presented when $\delta_{1}=\Omega_{(q,m)}+1$ and $\Delta_{2}+1\le\delta_{2}\le\Delta_{1}$ (see Theorems~\ref{sec3 thm3} and \ref{sec3.2 thm4}).
\end{itemize}
Moreover, several linear codes with the best known parameters are given.

For the products of nonbinary antiprimitive BCH codes, we have studied some cases as discussed above. For the binary case, however, Lemma~6 is not directly applicable. Hence, it is necessary to find a new approach for determining the minimum coset leader modulo $2^m+1$ of all elements in a given set.

\appendix

\section{\texorpdfstring{Proof of Theorem~\ref{sec3.2 thm1}}{Proof of Theorem 4}}\label{apdx a}

\begin{proof}
	According to Theorem~\ref{sec3.2 thm0}, we need to prove that 
	\begin{equation}\label{sec3.2 thm1 equ1}
		T\left(\Omega_{(q,m)},\Delta_1\right)=\emptyset.
	\end{equation}
	and
	\begin{equation}\label{sec3.2 thm1 equ2}
		T\left(\Omega_{(q,m)}+1,\Delta_1\right)=T\left(\Omega_{(q,m)}+1,\Delta_2+1\right)\ne\emptyset.
	\end{equation}
	
	We first prove (\ref{sec3.2 thm1 equ1}). Suppose that $\delta_{1}=\Omega_{(q,m)}$ and $\delta_{2}=\Delta_1$, then $T_1=C_{1}\cup C_{2}\cup\cdots\cup C_{\Omega_{(q,m)}-1}$ and $T_2=C_{1}\cup C_{2}\cup\cdots\cup C_{\Delta_1-1}$, and so $I_1=\mathbb{Z}_n\setminus T_1$ and $I_2=C_0\cup C_{\Delta_1}$. Note that (\ref{sec3.2 thm1 equ1}) is equivalent to 
	\begin{equation*}
		I_1+I_2=\mathbb{Z}_n.
	\end{equation*}
	It is obvious that $I_{1}+I_{2}\subseteq\mathbb{Z}_{n}$. Next we prove that $\mathbb{Z}_{n}\subseteq I_{1}+I_{2}$, it is sufficient to show that $i\in I_{1}+I_{2}$ for any coset leader $i\in\mathbb{Z}_n$.
	\begin{itemize}
		\item If $i \in I_1$, then $i = i+0\in I_1+I_2$.
		\item If $i\in\mathbb{Z}_n\setminus I_1=T_1$,
			\begin{itemize}
			\item $\boldsymbol{m=2:}$ we have $i=(0,i_0\ne0)_q$. Note that $\delta_1=\left(0,\frac{q}{2}+1\right)_q$. It is easily seen that the largest coset leader in $T_1$ is $\left(0,\frac{q}{2}\right)_q$. Note that 
			\begin{equation*}
				C_{\Delta_1}=\left\{\left(\frac{q}{2}-1,\frac{q}{2}\right)_q,\left(\frac{q}{2}-1,\frac{q}{2}+1\right)_q,\left(\frac{q}{2},\frac{q}{2}+1\right)_q,\left(\frac{q}{2},\frac{q}{2}\right)_q\right\}.
			\end{equation*}
			Let
			\begin{equation*}
				a+i=\left(\frac{q}{2},\frac{q}{2}\right)_q+i=\left(\frac{q}{2},\frac{q}{2}+i_0\right)_q
			\end{equation*}
			and
			\begin{equation*}
				n-a=\left(\frac{q}{2}-1,\frac{q}{2}+1\right)_q.
			\end{equation*}
			One can easily check that $\min\limits_{i\in T_1}\CL(a+i)=\left(0,\frac{q}{2}+1\right)_q=\delta_{1}$. Thus, we have $a+i \in I_1$ and $i \in I_1+I_2$ for any $i \in T_1$.
			
			\item $\boldsymbol{m=3:}$ we have
			\begin{equation*}
				i=(\underbrace{0,0,\ldots,0}_{m-r-1},i_r\ne0,i_{r-1},\ldots,i_1,i_0\ne0)_q,
			\end{equation*}
			where $0\leq r\leq1$ as $$i<\delta_{1}=\frac{q^2-q}{2}=\big(0,\frac{q}{2}-1,\frac{q}{2}\big)_q.$$ With the same notation as in Lemma~\ref{sec3 lem0}, take $X := a + T_1$. We will show that there always exists an integer $a\in C_{\Delta_1}$ such that $a+i \in I_1$ for any $i \in T_1$, which implies that $i\equiv(a+i)+(n-a)\bmod n$ and $i\in I_1+I_2$ since $n-a\in I_2$. Actually, we only need to prove that there always exists an integer $a\in C_{\Delta_1}$ such that $\min\limits_{i\in T_1}\CL(a+i)\ge\delta_{1}$. Through calculation, $\delta_1-1$ is a coset leader. Hence,the largest coset leader in $T_1$ is $$\delta_{1}-1=\big(0,\frac{q}{2}-1,\frac{q}{2}-1\big)_q.$$ 
			\begin{itemize}
				\item When $i_0 \ge \frac{q}{2}+1$, take $a:=n-\Delta_1$. Let
				\begin{equation*}
					a+i=n-\Delta_1+i=\left(\frac{q}{2},\frac{q}{2}-1+i_1,\frac{q}{2}+1+i_0\right)
				\end{equation*}
				and
				\begin{equation*}
					n-a=\Delta_{1}=\left(\frac{q}{2}-1,\frac{q}{2},\frac{q}{2}\right)_q.
				\end{equation*}
				Since $i_0\ge\frac{q}{2}+1$, $2\le x_0 \le \frac{q}{2}$ and $\frac{q}{2} \le x_1 \le q-2$ for any $x\in X$. Then $\max\limits_{1\le j\le 5}L_j=0$. By Lemma~\ref{sec3 lem0}, we have $\CL(x)\ge q^2>\delta_{1}$. Therefore, we have $a+i\in I_1$, and so $n+i=(a+i)+(n-a)$ and $i \in I_1+I_2$.
				\item When $i_0=\frac{q}{2}$, take $a:=\Delta_1$. Let
				\begin{equation*}
					a+i=\Delta_1+i=\left(\frac{q}{2}-1,\frac{q}{2}+1+i_1,0\right)_q
				\end{equation*}
				and
				\begin{equation*}
					n-a=n-\Delta_{1}=\left(\frac{q}{2},\frac{q}{2}-1,\frac{q}{2}+1\right)_q.
				\end{equation*}
				Since $i_0=\frac{q}{2}$, $i_1 \le \frac{q}{2}-2$, and so $\frac{q}{2}+1\le x_1 \le q-1$ for any $x\in X$. Then $L_1=1$ and $L_2=\dots=L_5=0$. Hence $\max\limits_{1\le j\le 5}L_j=L_1=1$, and so $\min\limits_{i\in T_1}\CL(a+i)=\min\limits_{x\in X}\CL(x)=\min R_1$. Note that
				\begin{equation*}
					S_1=\Big\{\Big(\frac{q}{2}-1,\frac{q}{2}+1+i_1,0\Big)\ \Big|\ i_1\le \frac{q}{2}-2\Big\}.
				\end{equation*}
				Hence, we have
				\begin{align*}
					\min R_1=&\min\{x\cdot q^{2m-1}\mid x\in S_1\}=\min\Big\{\Big(0,\frac{q}{2}-1,\frac{q}{2}+1+i_1\Big)\ \Big|\ i_1\le \frac{q}{2}-2\Big\}\\
					=&\Big(0,\frac{q}{2}-1,\frac{q}{2}+1\Big)_q>\delta_1.
				\end{align*}
				Thus, $a+i\in I_1$, which implies $i \in I_1+I_2$.
				\item When $i_0\le\frac{q}{2}-1$, take $a:=n-\Delta_1$. Let
				\begin{equation*}
					a+i=n-\Delta_1+i=\left(\frac{q}{2},\frac{q}{2}-1+i_1,\frac{q}{2}+1+i_0\right)
				\end{equation*}
				and
				\begin{equation*}
					n-a=\Delta_{1}=\left(\frac{q}{2}-1,\frac{q}{2},\frac{q}{2}\right)_q.
				\end{equation*}
				Since $i_0\le\frac{q}{2}-1$ and $i_1\le\frac{q}{2}-1$, we have $x_0=0$ or $\frac{q}{2}+1\le x_0\le q-1$, and $\frac{q}{2}-1\le x_1\le q-1$. Thus, $L_1=1$ and $L_2=\dots=L_5=0$. Hence $\max\limits_{1\le j\le 5}L_j=L_1=1$, and so $\min\limits_{i\in T_1}\CL(a+i)=\min\limits_{x\in X}\CL(x)=\min R_1$. Note that
				\begin{align*}
					S_1=&\Big\{\Big(\frac{q}{2},\frac{q}{2}-1+i_1,\frac{q}{2}+1+i_0\Big)\ \Big|\ i_0=\frac{q}{2}-1,\ i_1\le\frac{q}{2}-1\Big\}\\
					=&\Big\{\Big(\frac{q}{2},\frac{q}{2}+i_1,0\Big)\ \Big|\ i_1\le\frac{q}{2}-1\Big\}.
				\end{align*}
				Hence, we have
				\begin{align*}
					\min R_1=&\min\{x\cdot q^{2m-1}\mid x\in S_1\}=\min\Big\{\Big(0,\frac{q}{2},\frac{q}{2}+i_1\Big)\ \Big|\ i_1\le\frac{q}{2}-1\Big\}\\
					=&\Big(0,\frac{q}{2},\frac{q}{2}\Big)_q>\delta_{1}.
				\end{align*}
				Hence, $a+i\in I_1$ and $i \in I_1+I_2$.
			\end{itemize}
			To sum up, we have $a+i \in I_1$ and $n+i=(a+i)+(n-a)\in I_1+I_2$ for any $i \in T_1$.
			
			\item $\boldsymbol{m=4:}$ we have
			\begin{equation*}
				i=(\underbrace{0,0,\ldots,0}_{m-r-1},i_r\ne0,i_{r-1},\ldots,i_1,i_0\ne0)_q,
			\end{equation*}
			where $0\leq r\leq 2$ as $$i<\delta_{1}=\frac{q^3-q^2-q}{2}=\left(0,\frac{q}{2}-1,\frac{q}{2}-1,\frac{q}{2}\right)_q$$. With the same notation as in Lemma~\ref{sec3 lem0}, take $X := a + T_1$. We only need to prove that there always exists an integer $a\in C_{\Delta_1}$ such that $\min\limits_{i\in T_1}\CL(a+i)\ge\delta_{1}$. Through calculation, $\delta_1-1$ is a coset leader. Hence,the largest coset leader in $T_1$ is
			\begin{equation*}
				\delta_1-1=\left(0,\frac{q}{2}-1,\frac{q}{2}-1,\frac{q}{2}-1\right)_q.
			\end{equation*}
			\begin{itemize}
				\item When $i_1\ne\frac{q}{2}-1$ and $i_0\le \frac{q}{2}-1$, let
				\begin{equation*}
					a+i=\left(\frac{q}{2}-1,\frac{q}{2}-1+i_2,\frac{q}{2}+i_1,\frac{q}{2}+i_0\right)
				\end{equation*}
				and
				\begin{equation*}
					n-a=\left(\frac{q}{2},\frac{q}{2},\frac{q}{2}-1,\frac{q}{2}+1\right)_q,
				\end{equation*}
				then $\frac{q}{2}+1\le x_0\le q-1$ and $x_1\ne q-1$ for any $x \in X$. Since $i_2\le \frac{q}{2}-2$ if $i_1\ge\frac{q}{2}$, $x_2\notin\{0,q-1\}$. Hence $\max\limits_{1\le j\le 5}L_j=L_4=1$, and so $\min\limits_{i\in T_1}\CL(a+i)=\min\limits_{x\in X}\CL(x)=\min R_4$. Note that
				\begin{align*}
					S_4=&\Big\{\Big(\frac{q}{2}-1,\frac{q}{2}-1+i_2,\frac{q}{2}+i_1,\frac{q}{2}+i_0\Big)\ \Big|\ i_1=\frac{q}{2},\ 1\le i_0\le\frac{q}{2}-1,\ i_2\le\frac{q}{2}-2\Big\}\\
					=&\Big\{\Big(\frac{q}{2}-1,\frac{q}{2}+i_2,0,\frac{q}{2}+i_0\Big)\ \Big|\ 1\le i_0\le\frac{q}{2}-1,\ i_2\le\frac{q}{2}-2\Big\}.
				\end{align*}
				Hence, we have
				\begin{align*}
					\min R_4=&\min\{x\cdot q^{m-e-1}\mid x\in S_4,\ e=1\}\\
					=&\min\Big\{\Big(0,\frac{q}{2}+i_0-1,\frac{q}{2},\frac{q}{2}-i_2\Big)\ \Big|\ 1\le i_0\le\frac{q}{2}-1,\ i_2\le\frac{q}{2}-2\Big\}\\
					=&\Big(0,\frac{q}{2},\frac{q}{2},2\Big)_q>\delta_1.
				\end{align*}
				
				\item When $i_1=\frac{q}{2}-1$ and $i_0\le \frac{q}{2}-1$, let
				\begin{equation*}
					a+i=\left(\frac{q}{2},\frac{q}{2}-1+i_2,\frac{q}{2}-1+i_1,\frac{q}{2}+1+i_0\right)=\left(\frac{q}{2},\frac{q}{2}-1+i_2,q-2,\frac{q}{2}+1+i_0\right)
				\end{equation*}
				and
				\begin{equation*}
					n-a=\left(\frac{q}{2}-1,\frac{q}{2},\frac{q}{2},\frac{q}{2}\right)_q.
				\end{equation*}
				For any $x\in X$, we have $(x_1,x_0)=(q-1,0)$ if $i_0=\frac{q}{2}-1$, and $x_1=q-2$ and $\frac{q}{2}+2\le x_0 \le q-1$ if $i_0<\frac{q}{2}-1$. Note that $x_2\notin\{0,q-1\}$ since $i_2\le\frac{q}{2}-1$. Hence, $L_1=1$ and  $L_2=\dots=L_5=0$. Thus $\max\limits_{1\le j\le 5}L_j=L_1=1$, and so $\min\limits_{i\in T_1}\CL(a+i)=\min\limits_{x\in X}\CL(x)=\min R_1$. Note that
				\begin{align*}
					S_1=&\Big\{\Big(\frac{q}{2},\frac{q}{2}-1+i_2,q-2,\frac{q}{2}+1+i_0\Big)\ \Big|\ i_0=\frac{q}{2}-1,\ i_2\le\frac{q}{2}-1\Big\}\\
					=&\Big\{\Big(\frac{q}{2},\frac{q}{2}-1+i_2,q-1,0\Big)\ \Big|\ i_2\le\frac{q}{2}-1\Big\}.
				\end{align*}
				Hence, we have
				\begin{align*}
					\min R_1=&\min\{x\cdot q^{2m-1}\mid x\in S_1\}=\min\Big\{\Big(0,\frac{q}{2},\frac{q}{2}-1+i_2,q-1\Big)\ \Big|\ i_2\le\frac{q}{2}-1\Big\}\\
					=&\Big(0,\frac{q}{2},\frac{q}{2}-1,q-1\Big)_q>\delta_1.
				\end{align*}
				\item When $i_1\ne\frac{q}{2}-1$ and $i_0=\frac{q}{2}$, let
				\begin{equation*}
					a+i=\left(\frac{q}{2},\frac{q}{2}-1+i_2,\frac{q}{2}+i_1,\frac{q}{2}+i_0\right)_q=\left(\frac{q}{2},\frac{q}{2}-1+i_2,\frac{q}{2}+1+i_1,0\right)
				\end{equation*}
				and
				\begin{equation*}
					n-a=\left(\frac{q}{2}-1,\frac{q}{2},\frac{q}{2}-1,\frac{q}{2}+1\right)_q.
				\end{equation*}
				Then we have $x_1\ne0$ and $x_2\notin\{0,q-1\}$ for any $x \in X$. Thus, $L_1=1,L_2=0,L_3=1,L_4=0$ and $L_5=0$. Hence $\max\limits_{1\le j\le 5}L_j=1$, and so $\min\limits_{i\in T_1}\CL(a+i)=\min\limits_{x\in X}\CL(x)=\min (R_1\cup R_3)$. Note that
				\begin{align*}
					S_1=&\Big\{\Big(\frac{q}{2},\frac{q}{2}-1+i_2,\frac{q}{2}+1+i_1,0\Big)\ \Big|\ i_1\ne\frac{q}{2}-1,\ i_2\le\frac{q}{2}-1\Big\},\ \mathrm{and}\\
					S_3=&\Big\{\Big(\frac{q}{2},\frac{q}{2}-1+i_2,\frac{q}{2}+1+i_1,0\Big)\ \Big|\ i_1=\frac{q}{2},\ i_2\le\frac{q}{2}-2\Big\}\\
					=&\Big\{\Big(\frac{q}{2},\frac{q}{2}+i_2,1,0\Big)\ \Big|\ i_2\le\frac{q}{2}-2\Big\}.
				\end{align*}
				Hence, we have
				\begin{align*}
					\min R_1=&\min\{x\cdot q^{2m-1}\mid x\in S_1\}\\=&\min\Big\{\Big(0,\frac{q}{2},\frac{q}{2}-1+i_2,\frac{q}{2}+1+i_1\Big)\ \Big|\ i_1\ne\frac{q}{2}-1,i_2\le\frac{q}{2}-1\Big\}\\
					=&\Big(0,\frac{q}{2},\frac{q}{2}-1,\frac{q}{2}+1\Big)_q,\ \mathrm{and}\\
					\min R_3=&\min\{x\cdot q^{m-1-1}\mid x\in S_3\}=\min\Big\{\Big(0,q-1,\frac{q}{2}-1,\frac{q}{2}-i_2\Big)\ \Big|\ i_2\le\frac{q}{2}-2\Big\}\\
					=&\Big(0,q-1,\frac{q}{2}-1,2\Big)_q.
				\end{align*}
				Thus, $\min\limits_{i\in T_1}\CL(a+i)=\min R_1>\delta_1$.
				\item When $i_1=\frac{q}{2}-1$ and $i_0= \frac{q}{2}$, let
				\begin{equation*}
					a+i=\left(\frac{q}{2}-1,\frac{q}{2}+i_2,\frac{q}{2}-1+i_1,\frac{q}{2}+i_0\right)=\left(\frac{q}{2}-1,\frac{q}{2}+i_2,q-1,0\right)_q
				\end{equation*}
				and
				\begin{equation*}
					n-a=\left(\frac{q}{2},\frac{q}{2}-1,\frac{q}{2},\frac{q}{2}+1\right)_q.
				\end{equation*}
				Since $i_2\le\frac{q}{2}-2$, $x_2\notin\{0,q-1\}$ for any $x\in X$. Then we get $L_1=1$ and $L_2=\dots=L_5=0$. Hence $\max\limits_{1\le j\le 5}L_j=L_1=1$, and so $\min\limits_{i\in T_1}\CL(a+i)=\min\limits_{x\in X}\CL(x)=\min R_1$. Note that
				\begin{align*}
					S_1=\Big\{\Big(\frac{q}{2}-1,\frac{q}{2}+i_2,q-1,0\Big)\ \Big|\ i_2\le\frac{q}{2}-2\Big\}.
				\end{align*}
				Hence, we have
				\begin{align*}
					\min R_1=&\min\{x\cdot q^{2m-1}\mid x\in S_1\}=\min\Big\{\Big(0,\frac{q}{2}-1,\frac{q}{2}+i_2,q-1\Big)\ \Big|\ i_2\le\frac{q}{2}-2\Big\}\\
					=&\Big(0,\frac{q}{2}-1,\frac{q}{2},q-1\Big)_q>\delta_1.
				\end{align*} 
				\item When $i_1\ne\frac{q}{2}$ and $i_0\ge\frac{q}{2}+1$, let
				\begin{equation*}
					a+i=\left(\frac{q}{2},\frac{q}{2}-1+i_2,\frac{q}{2}-1+i_1,\frac{q}{2}+1+i_0\right)=\left(\frac{q}{2},\frac{q}{2}-1+i_2,\frac{q}{2}+i_1,i_0+1-\frac{q}{2}\right)
				\end{equation*}
				and
				\begin{equation*}
					n-a=\left(\frac{q}{2}-1,\frac{q}{2},\frac{q}{2},\frac{q}{2}\right)_q,
				\end{equation*}
				then $x_1\ne0$ and $2\le x_0\le \frac{q}{2}$ for any $x\in X$. Note that $x_2\notin\{0,q-1\}$. Thus $L_1=\dots=L_4=0$ and $L_5=1$. Hence $\max\limits_{1\le j\le 5}L_j=L_5=1$, and so $\min\limits_{i\in T_1}\CL(a+i)=\min\limits_{x\in X}\CL(x)=\min R_5$. Note that
				\begin{align*}
					S_5=&\Big\{\Big(\frac{q}{2},\frac{q}{2}-1+i_2,\frac{q}{2}+i_1,i_0+1-\frac{q}{2}\Big)\ \Big|\\
					&\quad i_1=\frac{q}{2}-1,\ \frac{q}{2}+1\le i_0\le q-1,\ i_2\le\frac{q}{2}-2\Big\}\\
					=&\Big\{\Big(\frac{q}{2},\frac{q}{2}-1+i_2,q-1,i_0+1-\frac{q}{2}\Big)\ \Big|\\
					&\quad \frac{q}{2}+1\le i_0\le q-1,\ i_2\le\frac{q}{2}-2\Big\}.
				\end{align*}
				Hence, we have
				\begin{align*}
					\min R_5=&\min\{x\cdot q^{2m-1-1}\mid x\in S_5,\ f=1\}\\
					=&\min\Big\{\Big(0,\frac{3q}{2}-1-i_0,\frac{q}{2},\frac{q}{2}+i_2\Big)\ \Big|\ \frac{q}{2}+1\le i_0\le q-1,\ i_2\le\frac{q}{2}-2\Big\}\\
					=&\Big(0,\frac{q}{2},\frac{q}{2},\frac{q}{2}\Big)_q>\delta_1.
				\end{align*}
				\item When $i_1=\frac{q}{2}$ and $i_0\ge\frac{q}{2}+1$, let
				\begin{equation*}
					a+i=\left(\frac{q}{2},\frac{q}{2}-1+i_2,\frac{q}{2}+i_1,\frac{q}{2}+1+i_0\right)_q=\left(\frac{q}{2},\frac{q}{2}+i_2,1,i_0+1-\frac{q}{2}\right)_q
				\end{equation*}
				and
				\begin{equation*}
					y=\left(\frac{q}{2}-1,\frac{q}{2},\frac{q}{2}-1,\frac{q}{2}\right)_q.
				\end{equation*}
				Note that $2\le x_0\le \frac{q}{2}$ and $x_2\notin\{0,q-1\}$ for any $x\in X$, and so $\max\limits_{1\le j\le 5}L_j=0$. In accordance with Lemma~\ref{sec3 lem0}, $\min\limits_{i\in T_1}\CL(a+i)\ge q^3>\delta_1.$
			\end{itemize}
			In summary, we have $a+i \in I_1$, and so $n+i=(a+i)+(n-a)\in I_1+I_2$ for any $i \in T_1$.
			
			\item $\boldsymbol{m=5:}$ we have
			\begin{equation*}
				i=(\underbrace{0,0,\ldots,0}_{m-r-1},i_r\ne0,i_{r-1},\ldots,i_1,i_0\ne0)_q,
			\end{equation*}
			where $0\leq r\leq 2$ as $$i<\delta_{1}=\frac{2q^3-3q^2+q}{2}-1=\left(0,0,q-2,\frac{q}{2},\frac{q}{2}-1\right)_q.$$ With the same notation as in Lemma~\ref{sec3 lem0}, take $X := a + T_1$. We only need to prove that there always exists an integer $a\in C_{\Delta_1}$ such that $\min\limits_{i\in T_1}\CL(a+i)\ge\delta_{1}$. Through calculation, the largest coset leader in $T_1$ is
			\begin{equation*}
				\left\{
				\begin{aligned}
					&\left(0,0,q-2,\frac{q}{2},\frac{q}{2}-2\right)_q && \mathrm{if}\ q>4;\\
					&\left(0,0,q-2,\frac{q}{2}-1,q-1\right)_q && \mathrm{if}\ q=4.
				\end{aligned}\right.
			\end{equation*}
			\begin{itemize}
				\item When $i_1\le\frac{q}{2}-2$, or $i_2 \ne \frac{q}{2}-1$ and $i_1=\frac{q}{2}-1$, take $a:=n-\Delta_{1}$. Let
				\begin{equation*}
					a+i=n-\Delta_{1}+i=\left(\frac{q}{2},\frac{q}{2}-1,\frac{q}{2}+i_2,\frac{q}{2}-1+i_1,\frac{q}{2}+1+i_0\right)
				\end{equation*}
				and
				\begin{equation*}
					n-a=\Delta_{1}=\left(\frac{q}{2}-1,\frac{q}{2},\frac{q}{2}-1,\frac{q}{2},\frac{q}{2}\right)_q.
				\end{equation*}
				For any $x\in X$, we have 
				\begin{equation*}
					\left\{
					\begin{aligned}
						&\frac{q}{2}-1\le x_1\le q-2 && \mathrm{if}\ i_1\le\frac{q}{2}-2;\\
						&q-2\le x_1\le q-1\ \mathrm{and}\ x_2\ne q-1 && \mathrm{if}\ i_2 \ne \frac{q}{2}-1\ \mathrm{and}\ i_1=\frac{q}{2}-1.
					\end{aligned}\right.
				\end{equation*} 
				When $x_0=0$, $2\le\frac{q}{2}\le x_1\le q-1$, i.e., $(x_1,x_0)\ne(1,0)$. Thus, $L_1=1,L_2=1,L_3=0,L_4=1$ and $L_5= 1$. Then $\max\limits_{1\le j\le 5}L_j=1$. It then follows from Lemma~\ref{sec3 lem0} that $\min\limits_{i\in T_1}\CL(x)\ge q^3 > \frac{2q^3-3q^2+q}{2}-1$.
				\item When $i_2 = \frac{q}{2}-1$, $i_1=\frac{q}{2}-1$ and $i_0 \notin \{\frac{q}{2},\frac{q}{2}+1\}$, take $a:=\Delta_{1}$. Let
				\begin{equation*}
					a+i=\Delta_{1}+i=\left(\frac{q}{2}-1,\frac{q}{2},q-2,q-1,\frac{q}{2}+i_0\right)
				\end{equation*}
				and
				\begin{equation*}
					n-a=n-\Delta_{1}=\left(\frac{q}{2},\frac{q}{2}-1,\frac{q}{2},\frac{q}{2}-1,\frac{q}{2}+1\right)_q.
				\end{equation*}
				Then for any $x \in X$, $(x_2,x_1)$ equals $(q-1,0)$ if $i_0>\frac{q}{2}$, and $(q-2,q-1)$ if $i_0<\frac{q}{2}$. Since $x_0\notin\{0,1\}$, $L_1=L_2=L_3=0$ and  $L_4=L_5=1$. It follows from Lemma~\ref{sec3 lem0} that $\min\limits_{i\in T_1}\CL(x)\ge q^3 > \delta_1$.
				\item When $i_2 = \frac{q}{2}-1$, $i_1=\frac{q}{2}-1$ and $i_0 \in \{\frac{q}{2},\frac{q}{2}+1\}$, take $a:=n-\Delta_{1}$. Let
				\begin{equation*}
					a+i=n-\Delta_{1}+i=\left(\frac{q}{2},\frac{q}{2}-1,q-1,q-1,\frac{q}{2}+1+i_0-q\right)_q
				\end{equation*}
				and
				\begin{equation*}
					n-a=\Delta_{1}=\left(\frac{q}{2}-1,\frac{q}{2},\frac{q}{2}-1,\frac{q}{2},\frac{q}{2}\right)_q.
				\end{equation*}
				Then $x_0=1$ or $2$ for any $x\in X$. Thus, $L_1=L_3=L_4=0$, $L_2=1$ and $L_5=2$. Hence $\max\limits_{1\le j\le 5}L_j=L_5=2$, and so $\min\limits_{i\in T_1}\CL(a+i)=\min\limits_{x\in X}\CL(x)=\min R_5$. Note that
				\begin{align*}
					S_5=					=&\Big\{\Big(\frac{q}{2},\frac{q}{2}-1,q-1,q-1,i_0+1-\frac{q}{2}\Big)\ \Big|\ i_0\in\Big\{\frac{q}{2},\frac{q}{2}+1\Big\}\Big\}\\
					=&\Big\{\Big(\frac{q}{2},\frac{q}{2}-1,q-1,q-1,1\Big)_q,\Big(\frac{q}{2},\frac{q}{2}-1,q-1,q-1,2\Big)_q\Big\}.
				\end{align*}
				Hence, we have
				\begin{align*}
					\min R_5=&\min\{x\cdot q^{2m-1-2}\mid x\in S_5,\ f=1\}\\
					=&\min\Big\{\Big(0,0,q-2,\frac{q}{2},\frac{q}{2}\Big)_q, \Big(0,0,q-1,\frac{q}{2},\frac{q}{2}\Big)_q\Big\}\\
					=&\Big(0,0,q-2,\frac{q}{2},\frac{q}{2}\Big)_q>\delta_1.
				\end{align*}
				\item When $i_1=\frac{q}{2}$ and $i_0\in\{\frac{q}{2},\frac{q}{2}-1\}$, take $a:=\Delta_{1}$. Let
				\begin{equation*}
					a+i=\Delta_{1}+i=\left(\frac{q}{2}-1,\frac{q}{2},\frac{q}{2}+i_2,0,\frac{q}{2}+i_0\right)
				\end{equation*}
				and
				\begin{equation*}
					n-a=n-\Delta_{1}=\left(\frac{q}{2},\frac{q}{2}-1,\frac{q}{2},\frac{q}{2}-1,\frac{q}{2}+1\right)_q.
				\end{equation*}
				In this case, we have $i_2\le q-3$, and
				\begin{equation*}
					\left\{\begin{aligned}
						&(x_2,x_1,x_0)=\Big(\frac{q}{2}+i_2,0,q-1\Big), && \mathrm{if}\ i_0 = \frac{q}{2}-1;\\
						&(x_2,x_1,x_0)=\Big(\frac{q}{2}+i_2,1,0\Big), && \mathrm{if}\ i_0 = \frac{q}{2}.
					\end{aligned}
					\right.
				\end{equation*}
				If $q=4$, $L_2=0,L_1=L_3=L_4=L_5=1$, and so $\min\limits_{i\in T_1}\CL(a+i)\ge q^3>\delta_1$ by Lemma~\ref{sec3 lem0}. If $q>4$, then $L_1=1,L_2=0,L_3=2,L_4=2,L_5=1$. Hence $\max\limits_{1\le j\le 5}L_j=2$, and so $\min\limits_{i\in T_1}\CL(a+i)=\min\limits_{x\in X}\CL(x)=\min (R_3\cup R_4)$. Note that
				\begin{align*}
					S_3=&\Big\{\Big(\frac{q}{2}-1,\frac{q}{2},\frac{q}{2}+i_2,0,\frac{q}{2}+i_0\Big)\ \Big|\ i_2=i_0=\frac{q}{2}\Big\}=\Big\{\Big(\frac{q}{2}-1,\frac{q}{2}+1,0,1,0\Big)_q\Big\},\ \mathrm{and}\\
					S_4=&\Big\{\Big(\frac{q}{2}-1,\frac{q}{2},\frac{q}{2}+i_2,0,\frac{q}{2}+i_0\Big)\ \Big|\ i_2=\frac{q}{2},\ i_0=\frac{q}{2}-1\Big\}=\Big\{\Big(\frac{q}{2}-1,\frac{q}{2}+1,0,0,q-1\Big)_q\Big\}.
				\end{align*}
				Hence, we have
				\begin{align*}
					\min R_3=&\min\{x\cdot q^{m-1-2}\mid x\in S_3\}=\Big(0,0,q-1,\frac{q}{2},\frac{q}{2}-1\Big)_q>\delta_1,\ \mathrm{and}\\
					\min R_4=&\min\{x\cdot q^{m-e-2}\mid x\in S_4,\ e=1\}=\Big(0,0,q-2,\frac{q}{2},\frac{q}{2}-1\Big)_q=\delta_1.
				\end{align*}
				Hence, $\min\limits_{i\in T_1}\CL(a+i)=\delta_1$.
				\item When $i_2\ne\frac{q}{2}-1$, $i_1=\frac{q}{2}$ and $i_0\notin\{\frac{q}{2},\frac{q}{2}-1\}$, take $a:=n-\Delta_{1}$. Let
				\begin{equation*}
					a+i=n-\Delta_{1}+i=\left(\frac{q}{2},\frac{q}{2}-1,\frac{q}{2}+i_2,q-1,\frac{q}{2}+1+i_0\right)
				\end{equation*}
				and
				\begin{equation*}
					n-a=\Delta_{1}=\left(\frac{q}{2}-1,\frac{q}{2},\frac{q}{2}-1,\frac{q}{2},\frac{q}{2}\right)_q.
				\end{equation*}
				For any $x\in X$, we have
				\begin{equation*}
					\left\{
					\begin{aligned}
						&x_2\ne 0, x_1=0\ \mathrm{and}\ 2\le x_0\le \frac{q}{2} && \mathrm{if}\ i_0\ge \frac{q}{2}+1;\\
						&x_2\ne q-1, x_1=q-1\ \mathrm{and}\ \frac{q}{2}+2\le x_0\le q-1 && \mathrm{if}\ i_0\le \frac{q}{2}-2.
					\end{aligned}\right.
				\end{equation*}
				That is to say, $L_1=L_2=L_3=0$ and $L_4=L_5=1$. Thus, $\min\limits_{i\in T_1}\CL(a+i)\ge q^3 > \delta_1$ by Lemma~\ref{sec3 lem0}.
				\item When $i_2=\frac{q}{2}-1$, $i_1=\frac{q}{2}$ and $i_0\notin\{\frac{q}{2},\frac{q}{2}-1\}$, take $a:=\Delta_1$. Let
				\begin{equation*}
					a+i=\Delta_{1}+i=\left(\frac{q}{2}-1,\frac{q}{2},q-1,0,\frac{q}{2}+i_0\right)
				\end{equation*}
				and
				\begin{equation*}
					n-a=n-\Delta_{1}=\left(\frac{q}{2},\frac{q}{2}-1,\frac{q}{2},\frac{q}{2}-1,\frac{q}{2}+1\right)_q.
				\end{equation*}
				Then for any $x\in X$, we have
				\begin{equation*}
					\left\{
					\begin{aligned}
						&x_2=q-1, x_1=1\ \mathrm{and}\ 1\le x_0\le \frac{q}{2}-1 && \mathrm{if}\ i_0\ge \frac{q}{2}+1;\\
						&x_2=q-1, x_1=0\ \mathrm{and}\ \frac{q}{2}+1\le x_0\le q-2 && \mathrm{if}\ i_0\le \frac{q}{2}-2.
					\end{aligned}\right.
				\end{equation*}
				Thus, $L_1=L_3=0$ and $L_2=L_4=L_5=1$. By Lemma~\ref{sec3 lem0}, we have $\min\limits_{i\in T_1}\CL(a+i)\ge q^3 > \delta_1$.
				\item When $i_1\ge\frac{q}{2}+1$, take $a := \Delta_{1}$. Let
				\begin{equation*}
					a+i=\Delta_{1}+i=\left(\frac{q}{2}-1,\frac{q}{2},\frac{q}{2}+i_2,\frac{q}{2}+i_1-q,\frac{q}{2}+i_0\right)
				\end{equation*}
				and
				\begin{equation*}
					n-a=n-\Delta_{1}=\left(\frac{q}{2},\frac{q}{2}-1,\frac{q}{2},\frac{q}{2}-1,\frac{q}{2}+1\right)_q.
				\end{equation*}
				Then we have $1\le x_1\le \frac{q}{2}<q-1$ for any $x\in X$. Hence, $\max\limits_{1\le j\le 5}L_j=1$. It then follows from Lemma~\ref{sec3 lem0} that $\min\limits_{i\in T_1}\CL(a+i)\ge q^3 > \delta_1$.
			\end{itemize}
			To sum up, we have $a+i \in I_1$ and $n+i=(a+i)+(n-a)\in I_1+I_2$ for any $i \in T_1$. 
			
		\item $\boldsymbol{m=6:}$ we have
	\begin{equation*}
		i=(\underbrace{0,0,\ldots,0}_{m-r-1},i_r\ne0,i_{r-1},\ldots,i_1,i_0\ne0)_q,
	\end{equation*}
	where $0\leq r\leq 3$ as $$i<\delta_{1}=\frac{2q^4-2q^3-q^2+q}{2}=\left(0,0,q-2,q-1,\frac{q}{2},\frac{q}{2}\right)_q.$$ With the same notation as in Lemma~\ref{sec3 lem0}, take $X := a + T_1$. We only need to prove that there always exists an integer $a\in C_{\Delta_1}$ such that $\min\limits_{i\in T_1}\CL(a+i)\ge\delta_{1}$. Through calculation, the largest coset leader in $T_1$ is
	\begin{equation*}
		\left(0,0,q-2,q-1,\frac{q}{2},\frac{q}{2}-1\right)_q.
	\end{equation*}
	\begin{itemize}
		\item When $i_3\ne \frac{q}{2}-1$ and $i_1\le\frac{q}{2}-2$, let
		\begin{equation*}
			a+i=\left(\frac{q}{2}-1,\frac{q}{2}-1,\frac{q}{2}+i_3,\frac{q}{2}+i_2,\frac{q}{2}-1+i_1,\frac{q}{2}+i_0\right)
		\end{equation*}
		and
		\begin{equation*}
			n-a=\left(\frac{q}{2},\frac{q}{2},\frac{q}{2}-1,\frac{q}{2}-1,\frac{q}{2},\frac{q}{2}+1\right)_q,
		\end{equation*}
		then $\frac{q}{2}-1\le x_1\le q-2$, i.e., $x_1\notin\{0,q-1\}$ for any $x\in X$. Since $i_3\ne\frac{q}{2}-1$, $(x_3,x_2)\notin\{(0,0),(q-1,q-1)\}$. Note that $x_1\ge\frac{q}{2}>1$ if $x_0=0$, which implies that $(x_2,x_1,x_0)\ne(0,1,0)$. Thus we have $\max\limits_{1\le j\le 5}L_j=1$. According to Lemma~\ref{sec3 lem0}, $\min\limits_{i\in T_1}\CL(a+i)\ge q^4>\delta_1$.
		
		\item When $i_3=\frac{q}{2}-1$, $i_2\ne\frac{q}{2}$ and $i_1\le \frac{q}{2}-2$, let
		\begin{align*}
			a+i=&\left(\frac{q}{2},\frac{q}{2},\frac{q}{2}-1+i_3,\frac{q}{2}-1+i_2,\frac{q}{2}+i_1,\frac{q}{2}+1+i_0\right)\\
			=&\left(\frac{q}{2},\frac{q}{2},q-2,\frac{q}{2}-1+i_2,\frac{q}{2}+i_1,\frac{q}{2}+1+i_0\right)
		\end{align*}
		and
		\begin{equation*}
			n-a=\left(\frac{q}{2}-1,\frac{q}{2}-1,\frac{q}{2},\frac{q}{2},\frac{q}{2}-1,\frac{q}{2}\right)_q,
		\end{equation*}
		then $1<\frac{q}{2}\le x_1\le q-1$ and $x_2\ne q-1$ for any $x\in X$. If $x_2=0$, then $x_3=q-1$, and so $(x_3,x_2)\ne(0,0)$. Hence, $\max\limits_{1\le j\le 5}L_j=1$. By Lemma~\ref{sec3 lem0}, $\min\limits_{i\in T_1}\CL(a+i)\ge q^4>\delta_1$.
		
		\item When $i_3=\frac{q}{2}-1$, $i_2=\frac{q}{2}$ and $i_1\le \frac{q}{2}-2$, let
		\begin{equation*}
			a+i=\left(\frac{q}{2},\frac{q}{2}-1,\frac{q}{2}-1+i_3,\frac{q}{2}+i_2,\frac{q}{2}+i_1,\frac{q}{2}+i_0\right)=\left(\frac{q}{2},\frac{q}{2}-1,q-1,0,\frac{q}{2}+i_1,\frac{q}{2}+i_0\right)
		\end{equation*}
		and
		\begin{equation*}
			n-a=\left(\frac{q}{2}-1,\frac{q}{2},\frac{q}{2},\frac{q}{2}-1,\frac{q}{2}-1,\frac{q}{2}+1\right)_q.
		\end{equation*}
		Note that $1<\frac{q}{2}\le x_1\le q-1$ for any $x\in X$. It is easily seen that $\max\limits_{1\le j\le 5}L_j=1$. Again by Lemma~\ref{sec3 lem0}, $\min\limits_{i\in T_1}\CL(a+i)\ge q^4>\delta_1$.
		
		\item When $i_2\ne \frac{q}{2}-1$, $i_1=\frac{q}{2}-1$ and $i_0\ge\frac{q}{2}$, let
		\begin{align*}
			a+i=&\left(\frac{q}{2}-1,\frac{q}{2}-1,\frac{q}{2}+i_3,\frac{q}{2}+i_2,\frac{q}{2}-1+i_1,\frac{q}{2}+i_0\right)\\
			=&\left(\frac{q}{2}-1,\frac{q}{2}-1,\frac{q}{2}+i_3,\frac{q}{2}+i_2,q-1,i_0-\frac{q}{2}\right)
		\end{align*}
		and
		\begin{equation*}
			n-a=\left(\frac{q}{2},\frac{q}{2},\frac{q}{2}-1,\frac{q}{2}-1,\frac{q}{2},\frac{q}{2}+1\right)_q,
		\end{equation*}
		then $x_2\ne q-1$ and $0\le x_0\le \frac{q}{2}-1$ for any $x \in X$. Thus we have $L_1=L_2=1,L_3=0,L_4=2$ and $L_5=1$. Hence $\max\limits_{1\le j\le 5}L_j=L_4=2$, and so $\min\limits_{i\in T_1}\CL(a+i)=\min\limits_{x\in X}\CL(x)=\min R_4$. Note that
		\begin{align*}
			S_4=&\Big\{\Big(\frac{q}{2}-1,\frac{q}{2}-1,\frac{q}{2}+i_3,\frac{q}{2}+i_2,q-1,i_0-\frac{q}{2}\Big)\ \Big|\ i_3=\frac{q}{2}-1,\ i_2=\frac{q}{2},\ \frac{q}{2}\le i_0\le q-1\Big\}\\
			=&\Big\{\Big(\frac{q}{2}-1,\frac{q}{2},0,0,q-1,i_0-\frac{q}{2}\Big)\ \Big|\ \frac{q}{2}\le i_0\le q-1\Big\}.
		\end{align*}
		Hence, we have
		\begin{align*}
			\min R_4=&\min\{x\cdot q^{m-e-2}\mid x\in S_4,\ e=2\}\\
			=&\min\Big\{\Big(0,0,q-1,i_0-\frac{q}{2}-1,\frac{q}{2},\frac{q}{2}\Big)\ \Big|\ \frac{q}{2}\le i_0\le q-1\Big\}\\
			=&\Big(0,0,q-2,q-1,\frac{q}{2},\frac{q}{2}\Big)_q=\delta_1.
		\end{align*}
		
		\item When $i_2\ne \frac{q}{2}-1$, $i_1=\frac{q}{2}-1$ and $i_0\le\frac{q}{2}-1$, let
		\begin{equation*}
			a+i=\left(\frac{q}{2},\frac{q}{2}-1,\frac{q}{2}-1+i_3,\frac{q}{2}+i_2,\frac{q}{2}+i_1,\frac{q}{2}+i_0\right)=\left(\frac{q}{2},\frac{q}{2}-1,\frac{q}{2}-1+i_3,\frac{q}{2}+i_2,q-1,\frac{q}{2}+i_0\right)
		\end{equation*}
		and
		\begin{equation*}
			n-a=\left(\frac{q}{2}-1,\frac{q}{2},\frac{q}{2},\frac{q}{2}-1,\frac{q}{2}-1,\frac{q}{2}+1\right)_q,
		\end{equation*}
		Then for any $x\in X$, $1<\frac{q}{2}+1\le x_0\le q-1$, $x_1=q-1$ and $x_2\ne q-1$. Hence, $L_1=L_2=L_3=0,L_4=2$ and $L_5=1$. Thus $\max\limits_{1\le j\le 5}L_j=L_4=2$, and so $\min\limits_{i\in T_1}\CL(a+i)=\min\limits_{x\in X}\CL(x)=\min R_4$. Note that
		\begin{align*}
			S_4=&\Big\{\Big(\frac{q}{2},\frac{q}{2}-1,\frac{q}{2}-1+i_3,\frac{q}{2}+i_2,q-1,\frac{q}{2}+i_0\Big)\ \Big|\ i_3=i_2=\frac{q}{2},\ 1\le i_0\le\frac{q}{2}-1\Big\}\\
			=&\Big\{\Big(\frac{q}{2},\frac{q}{2},0,0,q-1,\frac{q}{2}+i_0\Big)\ \Big|\ 1\le i_0\le\frac{q}{2}-1\Big\}.
		\end{align*}
		Hence, we have
		\begin{align*}
			\min R_4=&\min\{x\cdot q^{m-2-2}\mid x\in S_4,\ e=2\}\\
			=&\min\Big\{\Big(0,0,q-1,\frac{q}{2}+i_0-1,\frac{q}{2}-1,\frac{q}{2}\Big)\ \Big|\ 1\le i_0\le\frac{q}{2}-1\Big\}\\
			=&\Big(0,0,q-1,\frac{q}{2},\frac{q}{2}-1,\frac{q}{2}\Big)_q>\delta_1.
		\end{align*}
		
		\item When $i_2=i_1=\frac{q}{2}-1$, let
		\begin{align*}
			a+i=&\left(\frac{q}{2}-1,\frac{q}{2},\frac{q}{2}+i_3,\frac{q}{2}-1+i_2,\frac{q}{2}-1+i_1,\frac{q}{2}+1+i_0\right)\\
			=&\left(\frac{q}{2}-1,\frac{q}{2},\frac{q}{2}+i_3,q-2,q-2,\frac{q}{2}+1+i_0\right)
		\end{align*}
		and
		\begin{equation*}
			n-a=\left(\frac{q}{2},\frac{q}{2}-1,\frac{q}{2}-1,\frac{q}{2},\frac{q}{2},\frac{q}{2}\right)_q
		\end{equation*}
		Note that  $x_1\ne0$ and $x_2= q-2$ for any $x\in X$, which leads to $\max\limits_{1\le j\le 5}L_j=1$. Again by Lemma~\ref{sec3 lem0}, $\min\limits_{i\in T_1}\CL(a+i)\ge q^4>\delta_1$.
		
		\item When $i_2\ne \frac{q}{2}$, $i_1=\frac{q}{2}$ and $i_0\le \frac{q}{2}$, let
		\begin{equation*}
			a+i=\left(\frac{q}{2},\frac{q}{2},\frac{q}{2}-1+i_3,\frac{q}{2}-1+i_2,\frac{q}{2}+i_1,\frac{q}{2}+1+i_0\right)=\left(\frac{q}{2},\frac{q}{2},\frac{q}{2}-1+i_3,\frac{q}{2}+i_2,0,\frac{q}{2}+1+i_0\right)
		\end{equation*}
		and
		\begin{equation*}
			n-a=\left(\frac{q}{2}-1,\frac{q}{2}-1,\frac{q}{2},\frac{q}{2},\frac{q}{2}-1,\frac{q}{2}\right)_q,
		\end{equation*}
		then $x_2\ne 0$ for any $x\in X$. If $i_0\in\{\frac{q}{2}-1,\frac{q}{2}\}$, then $x_1=1$, that is, $(x_1,x_0)\notin\{(0,0),(0,1)\}$. Hence, $L_1=\dots=L_4=1$ and $L_5=2$. Thus $\max\limits_{1\le j\le 5}L_j=L_5=2$, and so $\min\limits_{i\in T_1}\CL(a+i)=\min\limits_{x\in X}\CL(x)=\min R_5$. Note that
		\begin{align*}
			S_5=&\Big\{\Big(\frac{q}{2},\frac{q}{2},\frac{q}{2}-1+i_3,\frac{q}{2}+i_2,0,\frac{q}{2}+1+i_0\Big)\ \Big|\ i_3=\frac{q}{2},\ i_2=\frac{q}{2}-1,\ 1\le i_0\le\frac{q}{2}\Big\}\\
			=&\Big\{\Big(\frac{q}{2},\frac{q}{2},q-1,q-1,0,\frac{q}{2}+1+i_0\Big)\ \Big|\ 1\le i_0\le\frac{q}{2}\Big\}.
		\end{align*}
		Hence, we have
		\begin{align*}
			\min R_5=&\min\{x\cdot q^{2m-2-2}\mid x\in S_5,\ f=2\}\\
			=&\min\Big\{\Big(0,0,q-1,\frac{q}{2}-i_0-1,\frac{q}{2},\frac{q}{2}+1\Big)\ \Big|\ 1\le i_0\le\frac{q}{2}\Big\}\\
			=&\Big(0,0,q-2,q-1,\frac{q}{2},\frac{q}{2}+1\Big)_q>\delta_1.
		\end{align*}
		
		\item When $i_2\ne \frac{q}{2}$, $i_1=\frac{q}{2}$ and $i_0\ge \frac{q}{2}+1$, let
		\begin{equation*}
			a+i=\left(\frac{q}{2}-1,\frac{q}{2},\frac{q}{2}+i_3,\frac{q}{2}-1+i_2,\frac{q}{2}-1+i_1,\frac{q}{2}+1+i_0\right)=\left(\frac{q}{2}-1,\frac{q}{2},\frac{q}{2}+i_3,\frac{q}{2}+i_2,0,i_0+1-\frac{q}{2}\right)
		\end{equation*}
		and
		\begin{equation*}
			n-a=\left(\frac{q}{2},\frac{q}{2}-1,\frac{q}{2}-1,\frac{q}{2},\frac{q}{2},\frac{q}{2}\right)_q,
		\end{equation*}
		Then $2\le x_0\le \frac{q}{2}$, $x_1=0$ and $x_2\ne 0$ for any $x\in X$, which implies that $L_1=L_2=L_3=0,L_4=1$ and $L_5=2$. Hence $\max\limits_{1\le j\le 5}L_j=L_5=2$, and so $\min\limits_{i\in T_1}\CL(a+i)=\min\limits_{x\in X}\CL(x)=\min R_5$. Note that
		\begin{align*}
			S_5=&\Big\{\Big(\frac{q}{2}-1,\frac{q}{2},\frac{q}{2}+i_3,\frac{q}{2}+i_2,0,i_0+1-\frac{q}{2}\Big)\ \Big|\ i_3=i_2=\frac{q}{2}-1,\ \frac{q}{2}+1\le i_0\le q-1\Big\}\\
			=&\Big\{\Big(\frac{q}{2}-1,\frac{q}{2},q-1,q-1,0,i_0-\frac{q}{2}+1\Big)\ \Big|\ \frac{q}{2}+1\le i_0\le q-1\Big\}.
		\end{align*}
		Hence, we have
		\begin{align*}
			\min R_5=&\min\{x\cdot q^{2m-2-2}\mid x\in S_5,\ f=2\}\\
			=&\min\Big\{\Big(0,0,q-1,\frac{3q}{2}-i_0-1,\frac{q}{2}-1,\frac{q}{2}+1\Big)\ \Big|\ \frac{q}{2}+1\le i_0\le q-1\Big\}\\
			=&\Big(0,0,q-1,\frac{q}{2},\frac{q}{2}-1,\frac{q}{2}+1\Big)_q>\delta_1.
		\end{align*}
		
		\item When $i_2=\frac{q}{2}$ and $i_1\ge\frac{q}{2}$, let
		\begin{equation*}
			a+i=\left(\frac{q}{2},\frac{q}{2}-1,\frac{q}{2}-1+i_3,\frac{q}{2}+i_2,\frac{q}{2}+i_1,\frac{q}{2}+i_0\right)=\left(\frac{q}{2},\frac{q}{2}-1,\frac{q}{2}+i_3,1,i_1-\frac{q}{2},\frac{q}{2}+i_0\right)
		\end{equation*}
		and
		\begin{equation*}
			n-a=\left(\frac{q}{2}-1,\frac{q}{2},\frac{q}{2},\frac{q}{2}-1,\frac{q}{2}-1,\frac{q}{2}+1\right)_q,
		\end{equation*}
		then $0\le x_1\le \frac{q}{2}$ and $x_2=1$ for any $x\in X$. Since $x_1\ge1$ if $i_0\ge\frac{q}{2}$, $(x_1,x_0)\notin\{(0,0),(0,1)\}$. Hence, we have $\max\limits_{1\le j\le 5}L_j=1$. By Lemma~\ref{sec3 lem0}, $\min\limits_{i\in T_1}\CL(a+i)\ge q^4>\delta_1.$
		
		\item When $i_2\notin\{\frac{q}{2}-1,\frac{q}{2}\}$ and $i_1\ge\frac{q}{2}+1$, let 
		\begin{equation*}
			a+i=\left(\frac{q}{2},\frac{q}{2},\frac{q}{2}-1+i_3,\frac{q}{2}-1+i_2,\frac{q}{2}+i_1,\frac{q}{2}+1+i_0\right)=\left(\frac{q}{2},\frac{q}{2},\frac{q}{2}-1+i_3,\frac{q}{2}+i_2,i_1-\frac{q}{2},\frac{q}{2}+1+i_0\right)
		\end{equation*}
		and
		\begin{equation*}
			n-a=\left(\frac{q}{2}-1,\frac{q}{2}-1,\frac{q}{2},\frac{q}{2},\frac{q}{2}-1,\frac{q}{2}\right)_q,
		\end{equation*}
		then $x_2\notin\{0,q-1\}$ and $1\le x_1\le \frac{q}{2}$ for any $x\in X$. Thus, we have $\max\limits_{1\le j\le 5}L_j=1$. It then follows from Lemma~\ref{sec3 lem0} that $\min\limits_{i\in T_1}\CL(a+i)\ge q^4>\delta_1.$
		
		\item When $i_3\ne\frac{q}{2}$, $i_2=\frac{q}{2}-1$ and $i_1\ge\frac{q}{2}+1$, let
		\begin{equation*}
			a+i=\left(\frac{q}{2},\frac{q}{2}-1,\frac{q}{2}-1+i_3,\frac{q}{2}+i_2,\frac{q}{2}+i_1,\frac{q}{2}+i_0\right)=\left(\frac{q}{2},\frac{q}{2}-1,\frac{q}{2}+i_3,0,i_1-\frac{q}{2},\frac{q}{2}+i_0\right)
		\end{equation*}
		and
		\begin{equation*}
			n-a=\left(\frac{q}{2}-1,\frac{q}{2},\frac{q}{2},\frac{q}{2}-1,\frac{q}{2}-1,\frac{q}{2}+1\right)_q,
		\end{equation*}
		then $x_3\ne0$, $x_2=0$ and $1\le x_1\le \frac{q}{2}$ for any $x\in X$. If $x_0=0$, then $x_1\ge2$, and so $(x_2,x_1,x_0)\ne(0,1,0)$. Hence, we have $\max\limits_{1\le j\le 5}L_j=1$. By Lemma~\ref{sec3 lem0}, we get  $\min\limits_{i\in T_1}\CL(a+i)\ge q^4>\delta_1.$ 
		
		\item When $i_3=\frac{q}{2}$, $i_2=\frac{q}{2}-1$ and $i_1\ge\frac{q}{2}+1$, let
		\begin{equation*}
			a+i=\left(\frac{q}{2}-1,\frac{q}{2},\frac{q}{2}+i_3,\frac{q}{2}-1+i_2,\frac{q}{2}-1+i_1,\frac{q}{2}+1+i_0\right)=\left(\frac{q}{2}-1,\frac{q}{2}+1,0,q-1,i_1-\frac{q}{2}-1,\frac{q}{2}+1+i_0\right)
		\end{equation*}
		and
		\begin{equation*}
			n-a=\left(\frac{q}{2},\frac{q}{2}-1,\frac{q}{2}-1,\frac{q}{2},\frac{q}{2},\frac{q}{2}\right)_q.
		\end{equation*}
		One can easily verify that $x_3=0,x_2=q-1$ and $0\le x_1\le \frac{q}{2}-1<q-1$ for any $x\in X$. If $x_0=0$ or 1, then $x_1\ge1$, i.e., $(x_1,x_0)\notin\{(0,0),(0,1)\}$. Hence, we have $\max\limits_{1\le j\le 5}L_j=1$. It follows from Lemma~\ref{sec3 lem0} that $\min\limits_{i\in T_1}\CL(a+i)\ge q^4>\delta_1.$ 
	\end{itemize}
	In summary, we have $a+i \in I_1$, and so $n+i=(a+i)+(n-a) \in I_1+I_2$ for any $i \in T_1$.
	
	\item $\boldsymbol{m=7:}$ we have
	\begin{equation*}
		i=(\underbrace{0,0,\ldots,0}_{m-r-1},i_r\ne0,i_{r-1},\ldots,i_1,i_0\ne0)_q,
	\end{equation*}
	where $0\leq r\leq 4$ as $$i<\delta_{1}=\frac{q^5-q^4+2q}{2}-1=\left(0,0,\frac{q}{2}-1,\frac{q}{2},0,0,q-1\right)_q.$$ With the same notation as in Lemma~\ref{sec3 lem0}, take $X := a + T_1$. We only need to prove that there always exists an integer $a\in C_{\Delta_1}$ such that $\min\limits_{i\in T_1}\CL(a+i)\ge\delta_{1}$. Through calculation, the largest coset leader in $T_1$ is
	\begin{equation*}
		\left(0,0,\frac{q}{2}-1,\frac{q}{2},0,0,q-2\right)_q.
	\end{equation*}
	\begin{itemize}
		\item When $i_2\le \frac{q}{2}-1$, take $a:=\Delta_1$. Let
		\begin{equation*}
			a+i=\Delta_1+i=\left(\frac{q}{2}-1,\frac{q}{2},\frac{q}{2}-1+i_4,\frac{q}{2}+i_3,\frac{q}{2}-1+i_2,\frac{q}{2}+i_1,\frac{q}{2}+i_0\right)
		\end{equation*}
		and
		\begin{equation*}
			y=n-\Delta_1=\left(\frac{q}{2},\frac{q}{2}-1,\frac{q}{2},\frac{q}{2}-1,\frac{q}{2},\frac{q}{2}-1,\frac{q}{2}+1\right)_q.
		\end{equation*}
		For any $x\in X$, we have $\frac{q}{2}-1\le x_2\le q-1$, which implies that $x_2\ne 0$ and $(x_2,x_1)\ne(q-1,q-1)$. If $x_1=0$, then $\frac{q}{2}\le x_2\le q-1$, i.e., $(x_2,x_1)\ne (1,0)$. Since $i_4\le \frac{q}{2}-1$, $(x_4,x_3)\ne (q-1,q-1)$ or $(0,0)$. Thus, we have $L_1=L_2=2,L_3=L_4=1$ and $L_5=2$. Hence $\max\limits_{1\le j\le 5}L_j=2$, and so $\min\limits_{i\in T_1}\CL(a+i)=\min\limits_{x\in X}\CL(x)=\min(R_1\cup R_2 \cup R_5)$. Note that
		\begin{align*}
			S_1=&\Big\{\Big(\frac{q}{2}-1,\frac{q}{2},\frac{q}{2}-1+i_4,\frac{q}{2}+i_3,\frac{q}{2}-1+i_2,\frac{q}{2}+i_1,\frac{q}{2}+i_0\Big)\ \Big|\\
			&\quad i_4,i_2\le\frac{q}{2}-1,\ i_3\le q-1,i_1=\frac{q}{2}-1,\ i_0=\frac{q}{2}\Big\}\\
			=&\Big\{\Big(\frac{q}{2}-1,\frac{q}{2},\frac{q}{2}-1+i_4,\frac{q}{2}+i_3,\frac{q}{2}+i_2,0,0\Big)\ \Big|\ i_4,i_2\le\frac{q}{2}-1,\ i_3\le q-1\Big\},\\
			S_2=&\Big\{\Big(\frac{q}{2}-1,\frac{q}{2},\frac{q}{2}-1+i_4,\frac{q}{2}+i_3,\frac{q}{2}-1+i_2,\frac{q}{2}+i_1,\frac{q}{2}+i_0\Big)\ \Big|\\
			&\quad i_4,i_2\le\frac{q}{2}-1,\ i_3\le q-1,\ i_1=\frac{q}{2}-1,\ i_0=\frac{q}{2}+1\Big\}\\
			=&\Big\{\Big(\frac{q}{2}-1,\frac{q}{2},\frac{q}{2}-1+i_4,\frac{q}{2}+i_3,\frac{q}{2}+i_2,0,1\Big)\ \Big|\ i_4,i_2\le\frac{q}{2}-1,\ i_3\le q-1\Big\},\ \mathrm{and}\\
			S_5=&\Big\{\Big(\frac{q}{2}-1,\frac{q}{2},\frac{q}{2}-1+i_4,\frac{q}{2}+i_3,\frac{q}{2}-1+i_2,\frac{q}{2}+i_1,\frac{q}{2}+i_0\Big)\ \Big|\ i_4\le\frac{q}{2}-1,\ i_3=i_2=\frac{q}{2}-1,\\
			&\quad \text{either } i_1\ge\frac{q}{2}\text{ and } 1\le i_0\le q-1, \text{ or } i_1=\frac{q}{2}-1\text{ and } \frac{q}{2}+1\le i_0\le q-1\}\\
			=&\Big\{\Big(\frac{q}{2}-1,\frac{q}{2},\frac{q}{2}-1+i_4,q-1,q-1,i_1-\frac{q}{2},\frac{q}{2}+i_0\Big)\ \Big|\ i_4\le\frac{q}{2}-1,\\
			&\quad \text{either } i_1\ge\frac{q}{2}\text{ and } 1\le i_0\le q-1, \text{ or } i_1=\frac{q}{2}-1\text{ and } \frac{q}{2}+1\le i_0\le q-1\Big\}.
		\end{align*}
		Hence, we have
		\begin{align*}
			\min R_1=&\min\{x\cdot q^{2m-2}\mid x\in S_1\}\\
			=&\min\Big\{\Big(0,0,\frac{q}{2}-1,\frac{q}{2},\frac{q}{2}-1+i_4,\frac{q}{2}+i_3,\frac{q}{2}+i_2\Big)\ \Big|\ i_4,i_2\le\frac{q}{2}-1,\ i_3\le q-1\Big\}\\
			=&\Big(0,0,\frac{q}{2}-1,\frac{q}{2},\frac{q}{2}-1,\frac{q}{2},\frac{q}{2}\Big)_q,\\
			\min R_2=&\min\{x\cdot q^{m-2}\mid x\in S_2\}\\
			=&\min\Big\{\Big(0,0,\frac{q}{2},\frac{q}{2}-1,\frac{q}{2}-i_4,\frac{q}{2}-1-i_3,\frac{q}{2}-i_2\Big)\ \Big|\ i_4,i_2\le\frac{q}{2}-1,\ i_3\le q-1\Big\},\ \mathrm{and}\\
			\min R_5=&\min\{x\cdot q^{2m-2-2}\mid x\in S_5,\ f=2\}\\
			=&\min\Big\{\Big(0,0,\frac{3q}{2}-1-i_1,\frac{q}{2}-i_0,\frac{q}{2}-1,\frac{q}{2},\frac{q}{2}+i_4\Big)\ \Big|\ i_4\le\frac{q}{2}-1,\\
			&\quad \text{either } i_1\ge\frac{q}{2}\text{ and } 1\le i_0\le q-1, \text{ or } i_1=\frac{q}{2}-1\text{ and } \frac{q}{2}+1\le i_0\le q-1\Big\}\\
			=&\Big(0,0,\frac{q}{2}-1,\frac{q}{2}+1,\frac{q}{2}-1,\frac{q}{2},\frac{q}{2}\Big)_q.
		\end{align*}
		It is easily seen that $\min R_1< \min R_5 < \min R_2$. Hence, $\min\limits_{i\in T_1}\CL(a+i)=\min R_1>\delta_1$.
		\item When $i_2=\frac{q}{2}$, take $a:=n-\Delta_1$. Let
		\begin{equation*}
			a+i=n-\Delta_1+i=\left(\frac{q}{2},\frac{q}{2}-1,\frac{q}{2}+i_4,\frac{q}{2}+i_3,0,\frac{q}{2}-1+i_1,\frac{q}{2}+1+i_0\right)
		\end{equation*}
		and
		\begin{equation*}
			n-a=\Delta_1=\left(\frac{q}{2}-1,\frac{q}{2},\frac{q}{2}-1,\frac{q}{2},\frac{q}{2}-1,\frac{q}{2},\frac{q}{2}\right)_q,
		\end{equation*}
		then $0\le x_2\le 1<q-1$ and $(x_2,x_1)\ne(0,0)$ for any $x\in X$. If $i_3=\frac{q}{2}$, then $i_4\le \frac{q}{2}-2$, and so $(x_4,x_3)\ne(0,0)$. Thus, we have $L_1=\dots=L_5=2$. Hence $\max\limits_{1\le j\le 5}L_j=2$, and so $\min\limits_{i\in T_1}\CL(a+i)=\min\limits_{x\in X}\CL(x)=\min\bigcup\limits_{1\le j\le 5}R_j$. Note that
		\begin{align*}
			S_1=&\Big\{\Big(\frac{q}{2},\frac{q}{2}-1,\frac{q}{2}+i_4,\frac{q}{2}+i_3,0,\frac{q}{2}-1+i_1,\frac{q}{2}+1+i_0\Big)\ \Big|\\
			&\quad i_4\le \frac{q}{2}-1,i_3\le q-1,i_1=\frac{q}{2},\ i_0=\frac{q}{2}-1\ \Big|\ i_4\le \frac{q}{2}-1,i_3\le q-1\Big\}\\
			=&\Big\{\Big(\frac{q}{2},\frac{q}{2}-1,\frac{q}{2}+i_4,\frac{q}{2}+i_3,1,0,0\Big)_q\Big\},\\
			S_2=&\Big\{\Big(\frac{q}{2},\frac{q}{2}-1,\frac{q}{2}+i_4,\frac{q}{2}+i_3,0,\frac{q}{2}-1+i_1,\frac{q}{2}+1+i_0\Big)_q\ i_4\le \frac{q}{2}-1,i_3\le q-1,i_1=i_0=\frac{q}{2}\Big\}\\
			=&\Big\{\Big(\frac{q}{2},\frac{q}{2}-1,\frac{q}{2}+i_4,\frac{q}{2}+i_3,1,0,1\Big)\ \Big|\ i_4\le \frac{q}{2}-1,i_3\le q-1\Big\},\\
			S_3=&\Big\{\Big(\frac{q}{2},\frac{q}{2}-1,\frac{q}{2}+i_4,\frac{q}{2}+i_3,0,\frac{q}{2}-1+i_1,\frac{q}{2}+1+i_0\Big)\ \Big|\ i_4\le \frac{q}{2}-2,i_3=i_1=\frac{q}{2},\ i_0=\frac{q}{2}-1\Big\}\\
			=&\Big\{\Big(\frac{q}{2},\frac{q}{2}-1,\frac{q}{2}+1+i_4,0,1,0,0\Big)\ \Big|\ i_4\le \frac{q}{2}-2\Big\},\\
			S_4=&\Big\{\Big(\frac{q}{2},\frac{q}{2}-1,\frac{q}{2}+i_4,\frac{q}{2}+i_3,0,\frac{q}{2}-1+i_1,\frac{q}{2}+1+i_0\Big)\ \Big|\\
			&\quad i_4\le \frac{q}{2}-2,i_3=\frac{q}{2},\ \mathrm{and\ either}\ i_1\le\frac{q}{2}-1,\ \mathrm{or}\ i_1=\frac{q}{2}\ \mathrm{and}\ 1\le i_0\le\frac{q}{2}-2\Big\}\\
			=&\Big\{\Big(\frac{q}{2},\frac{q}{2}-1,\frac{q}{2}+1+i_4,0,0,\frac{q}{2}-1+i_1,\frac{q}{2}+1+i_0\Big)\ \Big|\\
			&\quad i_4\le \frac{q}{2}-2,\mathrm{either}\ i_1\le\frac{q}{2}-1,\ \mathrm{or}\ i_1=\frac{q}{2}\ \mathrm{and}\ 1\le i_0\le\frac{q}{2}-2\Big\},\ \mathrm{and}\\
			S_5=&\Big\{\Big(\frac{q}{2},\frac{q}{2}-1,\frac{q}{2}+i_4,\frac{q}{2}+i_3,0,\frac{q}{2}-1+i_1,\frac{q}{2}+1+i_0\Big)\ \Big|\\
			&\quad i_4=i_3=\frac{q}{2}-1,i_1\le q-1,1\le i_0\le q-1\Big\}\\
			=&\Big\{\Big(\frac{q}{2},\frac{q}{2}-1,q-1,q-1,0,\frac{q}{2}-1+i_1,\frac{q}{2}+1+i_0\ \Big|\ i_1\le q-1,1\le i_0\le q-1\Big)_q\Big\}.
		\end{align*}
		Hence, we have
		\begin{align*}
			\min R_1=&\min\{x\cdot q^{2m-2}\mid x\in S_1\}\\
			=&\min\Big\{\Big(0,0,\frac{q}{2},\frac{q}{2}-1,\frac{q}{2}+i_4,\frac{q}{2}+i_3,1\Big)\ \Big|\ i_4\le \frac{q}{2}-1,i_3\le q-1\Big\},\\
			=&\Big(0,0,\frac{q}{2},\frac{q}{2}-1,\frac{q}{2},\frac{q}{2},1\Big)_q\\
			\min R_2=&\min\{x\cdot q^{m-2}\mid x\in S_2\}\\
			=&\min\Big\{\Big(0,0,\frac{q}{2}-1,\frac{q}{2},\frac{q}{2}-1-i_4,\frac{q}{2}-1-i_3,q-1\Big)\ \Big|\ i_4\le \frac{q}{2}-1,i_3\le q-1\Big\}\\
			=&\Big(0,0,\frac{q}{2}-1,\frac{q}{2},0,0,q-1\Big),\\
			\min R_3=&\min\{x\cdot q^{m-2-2}\mid x\in S_3\}\\
			=&\min\Big\{\Big(0,0,q-1,q-1,\frac{q}{2}-1,\frac{q}{2},\frac{q}{2}-1-i_4\Big)\ \Big|\ i_4\le \frac{q}{2}-2\Big\},\\
			\min R_4=&\min\{x\cdot q^{m-e-2}\mid x\in S_4,\ e= 2\}\\
			=&\min\Big\{\Big(0,0,\frac{q}{2}-1+i_1,\frac{q}{2}+i_0,\frac{q}{2}-1,\frac{q}{2},\frac{q}{2}-1-i_4\Big)\ \Big|\\
			&\quad i_4\le \frac{q}{2}-2,\mathrm{either}\ i_1\le\frac{q}{2}-1,\ \mathrm{or}\ i_1=\frac{q}{2}\ \mathrm{and}\ 1\le i_0\le\frac{q}{2}-2\Big\},\\
			=&\Big(0,0,\frac{q}{2}-1,\frac{q}{2},\frac{q}{2}-1,\frac{q}{2},1\Big)_q,\ \mathrm{and}\\
			\min R_5=&\min\{x\cdot q^{2m-f-2}\mid x\in S_5,\ f=2\}\\
			=&\min\Big\{\Big(0,0,q-1,\frac{q}{2}-i_1,\frac{q}{2}-1-i_0,\frac{q}{2},\frac{q}{2}\Big)\ \Big|\ i_1\le q-1,1\le i_0\le q-1\Big\}\\
			=&\Big(0,0,q-2,\frac{q}{2},\frac{q}{2},\frac{q}{2},\frac{q}{2}\Big)_q.
		\end{align*}
		Hence, we have $\min\limits_{i\in T_1}\CL(a+i)=\min R_2=\delta_1$.
		\item When $i_2\ge\frac{q}{2}+1$ and $i_1= \frac{q}{2}$, take $a := \Delta_1$. Let
		\begin{equation*}
			a+i=\Delta_1+i=\left(\frac{q}{2}-1,\frac{q}{2},\frac{q}{2}-1+i_4,\frac{q}{2}+1+i_3,i_2+\frac{q}{2}-q,0,\frac{q}{2}+i_0\right)
		\end{equation*}
		and
		\begin{equation*}
			n-a=n-\Delta_1=\left(\frac{q}{2},\frac{q}{2}-1,\frac{q}{2},\frac{q}{2}-1,\frac{q}{2},\frac{q}{2}-1,\frac{q}{2}+1\right)_q.
		\end{equation*}
		For any $x\in X$, we have $1\le x_2\le \frac{q}{2}-1$, and $x_1$ equals 1 if $0\le x_0\le \frac{q}{2}-1$ and 0 if $\frac{q}{2}+1\le x_0\le q-1$. That is, $x_2\notin\{0,q-1\}$, $x_1\ne q-1$ and $(x_1,x_0)\notin\{(0,0),(0,1)\}$. Since $i_4\le \frac{q}{2}-1$, $(x_4,x_3)\notin\{(0,0),(q-1,q-1)\}$. Then we have $L_1=\dots=L_5=1$, which leads to $\min\limits_{i\in T_1}\CL(a+i)\ge q^5>\delta_1$ by Lemma~\ref{sec3 lem0}.
		\item When $i_2\ge\frac{q}{2}+1$ and $i_1\ne \frac{q}{2}$, take $a := n-\Delta_{1}$. Let
		\begin{equation*}
			a+i=n-\Delta_1+i=\left(\frac{q}{2},\frac{q}{2}-1,\frac{q}{2}+i_4,\frac{q}{2}+i_3,i_2-\frac{q}{2},\frac{q}{2}-1+i_1,\frac{q}{2}+1+i_0\right)
		\end{equation*}
		and
		\begin{equation*}
			n-a=\Delta_1=\left(\frac{q}{2}-1,\frac{q}{2},\frac{q}{2}-1,\frac{q}{2},\frac{q}{2}-1,\frac{q}{2},\frac{q}{2}\right)_q.
		\end{equation*}
		Then for any $x\in X$, we have $1\le x_2\le \frac{q}{2}$, and so $x_2\not\in\{0,q-1\}$. Since $i_1\ne\frac{q}{2}$, $(x_1,x_0)\not\in\{(0,0),(0,1)\}$. Note that $(x_4,x_3)\ne(0,0)$ because $i_4\le \frac{q}{2}-2$ if $i_3=\frac{q}{2}$. Hence, we have $L_1=\dots=L_4=1$ and $L_5=2$. Thus $\max\limits_{1\le j\le 5}L_j=L_5=2$, and so $\min\limits_{i\in T_1}\CL(a+i)=\min\limits_{x \in X}\CL(x)=\min R_5$. Note that
		\begin{align*}
			S_5=&\Big\{\Big(\frac{q}{2},\frac{q}{2}-1,\frac{q}{2}+i_4,\frac{q}{2}+i_3,i_2-\frac{q}{2},\frac{q}{2}-1+i_1,\frac{q}{2}+1+i_0\Big)\ \Big|\\
			&\quad i_4=i_3=\frac{q}{2}-1,i_2\ge\frac{q}{2}+1,\ i_1\ne\frac{q}{2},\ 1\le i_0\le q-1\Big\}\\
			=&\Big\{\Big(\frac{q}{2},\frac{q}{2}-1,q-1,q-1,i_2-\frac{q}{2},\frac{q}{2}-1+i_1,\frac{q}{2}+1+i_0\Big)\ \Big|\\
			&\quad i_2\ge\frac{q}{2}+1,\ i_1\ne\frac{q}{2},1\le i_0\le q-1\Big\}.
		\end{align*}
		Hence, we have
		\begin{align*}
			\min R_5=&\min\{x\cdot q^{2m-f-2}\mid x\in S_5,\ f=2\}\\
			=&\min\Big\{\Big(0,0,\frac{3q}{2}-1-i_2,\frac{q}{2}-i_1,\frac{q}{2}-1-i_0,\frac{q}{2},\frac{q}{2}\Big)\ \Big|\ i_2\ge\frac{q}{2}+1,\ i_1\ne\frac{q}{2},1\le i_0\le q-1\Big\}\\
			=&\Big(0,0,\frac{3q}{2}-1-(q-1),\frac{q}{2}-(q-1),\frac{q}{2}-1-(q-1),\frac{q}{2},\frac{q}{2}\Big)\\
			=&\Big(0,0,\frac{q}{2}-1,\frac{q}{2},\frac{q}{2},\frac{q}{2},\frac{q}{2}\Big)_q>\delta_1.
		\end{align*}
	\end{itemize}
	In summary, we have $a+i \in I_1$, and so $n+i=(a+i)+(n-a) \in I_1+I_2$ for any $i \in T_1$.
	
	\item $\boldsymbol{m=8:}$ we have
	\begin{equation*}
		i=(\underbrace{0,0,\ldots,0}_{m-r-1},i_r\ne0,i_{r-1},\ldots,i_1,i_0\ne0)_q,
	\end{equation*}
	where $0\leq r\leq 5$ as $$i<\delta_{1}=\frac{2q^7-3q^3+q}{2(q+1)}=\left(0,0,q-1,0,q-2,q-1,\frac{q}{2},\frac{q}{2}\right)_q.$$ With the same notation as in Lemma~\ref{sec3 lem0}, take $X := a + T_1$. We only need to prove that there always exists an integer $a\in C_{\Delta_1}$ such that $\min\limits_{i\in T_1}\CL(a+i)\ge\delta_{1}$. Through calculation, the largest coset leader in $T_1$ is
	\begin{equation*}
		\left(0,0,q-1,0,q-2,q-1,\frac{q}{2},\frac{q}{2}-1\right)_q.
	\end{equation*}
	\begin{itemize}
		\item When $i_5\ne \frac{q}{2}-1,i_4\ne \frac{q}{2},i_3\ne \frac{q}{2}-1,i_2\ne \frac{q}{2}-1$ and $i_1\ne\frac{q}{2}$, let
		\begin{equation*}
			a+i=\left(\frac{q}{2}-1,\frac{q}{2}-1,\frac{q}{2}+i_5,\frac{q}{2}-1+i_4,\frac{q}{2}+i_3,\frac{q}{2}+i_2,\frac{q}{2}-1+i_1,\frac{q}{2}+i_0\right)
		\end{equation*}
		and
		\begin{equation*}
			n-a=\left(\frac{q}{2},\frac{q}{2},\frac{q}{2}-1,\frac{q}{2},\frac{q}{2}-1,\frac{q}{2}-1,\frac{q}{2},\frac{q}{2}+1\right)_q,
		\end{equation*}
		then we have $L_j=1$ for $1\le j\le 5$. According to Lemma~\ref{sec3 lem0}, $\min\limits_{i\in T_1}\CL(a+i)\ge q^6>\delta_1$.
		
		\item When $i_5=\frac{q}{2}-1,i_4\ne \frac{q}{2},i_3\ne \frac{q}{2}-1,i_2\ne \frac{q}{2}-1$ and $i_1\ne\frac{q}{2}$, let
		\begin{equation*}
			a+i=\left(\frac{q}{2},\frac{q}{2},q-2,\frac{q}{2}-1+i_4,\frac{q}{2}+i_3,\frac{q}{2}+i_2,\frac{q}{2}-1+i_1,\frac{q}{2}+1+i_0\right)
		\end{equation*}
		and
		\begin{equation*}
			n-a=\left(\frac{q}{2}-1,\frac{q}{2}-1,\frac{q}{2},\frac{q}{2},\frac{q}{2}-1,\frac{q}{2}-1,\frac{q}{2},\frac{q}{2}\right)_q.
		\end{equation*}
		Hence, we have $\max\limits_{1\le j\le 5}L_j=1$. By Lemma~\ref{sec3 lem0}, $\min\limits_{i\in T_1}\CL(a+i)\ge q^6>\delta_1$.
		
		\item When $i_4\notin\{\frac{q}{2}-1,\frac{q}{2}\},i_3= \frac{q}{2}-1,i_2\ne \frac{q}{2}-1$ and $i_1\ne\frac{q}{2}$, let
		\begin{equation*}
			a+i=\left(\frac{q}{2}-1,\frac{q}{2}-1,\frac{q}{2}+i_5,\frac{q}{2}+i_4,q-2,\frac{q}{2}+i_2,\frac{q}{2}-1+i_1,\frac{q}{2}+i_0\right)
		\end{equation*}
		and
		\begin{equation*}
			n-a=\left(\frac{q}{2},\frac{q}{2},\frac{q}{2}-1,\frac{q}{2}-1,\frac{q}{2},\frac{q}{2}-1,\frac{q}{2},\frac{q}{2}+1\right)_q.
		\end{equation*}
		It is clear that $L_1=L_2=\dots=L_5=1$. According to Lemma~\ref{sec3 lem0}, $\min\limits_{i\in T_1}\CL(a+i)\ge q^6>\delta_1$.
		
		\item When $i_4=\frac{q}{2}-1,i_3= \frac{q}{2}-1,i_2\ne \frac{q}{2}-1$ and $i_1\ne\frac{q}{2}$, let
		\begin{equation*}
			a+i=\left(\frac{q}{2}-1,\frac{q}{2},\frac{q}{2}+i_5,q-2,q-2,\frac{q}{2}+i_2,\frac{q}{2}-1+i_1,\frac{q}{2}+1+i_0\right)
		\end{equation*}
		and
		\begin{equation*}
			n-a=\left(\frac{q}{2},\frac{q}{2}-1,\frac{q}{2}-1,\frac{q}{2},\frac{q}{2},\frac{q}{2}-1,\frac{q}{2},\frac{q}{2}\right)_q.
		\end{equation*}
		Then we have $L_1=\dots=L_5=1$. It then follows from Lemma~\ref{sec3 lem0} that $\min\limits_{i\in T_1}\CL(a+i)\ge q^6>\delta_1$.
		
		\item When $i_4=\frac{q}{2},i_3=\frac{q}{2}-1$ and
		\begin{equation*}
			\left\{\begin{aligned}
				&i_5\ne\frac{q}{2}-1,i_2\ne \frac{q}{2}-1 \mathrm{\ and\ } i_1\ne\frac{q}{2}, \mathrm{\ or}\\
				&i_5=\frac{q}{2}-1, i_2\ge \frac{q}{2}+1 \mathrm{\ and\ } i_1\ne\frac{q}{2}, \mathrm{\ or}\\
				&i_5=\frac{q}{2}-1,i_2= \frac{q}{2} \mathrm{\ and\ } i_1\ge\frac{q}{2}+1, \mathrm{\ or}\\
				&i_5=\frac{q}{2}-1,i_2= \frac{q}{2},i_1=\frac{q}{2}-1, \mathrm{\ and\ } i_0\ge\frac{q}{2},
			\end{aligned}\right.
		\end{equation*} let
		\begin{equation*}
			a+i=\left(\frac{q}{2}-1,\frac{q}{2}-1,\frac{q}{2}+1+i_5,0,q-2,\frac{q}{2}+i_2,\frac{q}{2}-1+i_1,\frac{q}{2}+i_0\right)
		\end{equation*}
		and
		\begin{equation*}
			n-a=\left(\frac{q}{2},\frac{q}{2},\frac{q}{2}-1,\frac{q}{2}-1,\frac{q}{2},\frac{q}{2}-1,\frac{q}{2},\frac{q}{2}+1\right)_q.
		\end{equation*}
		Then we get that $L_1=L_2=L_3=L_5=1$ and $L_4=2$. Hence $\max\limits_{1\le j\le 5}L_j=L_4=2$, and so $\min\limits_{i\in T_1}\CL(a+i)=\min\limits_{x\in X}\CL(x)=\min R_4$. Note that
		\begin{align*}
			S_4=&\Big\{\Big(\frac{q}{2}-1,\frac{q}{2}-1,\frac{q}{2}+1+i_5,0,q-2,\frac{q}{2}+i_2,\frac{q}{2}-1+i_1,\frac{q}{2}+i_0\Big)\ \Big|\ i_5=\frac{q}{2}-1,\\
			\quad& \text{either } i_2\ge \frac{q}{2}+1, i_1\ne\frac{q}{2}\mathrm{\ and\ } 1\le i_0\le q-1,i_2= \frac{q}{2},i_1\ge\frac{q}{2}+1\mathrm{\ and\ } 1\le i_0\le q-1, \\
			\quad& \mathrm{or\ }i_2= \frac{q}{2},i_1=\frac{q}{2}-1 \mathrm{\ and\ } i_0\ge\frac{q}{2},\Big\}\\
			=&\Big\{\Big(\frac{q}{2}-1,\frac{q}{2},0,0,q-1,i_2-\frac{q}{2},\frac{q}{2}-1+i_1,\frac{q}{2}+i_0\Big)\ \Big|\\
			&\quad \text{either } i_2\ge \frac{q}{2}+1, i_1\ne\frac{q}{2}\mathrm{\ and\ } 1\le i_0\le q-1,\text{ or }i_2= \frac{q}{2},i_1\ge\frac{q}{2}+1\mathrm{\ and\ } 1\le i_0\le q-1, \\
			&\quad \mathrm{or\ }i_2= \frac{q}{2},i_1=\frac{q}{2}-1 \mathrm{\ and\ } i_0\ge\frac{q}{2},\Big\}.
		\end{align*}
		Hence, we have
		\begin{align*}
			\min R_4=&\min\{x\cdot q^{m-e-2}\mid x\in S_4,\ e=4\}\\
			=&\min\Big\{\Big(0,0,q-1,i_2-\frac{q}{2},\frac{q}{2}-1+i_1,\frac{q}{2}+i_0-1,\frac{q}{2},\frac{q}{2}\Big)\ \Big|\\
			&\quad \text{either } i_2\ge \frac{q}{2}+1, i_1\ne\frac{q}{2}\mathrm{\ and\ } 1\le i_0\le q-1,\text{ or }i_2= \frac{q}{2},i_1\ge\frac{q}{2}+1\mathrm{\ and\ } \\
			&\quad 1\le i_0\le q-1,\mathrm{or\ }i_2= \frac{q}{2},i_1=\frac{q}{2}-1 \mathrm{\ and\ } i_0\ge\frac{q}{2},\Big\}\\
			=&\Big(0,0,q-1,0,q-2,q-1,\frac{q}{2},\frac{q}{2}\Big)_q=\delta_1.
		\end{align*}
		
		\item When $i_5=\frac{q}{2}-1,i_4=\frac{q}{2},i_3=\frac{q}{2}-1,i_2\le \frac{q}{2}-2$ and $i_1\ne\frac{q}{2}$, let
		\begin{equation*}
			a+i=\left(\frac{q}{2}-1,\frac{q}{2},q-1,0,q-1,\frac{q}{2}-1+i_2,\frac{q}{2}-1+i_1,\frac{q}{2}+1+i_0\right)
		\end{equation*}
		and
		\begin{equation*}
			n-a=\left(\frac{q}{2},\frac{q}{2}-1,\frac{q}{2},\frac{q}{2}-1,\frac{q}{2}-1,\frac{q}{2},\frac{q}{2},\frac{q}{2}\right)_q.
		\end{equation*}
		Since $i_2\le\frac{q}{2}-2$, we have $x_2\notin\{q-1,0\}$, and so  $L_1=\dots=L_5=1$. By Lemma~\ref{sec3 lem0}, $\min\limits_{i\in T_1}\CL(a+i)\ge q^6>\delta_1$.
		
		\item When $i_5=\frac{q}{2}-1,i_4=\frac{q}{2},i_3=\frac{q}{2}-1,i_2= \frac{q}{2}$, and either $i_1\le\frac{q}{2}-2$, or $i_1=\frac{q}{2}-1$ and $i_0\le\frac{q}{2}-1$, let
		\begin{equation*}
			a+i=\left(\frac{q}{2},\frac{q}{2}-1,q-1,0,q-1,0,\frac{q}{2}+i_1,\frac{q}{2}+i_0\right)
		\end{equation*}
		and
		\begin{equation*}
			n-a=\left(\frac{q}{2}-1,\frac{q}{2},\frac{q}{2},\frac{q}{2}-1,\frac{q}{2},\frac{q}{2}-1,\frac{q}{2}-1,\frac{q}{2}+1\right)_q.
		\end{equation*}
		Note that $\max\limits_{1\le j\le 5}L_j=1$. Again by Lemma~\ref{sec3 lem0}, $\min\limits_{i\in T_1}\CL(a+i)\ge q^6>\delta_1$.
		
		\item When $i_4=\frac{q}{2}-1,i_2=\frac{q}{2}-1,i_1\ne\frac{q}{2}$ and either $i_3\le\frac{q}{2}-2$, or $i_5=\frac{q}{2}$ and $i_3\ge\frac{q}{2}$, let
		\begin{equation*}
			a+i=\left(\frac{q}{2},\frac{q}{2}-1,\frac{q}{2}+i_5,q-2,\frac{q}{2}+i_3,q-2,\frac{q}{2}-1+i_1,\frac{q}{2}+1+i_0\right)
		\end{equation*}
		and
		\begin{equation*}
			n-a=\left(\frac{q}{2}-1,\frac{q}{2},\frac{q}{2}-1,\frac{q}{2},\frac{q}{2}-1,\frac{q}{2},\frac{q}{2},\frac{q}{2}\right)_q,
		\end{equation*}
		then $x_3\ne q-1$ for any $x\in X$, and so $\max\limits_{1\le j\le 5}L_j=1$. In accordance with Lemma~\ref{sec3 lem0}, $\min\limits_{i\in T_1}\CL(a+i)\ge q^6>\delta_1$.
		
		\item When $i_5\ne\frac{q}{2},i_4=\frac{q}{2}-1,i_3\ge\frac{q}{2},i_2=\frac{q}{2}-1$ and $i_1\ge\frac{q}{2}+1$, let
		\begin{equation*}
			a+i=\left(\frac{q}{2},\frac{q}{2}-1,\frac{q}{2}+i_5,0,i_3-\frac{q}{2},q-1,i_1-\frac{q}{2},\frac{q}{2}+i_0\right)
		\end{equation*}
		and
		\begin{equation*}
			n-a=\left(\frac{q}{2}-1,\frac{q}{2},\frac{q}{2},\frac{q}{2}-1,\frac{q}{2}-1,\frac{q}{2},\frac{q}{2}-1,\frac{q}{2}+1\right)_q.
		\end{equation*}
		Then $x_1\notin\{0,q-1\},x_3\ne q-1$ and $x_5\ne0$ for any $x\in X$, which implies that $L_1=L_2=L_5=1,L_3=0$ and $L_4=2$. Hence $\max\limits_{1\le j\le 5}L_j=L_4=2$, and so $\min\limits_{i\in T_1}\CL(a+i)=\min\limits_{x\in X}\CL(x)=\min R_4$. Note that
		\begin{align*}
			S_4=&\Big\{\Big(\frac{q}{2},\frac{q}{2}-1,\frac{q}{2}+i_5,0,i_3-\frac{q}{2},q-1,i_1-\frac{q}{2},\frac{q}{2}+i_0\Big)\ \Big|\\
			&\quad i_5\ne\frac{q}{2},i_3=\frac{q}{2},i_1\ge\frac{q}{2}+1,1\le i_0\le q-1\Big\}\\
			=&\Big\{\Big(\frac{q}{2},\frac{q}{2}-1,\frac{q}{2}+i_5,0,0,q-1,i_1-\frac{q}{2},\frac{q}{2}+i_0\Big)\ \Big|\ i_5\ne\frac{q}{2},i_1\ge\frac{q}{2}+1,1\le i_0\le q-1\Big\}.
		\end{align*}
		Hence, we have
		\begin{align*}
			\min R_4=&\min\{x\cdot q^{m-e-2}\mid x\in S_4,\ e=3\}\\
			=&\min\Big\{\Big(0,0,q-1,i_1-\frac{q}{2},\frac{q}{2}-1+i_0,\frac{q}{2}-1,\frac{q}{2},\frac{q}{2}-i_5\Big)\ \Big|\\
			&\quad i_5\ne\frac{q}{2},i_1\ge\frac{q}{2}+1,1\le i_0\le q-1\Big\}\\
			=&\Big(0,0,q-1,1,\frac{q}{2},\frac{q}{2}-1,\frac{q}{2}-1,\frac{q}{2}+2\Big)>\delta_1.
		\end{align*}
		
		\item When $i_5\ne\frac{q}{2},i_4=\frac{q}{2}-1,i_3\ge\frac{q}{2},i_2=\frac{q}{2}-1$, and either $i_1\le\frac{q}{2}-2$, or $i_1=\frac{q}{2}-1$ and $i_0\in\{\frac{q}{2},\frac{q}{2}-1\}$, let
		\begin{equation*}
			a+i=\left(\frac{q}{2},\frac{q}{2},\frac{q}{2}-1+i_5,q-1,i_3-\frac{q}{2},q-1,\frac{q}{2}-1+i_1,\frac{q}{2}+1+i_0\right)
		\end{equation*}
		and
		\begin{equation*}
			n-a=\left(\frac{q}{2}-1,\frac{q}{2}-1,\frac{q}{2},\frac{q}{2},\frac{q}{2}-1,\frac{q}{2}-1,\frac{q}{2},\frac{q}{2}\right)_q,
		\end{equation*}
		then $x_1\ne0,x_3\ne q-1$ and $x_5\ne q-1$ for any $x\in X$. Hence, we have $L_1=L_2=L_4=1,L_3=0$ and $L_5=2$. Hence $\max\limits_{1\le j\le 5}L_j=L_5=2$, and so $\min\limits_{i\in T_1}\CL(a+i)=\min\limits_{x\in X}\CL(x)=\min R_5$. Note that
		\begin{align*}
			S_5=&\Big\{\Big(\frac{q}{2},\frac{q}{2},\frac{q}{2}-1+i_5,q-1,i_3-\frac{q}{2},q-1,\frac{q}{2}-1+i_1,\frac{q}{2}+1+i_0\Big)\ \Big|\\
			&\quad i_5\ne\frac{q}{2},i_3\ge\frac{q}{2},i_1=\frac{q}{2}-1,i_0=\frac{q}{2}\Big\}\\
			=&\Big\{\Big(\frac{q}{2},\frac{q}{2},\frac{q}{2}-1+i_5,q-1,i_3-\frac{q}{2},q-1,q-1,1\Big)\ \Big|\ i_5\ne\frac{q}{2},i_3\ge\frac{q}{2}\Big\}.
		\end{align*}
		Hence, we have
		\begin{align*}
			\min R_5=&\min\{x\cdot q^{2m-f-2}\mid x\in S_5,\ f=1\}\\
			=&\min\Big\{\Big(0,0,q-1,\frac{q}{2},\frac{q}{2},\frac{q}{2}-1+i_5,q-1,i_3-\frac{q}{2}+1\Big)\ \Big|\ i_5\ne\frac{q}{2},i_3\ge\frac{q}{2}\Big\}\\
			\ge&\Big(0,0,q-1,\frac{q}{2},\frac{q}{2},\frac{q}{2}-1,q-1,1\Big)>\delta_1.
		\end{align*}
		
		\item When $i_5\ne\frac{q}{2},i_4=\frac{q}{2}-1,i_3\ge\frac{q}{2},i_2=\frac{q}{2}-1,i_1=\frac{q}{2}-1$ and $i_0\notin\{\frac{q}{2},\frac{q}{2}-1\}$, let
		\begin{equation*}
			a+i=\left(\frac{q}{2},\frac{q}{2},\frac{q}{2}-1+i_5,q-1,i_3-\frac{q}{2},q-2,\frac{q}{2}+i_1,\frac{q}{2}+1+i_0\right)
		\end{equation*}
		and
		\begin{equation*}
			n-a=\left(\frac{q}{2}-1,\frac{q}{2}-1,\frac{q}{2},\frac{q}{2},\frac{q}{2}-1,\frac{q}{2},\frac{q}{2}-1,\frac{q}{2}\right)_q,
		\end{equation*}
		then $x_0\notin\{0,1\},x_3\ne q-1$ and $x_5\ne q-1$ for any $x\in X$. Thus, we have $L_1=L_2=L_3=0$ and $L_4=L_5=1$. It then follows from Lemma~\ref{sec3 lem0} that $\min\limits_{i\in T_1}\CL(a+i)\ge q^6>\delta_1.$
		
		\item When $i_2=\frac{q}{2}-1,i_1=\frac{q}{2}-1$ and 
		\begin{equation*}
			\left\{\begin{aligned}
				&i_5\ne\frac{q}{2},i_4\ne\frac{q}{2}-1,i_3\ne\frac{q}{2}-1,\mathrm{\ or\ }\\
				&i_5=\frac{q}{2},i_4\notin\{\frac{q}{2},\frac{q}{2}-1,\frac{q}{2}-2\},i_3\ne\frac{q}{2}-1,\mathrm{\ or\ }\\
				&i_5=\frac{q}{2},i_4=\frac{q}{2},i_3=\frac{q}{2},
			\end{aligned}\right.
		\end{equation*} let
		\begin{equation*}
			a+i=\left(\frac{q}{2}-1,\frac{q}{2},\frac{q}{2}-1+i_5,\frac{q}{2}+i_4,\frac{q}{2}+i_3,q-2,\frac{q}{2}-1+i_1,\frac{q}{2}+1+i_0\right)
		\end{equation*}
		and
		\begin{equation*}
			n-a=\left(\frac{q}{2},\frac{q}{2}-1,\frac{q}{2},\frac{q}{2}-1,\frac{q}{2}-1,\frac{q}{2},\frac{q}{2},\frac{q}{2}\right)_q,
		\end{equation*}
		Hence, we have $\max\limits_{1\le j\le 5}L_j=1$. By Lemma~\ref{sec3 lem0}, we get  $\min\limits_{i\in T_1}\CL(a+i)\ge q^6>\delta_1.$ 
		
		\item When $i_5=\frac{q}{2},i_2=\frac{q}{2}-1,i_1=\frac{q}{2}-1$, and either $i_4=\frac{q}{2}-2$ and $i_3\ne\frac{q}{2}-1$, or $i_4=\frac{q}{2}$ and $i_3\notin\{\frac{q}{2},\frac{q}{2}-1\}$, let
		\begin{equation*}
			a+i=\left(\frac{q}{2}-1,\frac{q}{2}+1,0,\frac{q}{2}-1+i_4,\frac{q}{2}+i_3,q-2,\frac{q}{2}-1+i_1,\frac{q}{2}+1+i_0\right)
		\end{equation*}
		and
		\begin{equation*}
			n-a=\left(\frac{q}{2},\frac{q}{2}-1,\frac{q}{2}-1,\frac{q}{2},\frac{q}{2}-1,\frac{q}{2},\frac{q}{2},\frac{q}{2}\right)_q,
		\end{equation*}
		One can easily check that $x_3\ne q-1$ and $(x_5,x_4),(x_4,x_3)\ne(0,0)$ for any $x\in X$. Hence, we have $\max\limits_{1\le j\le 5}L_j=1$. It follows from Lemma~\ref{sec3 lem0} that $\min\limits_{i\in T_1}\CL(a+i)\ge q^6>\delta_1.$ 
		
		\item When $i_5\ne\frac{q}{2}-1,i_3=\frac{q}{2}-1,i_2=\frac{q}{2}-1$ and $i_1\notin\{\frac{q}{2}+1,\frac{q}{2},\frac{q}{2}-1\}$, let
		\begin{equation*}
			a+i=\left(\frac{q}{2}-1,\frac{q}{2}-1,\frac{q}{2}+i_5,\frac{q}{2}+i_4,q-2,q-1,\frac{q}{2}-1+i_1,\frac{q}{2}+i_0\right)
		\end{equation*}
		and
		\begin{equation*}
			n-a=\left(\frac{q}{2},\frac{q}{2},\frac{q}{2}-1,\frac{q}{2}-1,\frac{q}{2},\frac{q}{2}-1,\frac{q}{2},\frac{q}{2}+1\right)_q.
		\end{equation*}
		Then $(x_5,x_4),(x_2,x_1)\notin\{(0,0),(q-1,q-1)\}$ for any $x\in X$, which implies that $L_1=L_2=\cdots=L_5=1$. According to Lemma~\ref{sec3 lem0}, $\min\limits_{i\in T_1}\CL(a+i)\ge q^6>\delta_1.$ 
		
		\item When $i_4\ne\frac{q}{2}-1,i_3=\frac{q}{2}-1,i_2=\frac{q}{2}-1$, and $i_1=\frac{q}{2}-1$, let
		\begin{equation*}
			a+i=\left(\frac{q}{2}-1,\frac{q}{2},\frac{q}{2}-1+i_5,\frac{q}{2}+i_4,q-1,q-2,q-2,\frac{q}{2}+1+i_0\right)
		\end{equation*}
		and
		\begin{equation*}
			n-a=\left(\frac{q}{2},\frac{q}{2}-1,\frac{q}{2},\frac{q}{2}-1,\frac{q}{2}-1,\frac{q}{2},\frac{q}{2},\frac{q}{2}\right)_q,
		\end{equation*}
		then $x_4\ne q-1$ for any $x\in X$. Hence, we have $L_1=L_2=L_5=1,L_3=0$ and $L_4=2$.  Hence $\max\limits_{1\le j\le 5}L_j=L_4=2$, and so $\min\limits_{i\in T_1}\CL(a+i)=\min\limits_{x\in X}\CL(x)=\min R_4$. Note that
		\begin{align*}
			S_4=&\Big\{\Big(\frac{q}{2}-1,\frac{q}{2},\frac{q}{2}-1+i_5,\frac{q}{2}+i_4,q-1,q-2,q-2,\frac{q}{2}+1+i_0\Big)\ \Big|\ i_5=i_4=\frac{q}{2},1\le i_0\le q-1\Big\}\\
			=&\Big\{\Big(\frac{q}{2}-1,\frac{q}{2}+1,0,0,q-1,q-2,q-2,\frac{q}{2}+1+i_0\Big)\ \Big|\ 1\le i_0\le q-1\Big\}.
		\end{align*}
		Hence, we have
		\begin{align*}
			\min R_4=&\min\{x\cdot q^{m-e-2}\mid x\in S_4,\ e=4\}\\
			=&\min\Big\{\Big(0,0,q-1,q-2,q-2,\frac{q}{2}+i_0,\frac{q}{2},\frac{q}{2}-1\Big)\ \Big|\ 1\le i_0\le q-1\Big\}\\
			=&\Big(0,0,q-1,q-2,q-2,\frac{q}{2}+1,\frac{q}{2},\frac{q}{2}-1\Big)_q>\delta_1.
		\end{align*}
		
		\item When $i_4=\frac{q}{2}-1,i_3=\frac{q}{2}-1,i_2=\frac{q}{2}-1$, and $i_1=\frac{q}{2}-1$, let
		\begin{equation*}
			a+i=\left(\frac{q}{2}-1,\frac{q}{2},\frac{q}{2}+i_5,q-2,q-1,q-2,q-2,\frac{q}{2}+1+i_0\right)
		\end{equation*}
		and
		\begin{equation*}
			n-a=\left(\frac{q}{2},\frac{q}{2}-1,\frac{q}{2}-1,\frac{q}{2},\frac{q}{2}-1,\frac{q}{2},\frac{q}{2},\frac{q}{2}\right)_q,
		\end{equation*}
		then we have $\max\limits_{1\le j\le 5}L_j=1$. It then follows from Lemma~\ref{sec3 lem0} that $\min\limits_{i\in T_1}\CL(a+i)\ge q^6>\delta_1.$
		
		\item When $i_5\ne\frac{q}{2}-1,i_3=\frac{q}{2}-1,i_2=\frac{q}{2}-1$, and $i_1=\frac{q}{2}+1$, let
		\begin{equation*}
			a+i=\left(\frac{q}{2}-1,\frac{q}{2},\frac{q}{2}+i_5,\frac{q}{2}-1+i_4,q-1,0,1,\frac{q}{2}+i_0\right)
		\end{equation*}
		and
		\begin{equation*}
			n-a=\left(\frac{q}{2},\frac{q}{2}-1,\frac{q}{2}-1,\frac{q}{2},\frac{q}{2},\frac{q}{2}-1,\frac{q}{2}-1,\frac{q}{2}+1\right)_q.
		\end{equation*}
		Then $(x_5,x_4)\notin\{(0,0),(q-1,q-1)\}$ for any $x\in X$, and so $L_1=L_2=L_4=1,L_3=0$ and $L_5=2$. Hence $\max\limits_{1\le j\le 5}L_j=L_5=2$, and so $\min\limits_{i\in T_1}\CL(a+i)=\min\limits_{x\in X}\CL(x)=\min R_5$. Note that
		\begin{align*}
			S_5=&\Big\{\Big(\frac{q}{2}-1,\frac{q}{2},\frac{q}{2}+i_5,\frac{q}{2}-1+i_4,q-1,0,1,\frac{q}{2}+i_0\Big)\ \Big|\ i_5\ne\frac{q}{2}-1,i_4=\frac{q}{2},1\le i_0\le q-1\Big\}\\
			=&\Big\{\Big(\frac{q}{2}-1,\frac{q}{2},\frac{q}{2}+i_5,q-1,q-1,0,1,\frac{q}{2}+i_0\Big)\ \Big|\ i_5\ne\frac{q}{2}-1,1\le i_0\le q-1\Big\}.
		\end{align*}
		Hence, we have
		\begin{align*}
			\min R_5=&\min\{x\cdot q^{2m-f-2}\mid x\in S_5,\ f=3\}\\
			=&\min\Big\{\Big(0,0,q-1,q-2,\frac{q}{2}-i_0,\frac{q}{2}-1,\frac{q}{2},\frac{q}{2}+1+i_5\Big)\ \Big|\ i_5\ne\frac{q}{2}-1,1\le i_0\le q-1\Big\}\\
			=&\Big(0,0,q-1,q-3,\frac{q}{2}+1,\frac{q}{2}-1,\frac{q}{2},\frac{q}{2}+1\Big)_q>\delta_1.
		\end{align*}		
		
		\item When $i_5=\frac{q}{2}-1,i_3=\frac{q}{2}-1,i_2=\frac{q}{2}-1$ and $i_1\ne\frac{q}{2}-1$, let
		\begin{equation*}
			a+i=\left(\frac{q}{2},\frac{q}{2},q-2,\frac{q}{2}+i_4,q-2,q-2,\frac{q}{2}+i_1,\frac{q}{2}+1+i_0\right)
		\end{equation*}
		and
		\begin{equation*}
			n-a=\left(\frac{q}{2}-1,\frac{q}{2}-1,\frac{q}{2},\frac{q}{2}-1,\frac{q}{2},\frac{q}{2},\frac{q}{2}-1,\frac{q}{2}\right)_q,
		\end{equation*}
		Hence, we have $\max\limits_{1\le j\le 5}L_j=1$. It follows from Lemma~\ref{sec3 lem0} that $\min\limits_{i\in T_1}\CL(a+i)\ge q^6>\delta_1.$ 
		
		\item When $i_4\ne\frac{q}{2}-1,i_1=\frac{q}{2}$ and 
		\begin{equation*}
			\left\{\begin{aligned}
				&i_5\ne\frac{q}{2}-1,i_3\ne\frac{q}{2},i_2\ne\frac{q}{2}, \mathrm{\ or\ }\\
				&i_3=\frac{q}{2},i_2\notin\{\frac{q}{2},\frac{q}{2}-1\},
			\end{aligned}\right.
		\end{equation*} let
		\begin{equation*}
			a+i=\left(\frac{q}{2},\frac{q}{2}-1,\frac{q}{2}+i_5,\frac{q}{2}+i_4,\frac{q}{2}-1+i_3,\frac{q}{2}+i_2,0,\frac{q}{2}+1+i_0\right)
		\end{equation*}
		and
		\begin{equation*}
			n-a=\left(\frac{q}{2}-1,\frac{q}{2},\frac{q}{2}-1,\frac{q}{2}-1,\frac{q}{2},\frac{q}{2},\frac{q}{2}-1,\frac{q}{2}\right)_q,
		\end{equation*}
		Note that $x_2\notin\{0,q-1\}$ when $i_3=\frac{q}{2}$. Hence, $L_1=L_2=L_3=1$ and $L_4=L_5=2$. Hence $\max\limits_{1\le j\le 5}L_j=2$, and so $\min\limits_{i\in T_1}\CL(a+i)=\min\limits_{x\in X}\CL(x)=\min (R_4\cup R_5)$. Note that
		\begin{align*}
			S_4=&\Big\{\Big(\frac{q}{2},\frac{q}{2}-1,\frac{q}{2}+i_5,\frac{q}{2}+i_4,\frac{q}{2}-1+i_3,\frac{q}{2}+i_2,0,\frac{q}{2}+1+i_0\Big)\ \Big|\\
			&\quad i_5=\frac{q}{2}-1,i_4=i_3=\frac{q}{2},\ i_2\le\frac{q}{2}-2,1\le i_0\le q-1\Big\}\\
			=&\Big\{\Big(\frac{q}{2},\frac{q}{2},0,0,q-1,\frac{q}{2}+i_2,0,\frac{q}{2}+1+i_0\Big)\ \Big|\ i_2\le\frac{q}{2}-2,1\le i_0\le q-1\Big\},\ \mathrm{and}\\
			S_5=&\Big\{\Big(\frac{q}{2},\frac{q}{2}-1,\frac{q}{2}+i_5,\frac{q}{2}+i_4,\frac{q}{2}-1+i_3,\frac{q}{2}+i_2,0,\frac{q}{2}+1+i_0\Big)\ \Big|\\
			&\quad i_5=\frac{q}{2}-1,\ i_4=\frac{q}{2}-2,\ i_3=\frac{q}{2},\ i_2\ge\frac{q}{2}+1,1\le i_0\le q-1\Big\}\\
			=&\Big\{\Big(\frac{q}{2},\frac{q}{2}-1,q-1,q-1,0,i_2-\frac{q}{2},0,\frac{q}{2}+1+i_0\Big)\ \Big|\ i_2\ge\frac{q}{2}+1,1\le i_0\le q-1\Big\}.
		\end{align*}
		Hence, we have
		\begin{align*}
			\min R_4=&\min\{x\cdot q^{m-e-2}\mid x\in S_4,\ e=4\}\\
			=&\min\Big\{\Big(0,0,q-1,\frac{q}{2}+i_2,0,\frac{q}{2}+i_0,\frac{q}{2}-1,\frac{q}{2}\Big)\ \Big|\ i_2\le\frac{q}{2}-2,1\le i_0\le q-1\Big\}\\
			=&\Big(0,0,q-1,\frac{q}{2},0,\frac{q}{2},\frac{q}{2}-1,\frac{q}{2}\Big)_q,\ \mathrm{and}\\
			\min R_5=&\min\{x\cdot q^{2m-f-2}\mid x\in S_5,\ f=4\}\\
			=&\min\Big\{\Big(0,0,q-1,\frac{3q}{2}-1-i_2,q-1,\frac{q}{2}-1-i_0,\frac{q}{2},\frac{q}{2}\Big)\ \Big|\ i_2\ge\frac{q}{2}+1,1\le i_0\le q-1\Big\}\\
			=&\Big(0,0,q-1,\frac{q}{2},q-2,\frac{q}{2},\frac{q}{2},\frac{q}{2}\Big)_q.
		\end{align*}
		Thus, $\min\limits_{i\in T_1}\CL(a+i)=\min R_4>\delta_1$.
		\item $i_5=\frac{q}{2}-1,i_4\ne\frac{q}{2}-1,i_3\ne\frac{q}{2},i_2\ne\frac{q}{2}$ and $i_1=\frac{q}{2}$, let
		\begin{equation*}
			a+i=\left(\frac{q}{2},\frac{q}{2},q-2,\frac{q}{2}+i_4,\frac{q}{2}-1+i_3,\frac{q}{2}+i_2,0,\frac{q}{2}+1+i_0\right)
		\end{equation*}
		and
		\begin{equation*}
			n-a=\left(\frac{q}{2}-1,\frac{q}{2}-1,\frac{q}{2},\frac{q}{2}-1,\frac{q}{2},\frac{q}{2},\frac{q}{2}-1,\frac{q}{2}\right)_q.
		\end{equation*}
		Hence, we have $L_1=\dots=L_5=1$. By Lemma~\ref{sec3 lem0}, $\min\limits_{i\in T_1}\CL(a+i)\ge q^6>\delta_1.$ 
		
		\item When $i_4\ne\frac{q}{2}-1,i_3=\frac{q}{2},i_2=\frac{q}{2}-1,i_1=\frac{q}{2}$ and $i_0\le\frac{q}{2}+1$, let
		\begin{equation*}
			a+i=\left(\frac{q}{2},\frac{q}{2}-1,\frac{q}{2}-1+i_5,\frac{q}{2}+1+i_4,0,q-1,0,\frac{q}{2}+i_0\right)
		\end{equation*}
		and
		\begin{equation*}
			n-a=\left(\frac{q}{2}-1,\frac{q}{2},\frac{q}{2},\frac{q}{2}-1,\frac{q}{2}-1,\frac{q}{2},\frac{q}{2}-1,\frac{q}{2}+1\right)_q,
		\end{equation*}
		then we have $x_4\ne0$, which implies that $L_1=\dots=L_4=1$ and $L_5=2$. Hence $\max\limits_{1\le j\le 5}L_j=L_5=2$, and so $\min\limits_{i\in T_1}\CL(a+i)=\min\limits_{x\in X}\CL(x)=\min R_5$. Note that
		\begin{align*}
			S_5=&\Big\{\Big(\frac{q}{2},\frac{q}{2}-1,\frac{q}{2}-1+i_5,\frac{q}{2}+1+i_4,0,q-1,0,\frac{q}{2}+i_0\Big)\ \Big|\ i_5=\frac{q}{2},i_4=\frac{q}{2}-2,1\le i_0\le \frac{q}{2}+1\Big\}\\
			=&\Big\{\Big(\frac{q}{2},\frac{q}{2}-1,q-1,q-1,0,q-1,0,\frac{q}{2}+i_0\Big)\ \Big|\ 1\le i_0\le \frac{q}{2}+1\Big\}.
		\end{align*}
		Hence, we have
		\begin{align*}
			\min R_5=&\min\{x\cdot q^{2m-f-2}\mid x\in S_5,\ f=4\}\\
			=&\min\Big\{\Big(0,0,q-1,0,q-1,\frac{q}{2}-i_0,\frac{q}{2},\frac{q}{2}\Big)\ \Big|\ 1\le i_0\le \frac{q}{2}+1\Big\}\\
			=&\Big(0,0,q-1,0,q-2,q-1,\frac{q}{2},\frac{q}{2}\Big)_q=\delta_1.
		\end{align*}
		
		\item When $i_4\ne\frac{q}{2}-1,i_3=\frac{q}{2},i_2=\frac{q}{2}-1,i_1=\frac{q}{2}$ and $i_0\ge\frac{q}{2}+2$, let
		\begin{equation*}
			a+i=\left(\frac{q}{2}-1,\frac{q}{2},\frac{q}{2}-1+i_5,\frac{q}{2}+1+i_4,0,q-1,0,i_0+1-\frac{q}{2}\right)
		\end{equation*}
		and
		\begin{equation*}
			n-a=\left(\frac{q}{2},\frac{q}{2}-1,\frac{q}{2},\frac{q}{2}-1,\frac{q}{2}-1,\frac{q}{2},\frac{q}{2},\frac{q}{2}\right)_q.
		\end{equation*}
		Since $x_4\ne0$ for any $x\in X$, $L_1=\dots=L_4=0$ and $L_5=2$. Hence $\max\limits_{1\le j\le 5}L_j=L_5=2$, and so $\min\limits_{i\in T_1}\CL(a+i)=\min\limits_{x\in X}\CL(x)=\min R_5$. Note that
		\begin{align*}
			S_5=&\Big\{\Big(\frac{q}{2}-1,\frac{q}{2},\frac{q}{2}-1+i_5,\frac{q}{2}+1+i_4,0,q-1,0,i_0+1-\frac{q}{2}\Big)\ \Big|\ i_5=\frac{q}{2},i_4=\frac{q}{2}-2,i_0\ge\frac{q}{2}+2\Big\}\\
			=&\Big\{\Big(\frac{q}{2}-1,\frac{q}{2},q-1,q-1,0,q-1,0,i_0+1-\frac{q}{2}\Big)\ \Big|\ i_0\ge\frac{q}{2}+2\Big\}.
		\end{align*}
		Hence, we have
		\begin{align*}
			\min R_5=&\min\{x\cdot q^{2m-f-2}\mid x\in S_5,\ f=4\}\\
			=&\min\Big\{\Big(0,0,q-1,0,q-1,\frac{3q}{2}-1-i_0,\frac{q}{2}-1,\frac{q}{2}+1\Big)\ \Big|\ i_0\ge\frac{q}{2}+2\Big\}\\
			=&\Big(0,0,q-1,0,q-1,\frac{q}{2},\frac{q}{2}-1,\frac{q}{2}+1\Big)_q>\delta_1.
		\end{align*}
		
		\item When $i_5\ne\frac{q}{2},i_4=\frac{q}{2}-1,i_3\ne\frac{q}{2}-1,i_2\ne\frac{q}{2}$ and $i_1=\frac{q}{2}$, let
		\begin{equation*}
			a+i=\left(\frac{q}{2},\frac{q}{2},\frac{q}{2}-1+i_5,q-2,\frac{q}{2}+i_3,\frac{q}{2}+i_2,0,\frac{q}{2}+1+i_0\right)
		\end{equation*}
		and
		\begin{equation*}
			n-a=\left(\frac{q}{2}-1,\frac{q}{2}-1,\frac{q}{2},\frac{q}{2},\frac{q}{2}-1,\frac{q}{2},\frac{q}{2}-1,\frac{q}{2}\right)_q.
		\end{equation*}
		Then we get $\max\limits_{1\le j\le 5}L_j=1$. It follows from Lemma~\ref{sec3 lem0} that $\min\limits_{i\in T_1}\CL(a+i)\ge q^6>\delta_1.$
		
		\item When $i_5\ne\frac{q}{2},i_4=\frac{q}{2}-1,i_3=\frac{q}{2}-1,i_2\le\frac{q}{2}-1$ and $i_1=\frac{q}{2}$, let
		\begin{equation*}
			a+i=\left(\frac{q}{2},\frac{q}{2},\frac{q}{2}-1+i_5,q-1,q-2,\frac{q}{2}+i_2,0,\frac{q}{2}+1+i_0\right)
		\end{equation*}
		and
		\begin{equation*}
			n-a=\left(\frac{q}{2}-1,\frac{q}{2}-1,\frac{q}{2},\frac{q}{2}-1,\frac{q}{2},\frac{q}{2},\frac{q}{2}-1,\frac{q}{2}\right)_q,
		\end{equation*}
		then $x_5\ne q-1$ and $\frac{q}{2}\le x_2\le q-1$ for any $x\in X$, and so $L_1=\dots=L_5=1$. Thus, $\min\limits_{i\in T_1}\CL(a+i)\ge q^6>\delta_1$ by Lemma~\ref{sec3 lem0}.

		\item When $i_5=\frac{q}{2},i_4=\frac{q}{2}-1,i_3=\frac{q}{2}-1,i_2\le\frac{q}{2}-1$ and $i_1=\frac{q}{2}$, let
		\begin{equation*}
			a+i=\left(\frac{q}{2},\frac{q}{2},0,q-1,q-2,\frac{q}{2}+i_2,0,\frac{q}{2}+1+i_0\right)
		\end{equation*}
		and
		\begin{equation*}
			n-a=\left(\frac{q}{2}-1,\frac{q}{2},\frac{q}{2}-1,\frac{q}{2}-1,\frac{q}{2},\frac{q}{2},\frac{q}{2}-1,\frac{q}{2}\right)_q.
		\end{equation*}
		Note that $\frac{q}{2}\le x_2\le q-1$ for any $x\in X$. Then $L_1=\dots=L_5=1$, and so $\min\limits_{i\in T_1}\CL(a+i)\ge q^6>\delta_1$ again by Lemma~\ref{sec3 lem0}.
		
		\item When $i_4=\frac{q}{2}-1,i_2\ge\frac{q}{2}+1,i_1=\frac{q}{2}$, and either $i_3=\frac{q}{2}-1$, or $i_5=\frac{q}{2}$ and $i_3=\frac{q}{2}$, let
		\begin{equation*}
			a+i=\left(\frac{q}{2}-1,\frac{q}{2},\frac{q}{2}+i_5,q-2,\frac{q}{2}+i_3,i_2+1-\frac{q}{2},0,\frac{q}{2}+i_0\right)
		\end{equation*}
		and
		\begin{equation*}
			n-a=\left(\frac{q}{2},\frac{q}{2}-1,\frac{q}{2}-1,\frac{q}{2},\frac{q}{2},\frac{q}{2}-1,\frac{q}{2}-1,\frac{q}{2}+1\right)_q,
		\end{equation*}
		then $2\le x_2\le \frac{q}{2}$ for any $x\in X$. If $i_3=\frac{q}{2}-1$, $(x_4,x_3)=(q-2,q-1)$. If $i_5=i_3=\frac{q}{2}$, $(x_5,x_4,x_3)=(0,q-1,0)$. Thus, $L_1=\dots=L_5=1$. According to  Lemma~\ref{sec3 lem0}, $\min\limits_{i\in T_1}\CL(a+i)\ge q^6>\delta_1$.
		
		\item When $i_5=\frac{q}{2},i_4=\frac{q}{2}-1,i_1=\frac{q}{2}$, and either $i_3\le\frac{q}{2}-2$ and $i_2\ne\frac{q}{2}$, or $i_3=\frac{q}{2},i_2\le\frac{q}{2}-1$ and $i_0\le\frac{q}{2}$, let
		\begin{equation*}
			a+i=\left(\frac{q}{2},\frac{q}{2},q-1,q-2,\frac{q}{2}+i_3,\frac{q}{2}+i_2,0,\frac{q}{2}+1+i_0\right)
		\end{equation*}
		and
		\begin{equation*}
			n-a=\left(\frac{q}{2}-1,\frac{q}{2}-1,\frac{q}{2},\frac{q}{2},\frac{q}{2}-1,\frac{q}{2},\frac{q}{2}-1,\frac{q}{2}\right)_q,
		\end{equation*}
		then $x_2\ne0$ and $(x_3,x_2)\ne(q-1,q-1)$ for any $x\in X$. Note that $x_4$ equals $q-2$ if $i_3\le\frac{q}{2}-2$, and $q-1$ if $i_3=\frac{q}{2}$. That is, $L_1=\dots=L_4=1$ and $L_5=2$. Hence $\max\limits_{1\le j\le 5}L_j=L_5=2$, and so $\min\limits_{i\in T_1}\CL(a+i)=\min\limits_{x\in X}\CL(x)=\min R_5$. Note that
		\begin{align*}
			S_5=&\Big\{\Big(\frac{q}{2},\frac{q}{2},q-1,q-2,\frac{q}{2}+i_3,\frac{q}{2}+i_2,0,\frac{q}{2}+1+i_0\Big)\ \Big|\ i_3=\frac{q}{2},i_2\le \frac{q}{2}-1,i_0\le\frac{q}{2}\Big\}\\
			=&\Big\{\Big(\frac{q}{2},\frac{q}{2},q-1,q-1,0,\frac{q}{2}+i_2,0,\frac{q}{2}+1+i_0\Big)\ \Big|\ i_2\le \frac{q}{2}-1,i_0\le\frac{q}{2}\Big\}.
		\end{align*}
		Hence, we have
		\begin{align*}
			\min R_5=&\min\{x\cdot q^{2m-f-2}\mid x\in S_5,\ f=4\}\\
			=&\min\Big\{\Big(0,0,q-1,\frac{q}{2}-1-i_2,q-1,\frac{q}{2}-1-i_0,\frac{q}{2},\frac{q}{2}+1\Big)\ \Big|\ i_2\le \frac{q}{2}-1,i_0\le\frac{q}{2}\Big\}\\
			=&\Big(0,0,q-1,0,q-2,q-1,\frac{q}{2},\frac{q}{2}+1\Big)_q>\delta_1.
		\end{align*}
		
		\item When $i_5=\frac{q}{2},i_4=\frac{q}{2}-1,i_1=\frac{q}{2}$, and either $i_3=\frac{q}{2},i_2\le\frac{q}{2}-1$ and $i_0\ge\frac{q}{2}+1$, or $i_3=\frac{q}{2}+1,i_2=\frac{q}{2}-1$ and $i_0\le\frac{q}{2}-2$, let
		\begin{equation*}
			a+i=\left(\frac{q}{2}-1,\frac{q}{2}+1,0,q-1,i_3-\frac{q}{2},\frac{q}{2}-1+i_2,q-1,\frac{q}{2}+1+i_0\right)
		\end{equation*}
		and
		\begin{equation*}
			n-a=\left(\frac{q}{2},\frac{q}{2}-1,\frac{q}{2}-1,\frac{q}{2},\frac{q}{2}-1,\frac{q}{2},\frac{q}{2},\frac{q}{2}\right)_q.
		\end{equation*}
		If $i_3=\frac{q}{2},i_2\le\frac{q}{2}-1$ and $i_0\ge\frac{q}{2}+1$, then $x_3=x_1=0,\frac{q}{2}\le x_2\le q-1$ and $2\le x_0\le \frac{q}{2}$ for any $x\in X$. If $i_3=\frac{q}{2}+1,i_2=\frac{q}{2}-1,i_0\le\frac{q}{2}-2$, then $(x_3,x_2,x_1)=(1,q-2,q-1)$ and $\frac{q}{2}+2\le x_0\le q-1$. Hence, we have $\max\limits_{1\le j\le 5}L_j=1$, and so $\min\limits_{i\in T_1}\CL(a+i)\ge q^6>\delta_1$ by Lemma~\ref{sec3 lem0}. 
		
		\item When $i_5=\frac{q}{2},i_4=\frac{q}{2}-1,i_1=\frac{q}{2}$, and
		\begin{equation*}
			\left\{\begin{aligned}
				&i_3=\frac{q}{2}+1,i_2=\frac{q}{2}-1,i_0\ge\frac{q}{2}-1,\mathrm{\ or\ }\\
				&i_3=\frac{q}{2}+1,i_2\ge\frac{q}{2}+1,\mathrm{\ or\ }\\
				&i_3\ge\frac{q}{2}+2,i_2\ne\frac{q}{2},\\
			\end{aligned}\right.
		\end{equation*} let
		\begin{equation*}
			a+i=\left(\frac{q}{2},\frac{q}{2},1,0,i_3-\frac{q}{2}-1,\frac{q}{2}+i_2,0,\frac{q}{2}+1+i_0\right)
		\end{equation*}
		and
		\begin{equation*}
			n-a=\left(\frac{q}{2}-1,\frac{q}{2},\frac{q}{2}-1,\frac{q}{2}-1,\frac{q}{2},\frac{q}{2},\frac{q}{2}-1,\frac{q}{2}\right)_q,
		\end{equation*}
		then $x_2\ne0$ for any $x\in X$. If $i_3\ge\frac{q}{2}+2$ or $i_3=\frac{q}{2}+1,i_2\ge\frac{q}{2}+1$, then $x_3\ne1$. If $i_3=\frac{q}{2}+1,i_2=\frac{q}{2}-1$, then $(x_3,x_2)=(0,q-1)$, which implies that $L_1=L_2=L_3=L_5=1$ and $L_4=2$. Hence $\max\limits_{1\le j\le 5}L_j=L_4=2$, and so $\min\limits_{i\in T_1}\CL(a+i)=\min\limits_{x\in X}\CL(x)=\min R_4$. Note that
		\begin{align*}
			S_4=&\Big\{\Big(\frac{q}{2},\frac{q}{2},1,0,i_3-\frac{q}{2}-1,\frac{q}{2}+i_2,0,\frac{q}{2}+1+i_0\Big)\ \Big|\ i_3=\frac{q}{2}+1,i_2=\frac{q}{2}-1,i_0\ge\frac{q}{2}-1\Big\}\\
			=&\Big\{\Big(\frac{q}{2},\frac{q}{2},1,0,0,q-1,1,i_0-\frac{q}{2}+1\Big)\ \Big|\ i_0\ge\frac{q}{2}-1\Big\}.
		\end{align*}
		Hence, we have
		\begin{align*}
			\min R_4=&\min\{x\cdot q^{m-e-2}\mid x\in S_4,\ e=3\}\\
			=&\min\Big\{\Big(0,0,q-1,1,i_0-\frac{q}{2},\frac{q}{2}-1,\frac{q}{2}-1,q-1\Big)\ \Big|\ i_0\ge\frac{q}{2}-1\Big\}\\
			=&\Big(0,0,q-1,0,q-1,\frac{q}{2}-1,\frac{q}{2}-1,q-1\Big)_q>\delta_1.
		\end{align*}
		
		\item When $i_5=\frac{q}{2},i_4=\frac{q}{2}-1,i_3=\frac{q}{2}+1,i_2\le\frac{q}{2}-2$ and $i_1=\frac{q}{2}$, let
		\begin{equation*}
			a+i=\left(\frac{q}{2}-1,\frac{q}{2}+1,0,q-1,0,\frac{q}{2}+1+i_2,0,\frac{q}{2}+i_0\right)
		\end{equation*}
		and
		\begin{equation*}
			n-a=\left(\frac{q}{2},\frac{q}{2}-1,\frac{q}{2}-1,\frac{q}{2},\frac{q}{2},\frac{q}{2}-1,\frac{q}{2}-1,\frac{q}{2}+1\right)_q,
		\end{equation*}
		then $\frac{q}{2}+1\le x_2\le q-1$ for any $x\in X$. Thus, $L_1=\dots=L_5=1$, and so $\min\limits_{i\in T_1}\CL(a+i)\ge q^6>\delta_1$ by Lemma~\ref{sec3 lem0}.
		
		\item When $i_5=\frac{q}{2}-1,i_4\ne\frac{q}{2},i_3\ne\frac{q}{2}-1,i_2=\frac{q}{2},i_1=\frac{q}{2}$ and $i_0\ge\frac{q}{2}-1$, let
		\begin{equation*}
			a+i=\left(\frac{q}{2},\frac{q}{2},q-2,\frac{q}{2}-1+i_4,\frac{q}{2}+1+i_3,0,1,i_0+1-\frac{q}{2}\right)
		\end{equation*}
		and
		\begin{equation*}
			n-a=\left(\frac{q}{2}-1,\frac{q}{2}-1,\frac{q}{2},\frac{q}{2},\frac{q}{2}-1,\frac{q}{2},\frac{q}{2}-1,\frac{q}{2}\right)_q.
		\end{equation*}
		Then $L_1=L_2=L_4=L_5=1$ and $L_3=2$. Hence $\max\limits_{1\le j\le 5}L_j=L_3=2$, and so $\min\limits_{i\in T_1}\CL(a+i)=\min\limits_{x\in X}\CL(x)=\min R_3$. Note that
		\begin{align*}
			S_3=&\Big\{\Big(\frac{q}{2},\frac{q}{2},q-2,\frac{q}{2}-1+i_4,\frac{q}{2}+1+i_3,0,1,i_0+1-\frac{q}{2}\Big)\ \Big|\ i_4\ne\frac{q}{2},i_3\ne\frac{q}{2}-1,i_0=\frac{q}{2}-1\Big\}\\
			=&\Big\{\Big(\frac{q}{2},\frac{q}{2},q-2,\frac{q}{2}-1+i_4,\frac{q}{2}+1+i_3,0,1,0\Big)\ \Big|\ i_4\ne\frac{q}{2},i_3\ne\frac{q}{2}-1\Big\}.
		\end{align*}
		Hence, we have
		\begin{align*}
			\min R_3=&\min\{x\cdot q^{m-1-2}\mid x\in S_3\}\\
			=&\min\Big\{\Big(0,0,q-1,\frac{q}{2}-1,\frac{q}{2}-1,1,\frac{q}{2}-i_4,\frac{q}{2}-1-i_3\Big)\ \Big|\ i_4\ne\frac{q}{2},i_3\ne\frac{q}{2}-1\Big\}\\
			=&\Big(0,0,q-1,\frac{q}{2}-1,\frac{q}{2}-1,0,\frac{q}{2},\frac{q}{2}\Big)_q>\delta_1.
		\end{align*}
		
		\item When $i_5=\frac{q}{2}-1,i_4\ne\frac{q}{2},i_3\ne\frac{q}{2}-1,i_2=\frac{q}{2},i_1=\frac{q}{2}$ and $i_0\le\frac{q}{2}-2$, let
		\begin{equation*}
			a+i=\left(\frac{q}{2},\frac{q}{2},q-2,\frac{q}{2}-1+i_4,\frac{q}{2}+1+i_3,0,q-1,\frac{q}{2}+1+i_0\right)
		\end{equation*}
		and
		\begin{equation*}
			n-a=\left(\frac{q}{2}-1,\frac{q}{2}-1,\frac{q}{2},\frac{q}{2},\frac{q}{2}-1,\frac{q}{2}-1,\frac{q}{2},\frac{q}{2}\right)_q.
		\end{equation*}
		One can easily check that $L_1=L_2=L_3=0$ and $L_4=L_5=1$. In accordance with Lemma~\ref{sec3 lem0}, $\min\limits_{i\in T_1}\CL(a+i)\ge q^6>\delta_1$.
		
		\item When $i_2=\frac{q}{2},i_1=\frac{q}{2}$, and
		\begin{equation*}
			\left\{\begin{aligned}
				&i_5\ne\frac{q}{2}-1,i_4\ne\frac{q}{2},\mathrm{\ or\ }\\
				&i_5=\frac{q}{2}-1,i_4\ne\frac{q}{2},i_3=\frac{q}{2}-1,\mathrm{\ or\ }\\
				&i_5=\frac{q}{2},i_4=\frac{q}{2},i_3\le\frac{q}{2}-2,\\
			\end{aligned}\right.
		\end{equation*} let
		\begin{equation*}
			a+i=\left(\frac{q}{2}-1,\frac{q}{2},\frac{q}{2}+i_5,\frac{q}{2}-1+i_4,\frac{q}{2}+i_3,1,0,\frac{q}{2}+i_0\right)
		\end{equation*}
		and
		\begin{equation*}
			n-a=\left(\frac{q}{2},\frac{q}{2}-1,\frac{q}{2}-1,\frac{q}{2},\frac{q}{2},\frac{q}{2}-1,\frac{q}{2}-1,\frac{q}{2}+1\right)_q,
		\end{equation*}
		Then we have $L_1=\dots=L_5=1$, and so $\min\limits_{i\in T_1}\CL(a+i)\ge q^6>\delta_1$ by Lemma~\ref{sec3 lem0}.
		
		\item When $i_4=\frac{q}{2},i_2=\frac{q}{2},i_1=\frac{q}{2}$, and either $i_5\ne\frac{q}{2}$, or $i_5=\frac{q}{2}$ and $i_3\ge\frac{q}{2}-1$, let
		\begin{equation*}
			a+i=\left(\frac{q}{2},\frac{q}{2}-1,\frac{q}{2}+i_5,0,\frac{q}{2}+i_3,1,0,\frac{q}{2}+i_0\right)
		\end{equation*}
		and
		\begin{equation*}
			n-a=\left(\frac{q}{2}-1,\frac{q}{2},\frac{q}{2},\frac{q}{2}-1,\frac{q}{2},\frac{q}{2}-1,\frac{q}{2}-1,\frac{q}{2}+1\right)_q,
		\end{equation*}
		then $(x_4,x_3)\ne(0,0)$ for any $x\in X$. Hence, $L_1=L_2=L_3=L_5=1$ and $L_4=2$. 
		Hence $\max\limits_{1\le j\le 5}L_j=L_4=2$, and so $\min\limits_{i\in T_1}\CL(a+i)=\min\limits_{x\in X}\CL(x)=\min R_4$. Note that
		\begin{align*}
			S_4=&\Big\{\Big(\frac{q}{2},\frac{q}{2}-1,\frac{q}{2}+i_5,0,\frac{q}{2}+i_3,1,0,\frac{q}{2}+i_0\Big)\ \Big|\ i_5=\frac{q}{2},i_3\ge\frac{q}{2}-1,1\le i_0\le q-1\Big\}\\
			=&\Big\{\Big(\frac{q}{2},\frac{q}{2},0,0,\frac{q}{2}+i_3,1,0,\frac{q}{2}+i_0\Big)\ \Big|\ i_3\ge\frac{q}{2}-1,1\le i_0\le q-1\Big\}.
		\end{align*}
		Hence, we have
		\begin{align*}
			\min R_4=&\min\{x\cdot q^{m-e-2}\mid x\in S_4,e=\}\\
			=&\min\Big\{\Big(0,0,\frac{q}{2}+i_3,1,0,\frac{q}{2}+i_0-1,\frac{q}{2}-1,\frac{q}{2}\Big)\ \Big|\ i_3\ge\frac{q}{2}-1,1\le i_0\le q-1\Big\}\\
			=&\Big(0,0,q-1,1,0,\frac{q}{2},\frac{q}{2}-1,\frac{q}{2}\Big)_q>\delta_1.
		\end{align*}
	\end{itemize}
	In summary, we have $a+i \in I_1$, and so $n+i=(a+i)+(n-a) \in I_1+I_2$ for any $i \in T_1$.
	
	\item $\boldsymbol{m=9:}$ we have
	\begin{equation*}
		i=(\underbrace{0,0,\ldots,0}_{m-r-1},i_r\ne0,i_{r-1},\ldots,i_1,i_0\ne0)_q,
	\end{equation*}
	where $0\leq r\leq 5$ as $$i<\delta_{1}=\frac{q^7-2q^3+q}{2(q+1)}=\left(0,0,0,\frac{q}{2}-1,\frac{q}{2},\frac{q}{2}-1,\frac{q}{2}-1,\frac{q}{2},\frac{q}{2}\right)_q.$$ With the same notation as in Lemma~\ref{sec3 lem0}, take $X := a + T_1$. We only need to prove that there always exists an integer $a\in C_{\Delta_1}$ such that $\min\limits_{i\in T_1}\CL(a+i)\ge\delta_{1}$. Through calculation, the largest coset leader in $T_1$ is
	\begin{equation*}
		\left(0,0,0,\frac{q}{2}-1,\frac{q}{2},\frac{q}{2}-1,\frac{q}{2}-1,\frac{q}{2},\frac{q}{2}-1\right)_q.
	\end{equation*}
	\begin{itemize}
		\item When $i_3\ne \frac{q}{2}-1$, take $a:=\Delta_1$. Let
		\begin{equation*}
			a+i=\Delta_1+i=\left(\frac{q}{2}-1,\frac{q}{2},\frac{q}{2}-1,\frac{q}{2}+i_5,\frac{q}{2}-1+i_4,\frac{q}{2}+i_3,\frac{q}{2}-1+i_2,\frac{q}{2}+i_1,\frac{q}{2}+i_0\right)
		\end{equation*}
		and
		\begin{equation*}
			y=n-\Delta_1=\left(\frac{q}{2},\frac{q}{2}-1,\frac{q}{2},\frac{q}{2}-1,\frac{q}{2},\frac{q}{2}-1,\frac{q}{2},\frac{q}{2}-1,\frac{q}{2}+1\right)_q.
		\end{equation*}
		For any $x\in X$, we have $(x_3,x_2)\notin \{(0,0),(q-1,q-1)\}$ and $(x_3,x_2,x_1,x_0)\ne (0,1,0,0)$. Since $i_5\le \frac{q}{2}-1$, $(x_5,x_4,x_3)\notin \{(0,0,0),(0,0,1)\}$. Thus, we have $L_1=L_2=L_5=3$ and $L_3=L_4=2$. Hence $\max\limits_{1\le j\le 5}L_j=3$, and so $\min\limits_{i\in T_1}\CL(a+i)=\min\limits_{x\in X}\CL(x)=\min (R_1\cup R_2\cup R_5)$. Note that
		\begin{align*}
			S_1=&\Big\{\Big(\frac{q}{2}-1,\frac{q}{2},\frac{q}{2}-1,\frac{q}{2}+i_5,\frac{q}{2}-1+i_4,\frac{q}{2}+i_3,\frac{q}{2}-1+i_2,\frac{q}{2}+i_1,\frac{q}{2}+i_0\Big)\ \Big|\\
			&\quad i_5\le \frac{q}{2}-1,i_4\le q-1,i_3\ne\frac{q}{2}-1,\ i_2=\frac{q}{2},\ i_1=\frac{q}{2}-1,\ i_0=\frac{q}{2}\Big\}\\
			=&\Big\{\Big(\frac{q}{2}-1,\frac{q}{2},\frac{q}{2}-1,\frac{q}{2}+i_5,\frac{q}{2}-1+i_4,\frac{q}{2}+1+i_3,0,0,0\Big)\ \Big|\\
			&\quad i_5\le \frac{q}{2}-1,i_4\le q-1,i_3\ne\frac{q}{2}-1\Big\},\\
			S_2=&\Big\{\Big(\frac{q}{2}-1,\frac{q}{2},\frac{q}{2}-1,\frac{q}{2}+i_5,\frac{q}{2}-1+i_4,\frac{q}{2}+i_3,\frac{q}{2}-1+i_2,\frac{q}{2}+i_1,\frac{q}{2}+i_0\Big)\ \Big|\\
			&\quad i_5\le \frac{q}{2}-1,i_4\le q-1,i_3\ne\frac{q}{2}-1,\ i_2=\frac{q}{2},\ i_1=\frac{q}{2}-1,\ i_0=\frac{q}{2}+1\Big\}\\
			=&\Big\{\Big(\frac{q}{2}-1,\frac{q}{2},\frac{q}{2}-1,\frac{q}{2}+i_5,\frac{q}{2}-1+i_4,\frac{q}{2}+1+i_3,0,0,1\Big)\ \Big|\\
			&\quad i_5\le \frac{q}{2}-1,i_4\le q-1, i_3\ne\frac{q}{2}-1\Big\},\ \mathrm{and}\\
			S_5=&\Big\{\Big(\frac{q}{2}-1,\frac{q}{2},\frac{q}{2}-1,\frac{q}{2}+i_5,\frac{q}{2}-1+i_4,\frac{q}{2}+i_3,\frac{q}{2}-1+i_2,\frac{q}{2}+i_1,\frac{q}{2}+i_0\Big)\ \Big|\\
			&\quad i_5=\frac{q}{2}-1,\ i_4=\frac{q}{2},\ i_3=\frac{q}{2}-2,\ \mathrm{and\ either}\ i_2\ge\frac{q}{2}+1,\ \mathrm{or}\\
			&\quad i_2=\frac{q}{2}\ \mathrm{and}\ i_1\ge\frac{q}{2},\ \mathrm{or}\ i_2=\frac{q}{2},\ i_1=\frac{q}{2}-1\ \mathrm{and}\ i_0>\frac{q}{2}\Big\}\\
			=&\Big\{\Big(\frac{q}{2}-1,\frac{q}{2},\frac{q}{2}-1,q-1,q-1,q-1,i_2-\frac{q}{2}-1,\frac{q}{2}+i_1,\frac{q}{2}+i_0\Big)\ \Big|\\
			&\quad \mathrm{\ either}\ i_2\ge\frac{q}{2}+1,\ \mathrm{or}\ i_2=\frac{q}{2}\ \mathrm{and}\ i_1\ge\frac{q}{2},\ \mathrm{or}\ i_2=\frac{q}{2},\ i_1=\frac{q}{2}-1\ \mathrm{and}\ i_0>\frac{q}{2}\Big\}.
		\end{align*}
		Hence, we have
		\begin{align*}
			\min R_1=&\min\{x\cdot q^{2m-3}\mid x\in S_1\}\\
			=&\min\Big\{\Big(0,0,0,\frac{q}{2}-1,\frac{q}{2},\frac{q}{2}-1,\frac{q}{2}+i_5,\frac{q}{2}-1+i_4,\frac{q}{2}+1+i_3\Big)\ \Big|\\
			&\quad i_5\le \frac{q}{2}-1,i_4\le q-1,i_3\ne\frac{q}{2}-1\Big\}\\
			=&\Big(0,0,0,\frac{q}{2}-1,\frac{q}{2},\frac{q}{2}-1,\frac{q}{2},\frac{q}{2}-1,\frac{q}{2}+1\Big)_q,\\
			\min R_2=&\min\{x\cdot q^{m-3}\mid x\in S_2\}\\
			=&\min\Big\{\Big(0,0,0,\frac{q}{2},\frac{q}{2}-1,\frac{q}{2},\frac{q}{2}-1-i_5,\frac{q}{2}-i_4,\frac{q}{2}-1-i_3\Big)\ \Big|\\
			&\quad i_5\le \frac{q}{2}-1,i_4\le q-1,i_3\ne\frac{q}{2}-1\Big\}\\
			=&\Big(0,0,0,\frac{q}{2},\frac{q}{2}-1,\frac{q}{2}-1,0,0,1\Big)_q,\  \mathrm{and}\\
			\min R_5=&\min\{x\cdot q^{2m-f-3}\mid x\in S_5,\ f=3\}\\
			=&\min\Big\{\Big(0,0,0,\frac{3q}{2}-i_2,\frac{q}{2}-1-i_1,\frac{q}{2}-i_0,\frac{q}{2}-1,\frac{q}{2},\frac{q}{2}\Big)\ \Big|\\
			&\quad \mathrm{either}\ i_2\ge\frac{q}{2}+1,\ \mathrm{or}\ i_2=\frac{q}{2}\ \mathrm{and}\ i_1\ge\frac{q}{2},\ \mathrm{or}\ i_2=\frac{q}{2},\ i_1=\frac{q}{2}-1\ \mathrm{and}\ i_0>\frac{q}{2}\Big\}\\
			=&\Big(0,0,0,\frac{3q}{2}-(q-1),\frac{q}{2}-1-(q-1),\frac{q}{2}-(q-1),\frac{q}{2}-1,\frac{q}{2},\frac{q}{2}\Big)\\
			=&\Big(0,0,0,\frac{q}{2},\frac{q}{2}-1,\frac{q}{2}+1,\frac{q}{2}-1,\frac{q}{2},\frac{q}{2}\Big)_q.
		\end{align*}
		It is easily seen that $\min R_1< \min R_5 < \min R_2$. Hence, $\min\limits_{i\in T_1}\CL(a+i)=\min R_1>\delta_1$.
		\item When $i_3=\frac{q}{2}-1$ and $i_2\ne\frac{q}{2}-1$, take $a:=n-\Delta_1$. Let
		\begin{equation*}
			a+i=n-\Delta_1+i=\left(\frac{q}{2},\frac{q}{2}-1,\frac{q}{2},\frac{q}{2}-1+i_5,\frac{q}{2}+i_4,q-2,\frac{q}{2}+i_2,\frac{q}{2}-1+i_1,\frac{q}{2}+1+i_0\right)
		\end{equation*}
		and
		\begin{equation*}
			n-a=\Delta_1=\left(\frac{q}{2}-1,\frac{q}{2},\frac{q}{2}-1,\frac{q}{2},\frac{q}{2}-1,\frac{q}{2},\frac{q}{2}-1,\frac{q}{2},\frac{q}{2}\right)_q,
		\end{equation*}
		then $(x_2,x_1)\notin\{(0,0),(q-1,q-1)\}$ and $(x_2,x_1,x_0)\ne(0,1,0)$ for any $x\in X$. Since $i_5\le\frac{q}{2}-1$, 
		$x_5\ne 0$ and $(x_5,x_4)\ne(q-1,q-1)$. Hence, we have $L_1=L_2=L_5=2$ and $L_3=L_4=1$. According to Lemma~\ref{sec3 lem0},
		$\min\limits_{i\in T_1}\CL(a+i)=\min\limits_{x\in X}\CL(x)\ge q^7>\delta_1$.
		\item When $i_3=i_2=\frac{q}{2}-1$, and either $i_5\ne \frac{q}{2}-1$ or $i_4\ne\frac{q}{2}$, take $a := \Delta_1$. Let
		\begin{equation*}
			a+i=\Delta_1+i=\left(\frac{q}{2}-1,\frac{q}{2},\frac{q}{2}-1,\frac{q}{2}+i_5,\frac{q}{2}-1+i_4,q-1,q-2,\frac{q}{2}+i_1,\frac{q}{2}+i_0\right)
		\end{equation*}
		and
		\begin{equation*}
			n-a=n-\Delta_1=\left(\frac{q}{2},\frac{q}{2}-1,\frac{q}{2},\frac{q}{2}-1,\frac{q}{2},\frac{q}{2}-1,\frac{q}{2},\frac{q}{2}-1,\frac{q}{2}+1\right)_q.
		\end{equation*}
		For any $x\in X$, we have $(x_5,x_4)\ne(q-1,q-1)$. Then we have $L_1=L_2=2,L_3=L_4=1$ and $L_5=3$. Hence $\max\limits_{1\le j\le 5}L_j=L_5=3$, and so $\min\limits_{i\in T_1}\CL(a+i)=\min\limits_{x\in X}\CL(x)=\min R_5$. Note that
		\begin{align*}
			S_5=&\Big\{\Big(\frac{q}{2}-1,\frac{q}{2},\frac{q}{2}-1,\frac{q}{2}+i_5,\frac{q}{2}-1+i_4,q-1,q-2,\frac{q}{2}+i_1,\frac{q}{2}+i_0\Big)\ \Big|\\
			&\quad i_5\ne\frac{q}{2}-1,i_4=\frac{q}{2},\ \mathrm{and\ either\ }i_1\ge\frac{q}{2},\ \mathrm{or}\ i_1=\frac{q}{2}-1\mathrm{\ and\ } i_0>\frac{q}{2}\Big\}\\
			=&\Big\{\Big(\frac{q}{2}-1,\frac{q}{2},\frac{q}{2}-1,\frac{q}{2}+i_5,q-1,q-1,q-1,i_1-\frac{q}{2},\frac{q}{2}+i_0\Big)\ \Big|\\
			&\quad i_5\ne\frac{q}{2}-1,\ \mathrm{and\ either\ }i_1\ge\frac{q}{2},\ \mathrm{or}\ i_1=\frac{q}{2}-1\mathrm{\ and\ } i_0>\frac{q}{2}\Big\}.
		\end{align*}
		Hence, we have
		\begin{align*}
			\min R_5=&\min\{x\cdot q^{2m-f-3}\mid x\in S_5,\ f=2\}\\
			=&\min\Big\{\Big(0,0,0,\frac{3q}{2}-1-i_1,\frac{q}{2}-i_0,\frac{q}{2}-1,\frac{q}{2},\frac{q}{2}-1,\frac{q}{2}+i_5+1\Big) \ \Big|\\
			&\quad i_5\ne\frac{q}{2}-1,\ \mathrm{and\ either\ }i_1\ge\frac{q}{2},\ \mathrm{or}\ i_1=\frac{q}{2}-1\mathrm{\ and\ } i_0>\frac{q}{2} \Big\}\\
			=&\Big(0,0,0,\frac{3q}{2}-1-(q-1),\frac{q}{2}-(q-1),\frac{q}{2}-1,\frac{q}{2},\frac{q}{2}-1,\frac{q}{2}+1\Big)\\
			=&\Big(0,0,0,\frac{q}{2}-1,\frac{q}{2}+1,\frac{q}{2}-1,\frac{q}{2},\frac{q}{2}-1,\frac{q}{2}+1\Big)_q>\delta_1.
		\end{align*}
		\item When $i_5= \frac{q}{2}-1,i_4=\frac{q}{2}$ and  $i_3=i_2=\frac{q}{2}-1$, take $a := n-\Delta_{1}$. Let
		\begin{equation*}
			a+i=n-\Delta_1+i=\left(\frac{q}{2},\frac{q}{2}-1,\frac{q}{2},q-1,0,q-2,q-1,\frac{q}{2}-1+i_1,\frac{q}{2}+1+i_0\right)
		\end{equation*}
		and
		\begin{equation*}
			n-a=\Delta_1=\left(\frac{q}{2}-1,\frac{q}{2},\frac{q}{2}-1,\frac{q}{2},\frac{q}{2}-1,\frac{q}{2},\frac{q}{2}-1,\frac{q}{2},\frac{q}{2}\right)_q.
		\end{equation*}
		Note that $i_1<\frac{q}{2}$ if $i_0=\frac{q}{2}$, and so $(x_1,x_0)\ne(0,1)$ for any $x\in X$. Hence, we have $L_1=3,L_2=L_3=L_4=1$ and $L_5=2$. Hence $\max\limits_{1\le j\le 5}L_j=L_1=3$, and so $\min\limits_{i\in T_1}\CL(a+i)=\min\limits_{x\in X}\CL(x)=\min R_1$. Note that
		\begin{align*}
			S_1=&\Big\{\Big(\frac{q}{2},\frac{q}{2}-1,\frac{q}{2},q-1,0,q-2,q-1,\frac{q}{2}-1+i_1,\frac{q}{2}+1+i_0\Big)\ \Big|\\
			&\qquad i_1=\frac{q}{2},\ i_0=\frac{q}{2}-1\Big\}\\
			=&\Big\{\Big(\frac{q}{2},\frac{q}{2}-1,\frac{q}{2},q-1,0,q-1,0,0,0\Big)_q\Big\}.
		\end{align*}
		Hence, we have
		\begin{align*}
			\min R_1=\min\{x\cdot q^{2m-3}\mid x\in S_1\}=\Big(0,0,0,\frac{q}{2},\frac{q}{2}-1,\frac{q}{2},q-1,0,q-1\Big)_q>\delta_1.
		\end{align*}
	\end{itemize}
	In summary, we have $a+i \in I_1$, and so $n+i=(a+i)+(n-a) \in I_1+I_2$ for any $i \in T_1$.
	
	\item $\boldsymbol{m=10:}$ we have
	\begin{equation*}
		i=(\underbrace{0,0,\ldots,0}_{m-r-1},i_r\ne0,i_{r-1},\ldots,i_1,i_0\ne0)_q,
	\end{equation*}
	where $0\leq r\leq 6$ as $$i<\delta_{1}=\frac{2q^7-2q^6+q^5+q^4+q^3-2q^2-2q}{2}+1=\left(0,0,0,q-1,0,\frac{q}{2},\frac{q}{2},\frac{q}{2}-2,q-1,1\right)_q.$$ With the same notation as in Lemma~\ref{sec3 lem0}, take $X := a + T_1$. We only need to prove that there always exists an integer $a\in C_{\Delta_1}$ such that $\min\limits_{i\in T_1}\CL(a+i)\ge\delta_{1}$. Through calculation, the largest coset leader in $T_1$ is
	\begin{equation*}
		\left(0,0,0,q-1,0,\frac{q}{2},\frac{q}{2},\frac{q}{2}-2,q-2,q-1\right)_q.
	\end{equation*}
	\begin{itemize}
		\item When $i_5=\frac{q}{2},i_3\ne\frac{q}{2}$ and $i_2\ne\frac{q}{2}$, let
		\begin{equation*}
			a+i=\left(\frac{q}{2},\frac{q}{2},\frac{q}{2}-1,\frac{q}{2}+i_6,0,\frac{q}{2}+i_4,\frac{q}{2}-1+i_3,\frac{q}{2}-1+i_2,\frac{q}{2}+i_1,\frac{q}{2}+1+i_0\right)
		\end{equation*}
		and
		\begin{equation*}
			n-a=\left(\frac{q}{2}-1,\frac{q}{2}-1,\frac{q}{2},\frac{q}{2},\frac{q}{2}-1,\frac{q}{2}-1,\frac{q}{2},\frac{q}{2},\frac{q}{2}-1,\frac{q}{2}\right)_q.
		\end{equation*}
		Then we have $\max\limits_{1\le j\le 5}L_j=2$. According to Lemma~\ref{sec3 lem0}, $\min\limits_{i\in T_1}\CL(a+i)\ge q^7>\delta_1$.
		
		\item When $i_5=\frac{q}{2},i_3\ne\frac{q}{2}$ and $i_2=\frac{q}{2}$, let
		\begin{equation*}
			a+i=\left(\frac{q}{2},\frac{q}{2}-1,\frac{q}{2}-1,\frac{q}{2}+1+i_6,0,\frac{q}{2}-1+i_4,\frac{q}{2}+i_3,0,\frac{q}{2}+i_1,\frac{q}{2}+i_0\right)
		\end{equation*}
		and
		\begin{equation*}
			n-a=\left(\frac{q}{2}-1,\frac{q}{2},\frac{q}{2},\frac{q}{2}-1,\frac{q}{2}-1,\frac{q}{2},\frac{q}{2},\frac{q}{2}-1,\frac{q}{2}-1,\frac{q}{2}+1\right)_q,
		\end{equation*}
		then $\max\limits_{1\le j\le 5}L_j=2$, and so $\min\limits_{i\in T_1}\CL(a+i)\ge q^7>\delta_1$ by Lemma~\ref{sec3 lem0}.
		
		\item When $i_5=\frac{q}{2},i_3=\frac{q}{2}$, and
		\begin{equation*}
			\left\{\begin{aligned}
				&i_4\ne\frac{q}{2},i_2\ne\frac{q}{2}-1,\mathrm{\ or\ }\\
				&i_4=\frac{q}{2},i_2=\frac{q}{2},i_1=\frac{q}{2},\mathrm{\ or\ }\\
				&i_4=\frac{q}{2}-1,i_2=\frac{q}{2}-1,i_1\notin\{\frac{q}{2}-1,\frac{q}{2}-2\},
			\end{aligned}\right.
		\end{equation*} let
		\begin{equation*}
			a+i=\left(\frac{q}{2},\frac{q}{2}-1,\frac{q}{2}-1,\frac{q}{2}+1+i_6,0,\frac{q}{2}-1+i_4,q-1,\frac{q}{2}+i_2,\frac{q}{2}+i_1,\frac{q}{2}+i_0\right)
		\end{equation*}
		and
		\begin{equation*}
			n-a=\left(\frac{q}{2}-1,\frac{q}{2},\frac{q}{2},\frac{q}{2}-1,\frac{q}{2}-1,\frac{q}{2},\frac{q}{2},\frac{q}{2}-1,\frac{q}{2}-1,\frac{q}{2}+1\right)_q.
		\end{equation*}
		Then we have $L_1=L_2=L_4=L_5=2$ and $L_3=3$. Hence $\max\limits_{1\le j\le 5}L_j=L_3=3$, and so $\min\limits_{i\in T_1}\CL(a+i)=\min\limits_{x\in X}\CL(x)=\min R_3$. Note that
		\begin{align*}
			S_3=&\Big\{\Big(\frac{q}{2},\frac{q}{2}-1,\frac{q}{2}-1,\frac{q}{2}+1+i_6,0,\frac{q}{2}-1+i_4,q-1,\frac{q}{2}+i_2,\frac{q}{2}+i_1,\frac{q}{2}+i_0\Big)\ \Big|\\
			&\quad i_6\le q-2,i_4=i_2=\frac{q}{2}-1,i_1=i_0=\frac{q}{2}\Big\}\\
			=&\Big\{\Big(\frac{q}{2},\frac{q}{2}-1,\frac{q}{2}-1,\frac{q}{2}+1+i_6,0,q-1,0,0,1,0\Big)\ \Big|\ i_6\le q-2\Big\}.
		\end{align*}
		Hence, we have
		\begin{align*}
			\min R_3=&\min\{x\cdot q^{m-1-3}\mid x\in S_3\}\\
			=&\min\Big\{\Big(0,0,0,q-1,\frac{q}{2}-1,\frac{q}{2},\frac{q}{2},\frac{q}{2}-2-i_6,q-1,1\Big)\ \Big|\ i_6\le q-2\Big\}\\
			=&\Big(0,0,0,q-1,\frac{q}{2}-1,\frac{q}{2},\frac{q}{2}-1,\frac{q}{2}-1,q-1,1\Big)_q>\delta_1.
		\end{align*}
		\item When $i_5=\frac{q}{2},i_4=\frac{q}{2},i_3=\frac{q}{2}$, and either $i_2\le\frac{q}{2}-2$, or $i_2=\frac{q}{2},i_1\ne\frac{q}{2}$, let
		\begin{equation*}
			a+i=\left(\frac{q}{2}-1,\frac{q}{2},\frac{q}{2},\frac{q}{2}+i_6,0,1,0,\frac{q}{2}-1+i_2,\frac{q}{2}-1+i_1,\frac{q}{2}+1+i_0\right)
		\end{equation*}
		and
		\begin{equation*}
			n-a=\left(\frac{q}{2},\frac{q}{2}-1,\frac{q}{2}-1,\frac{q}{2},\frac{q}{2},\frac{q}{2}-1,\frac{q}{2}-1,\frac{q}{2},\frac{q}{2},\frac{q}{2}\right)_q,
		\end{equation*}
		then $(x_2,x_1)\ne(0,0)$ for any $x \in X$. Thus we have $\max\limits_{1\le j\le 5}L_j=2$. It then follows from Lemma~\ref{sec3 lem0} that $\min\limits_{i\in T_1}\CL(a+i)\ge q^7>\delta_1$.
		
		\item When $i_5=\frac{q}{2},i_3=\frac{q}{2},$ and
		\begin{equation*}
			\left\{\begin{aligned}
				&i_4=\frac{q}{2},i_2\ge \frac{q}{2}+1,\mathrm{\ or\ }\\
				&i_4\ne\frac{q}{2}-1,i_2=\frac{q}{2}-1,\mathrm{\ or\ }\\
				&i_4=\frac{q}{2}-1,i_2=\frac{q}{2}-1,i_1\in\{\frac{q}{2}-1,\frac{q}{2}-2\},
			\end{aligned}\right.
		\end{equation*} let
		\begin{equation*}
			a+i=\left(\frac{q}{2},\frac{q}{2},\frac{q}{2}-1,\frac{q}{2}+i_6,0,\frac{q}{2}+i_4,q-1,\frac{q}{2}-1+i_2,\frac{q}{2}+i_1,\frac{q}{2}+1+i_0\right)
		\end{equation*}
		and
		\begin{equation*}
			n-a=\left(\frac{q}{2}-1,\frac{q}{2}-1,\frac{q}{2},\frac{q}{2},\frac{q}{2}-1,\frac{q}{2}-1,\frac{q}{2},\frac{q}{2},\frac{q}{2}-1,\frac{q}{2}\right)_q.
		\end{equation*}
		Then $L_1=L_2=L_3=L_4=2$ and $L_5=3$. Hence $\max\limits_{1\le j\le 5}L_j=L_5=3$, and so $\min\limits_{i\in T_1}\CL(a+i)=\min\limits_{x\in X}\CL(x)=\min R_5$. Note that
		\begin{align*}
			S_5=&\Big\{\Big(\frac{q}{2},\frac{q}{2},\frac{q}{2}-1,\frac{q}{2}+i_6,0,\frac{q}{2}+i_4,q-1,\frac{q}{2}-1+i_2,\frac{q}{2}+i_1,\frac{q}{2}+1+i_0\Big)\ \Big|\\
			&\qquad i_6\le q-2,i_4=i_2=i_1=\frac{q}{2}-1,i_0\ge\frac{q}{2}\}\Big\}\\
			=&\Big\{\Big(\frac{q}{2},\frac{q}{2},\frac{q}{2}-1,\frac{q}{2}+i_6,0,q-1,q-1,q-1,0,i_0-\frac{q}{2}+1\Big)\ \Big|\ i_6\le q-2,i_0\ge\frac{q}{2}\}\Big\}.
		\end{align*}
		Hence, we have
		\begin{align*}
			\min R_5=&\min\{x\cdot q^{2m-f-3}\mid x\in S_5,\ f=2\}\\
			=&\min\Big\{\Big(0,0,0,q-1,\frac{3q}{2}-i_0-1,\frac{q}{2},\frac{q}{2},\frac{q}{2}-1,\frac{q}{2}+i_6,1\Big)\ \Big|\ i_6\le q-2,i_0\ge\frac{q}{2} \Big\} \\
			=&\Big(0,0,0,q-1,\frac{q}{2},\frac{q}{2},\frac{q}{2},\frac{q}{2}-1,\frac{q}{2},1\Big)_q>\delta_1.
		\end{align*}
		
		\item When $i_5\ne\frac{q}{2},i_3\ne\frac{q}{2}-1$ and $i_2\ne\frac{q}{2}-1$, let
		\begin{equation*}
			a+i=\left(\frac{q}{2}-1,\frac{q}{2}-1,\frac{q}{2},\frac{q}{2}+i_6,\frac{q}{2}-1+i_5,\frac{q}{2}-1+i_4,\frac{q}{2}+i_3,\frac{q}{2}+i_2,\frac{q}{2}-1+i_1,\frac{q}{2}+i_0\right)
		\end{equation*}
		and
		\begin{equation*}
			n-a=\left(\frac{q}{2},\frac{q}{2},\frac{q}{2}-1,\frac{q}{2}-1,\frac{q}{2},\frac{q}{2},\frac{q}{2}-1,\frac{q}{2}-1,\frac{q}{2},\frac{q}{2}+1\right)_q,
		\end{equation*}
		then $\max\limits_{1\le j\le 5}L_j=2$. Again by Lemma~\ref{sec3 lem0}, $\min\limits_{i\in T_1}\CL(a+i)\ge q^7>\delta_1$.
		
		\item When $i_5\ne\frac{q}{2},i_3\ne\frac{q}{2}-1$ and $i_2=\frac{q}{2}-1$, let
		\begin{equation*}
			a+i=\left(\frac{q}{2}-1,\frac{q}{2},\frac{q}{2},\frac{q}{2}-1+i_6,\frac{q}{2}-1+i_5,\frac{q}{2}+i_4,\frac{q}{2}+i_3,q-2,\frac{q}{2}-1+i_1,\frac{q}{2}+1+i_0\right)
		\end{equation*}
		and
		\begin{equation*}
			n-a=\left(\frac{q}{2},\frac{q}{2}-1,\frac{q}{2}-1,\frac{q}{2},\frac{q}{2},\frac{q}{2}-1,\frac{q}{2}-1,\frac{q}{2},\frac{q}{2},\frac{q}{2}\right)_q.
		\end{equation*}
		It is easily seen that $\max\limits_{1\le j\le 5}L_j=2$. In accordance with Lemma~\ref{sec3 lem0}, $\min\limits_{i\in T_1}\CL(a+i)\ge q^7>\delta_1$.
		
		\item When $i_5\ne\frac{q}{2}-1,i_3=\frac{q}{2}-1$ and $i_2=\frac{q}{2}$, let
		\begin{equation*}
			a+i=\left(\frac{q}{2},\frac{q}{2}-1,\frac{q}{2}-1,\frac{q}{2}+i_6,\frac{q}{2}+i_5,\frac{q}{2}-1+i_4,q-1,0,\frac{q}{2}+i_1,\frac{q}{2}+i_0\right)
		\end{equation*}
		and
		\begin{equation*}
			n-a=\left(\frac{q}{2}-1,\frac{q}{2},\frac{q}{2},\frac{q}{2}-1,\frac{q}{2}-1,\frac{q}{2},\frac{q}{2},\frac{q}{2}-1,\frac{q}{2}-1,\frac{q}{2}+1\right)_q.
		\end{equation*}
		Then we have $\max\limits_{1\le j\le 5}L_j=2$. According to Lemma~\ref{sec3 lem0},  $\min\limits_{i\in T_1}\CL(a+i)\ge q^7>\delta_1$.
		
		\item When $i_3=\frac{q}{2}-1$, and either $i_5\ne\frac{q}{2}-1,i_2\ne\frac{q}{2}$, or $i_5=\frac{q}{2}-1,i_4=\frac{q}{2},i_2\ge\frac{q}{2}+1$, let
		\begin{equation*}
			a+i=\left(\frac{q}{2},\frac{q}{2},\frac{q}{2}-1,\frac{q}{2}-1+i_6,\frac{q}{2}+i_5,\frac{q}{2}+i_4,q-2,\frac{q}{2}-1+i_2,\frac{q}{2}+i_1,\frac{q}{2}+1+i_0\right)
		\end{equation*}
		and
		\begin{equation*}
			n-a=\left(\frac{q}{2}-1,\frac{q}{2}-1,\frac{q}{2},\frac{q}{2},\frac{q}{2}-1,\frac{q}{2}-1,\frac{q}{2},\frac{q}{2},\frac{q}{2}-1,\frac{q}{2}\right)_q.
		\end{equation*}
		Hence, we have $L_1=L_2=L_3=L_5=2$ and $L_4=3$. Hence $\max\limits_{1\le j\le 5}L_j=L_4=3$, and so $\min\limits_{i\in T_1}\CL(a+i)=\min\limits_{x\in X}\CL(x)=\min R_4$. Note that
		\begin{align*}
			S_4=&\Big\{\Big(\frac{q}{2},\frac{q}{2},\frac{q}{2}-1,\frac{q}{2}-1+i_6,\frac{q}{2}+i_5,\frac{q}{2}+i_4,q-2,\frac{q}{2}-1+i_2,\frac{q}{2}+i_1,\frac{q}{2}+1+i_0\Big)\ \Big|\\
			&\quad i_6=i_4=\frac{q}{2},i_5=\frac{q}{2}-1,i_2\ge\frac{q}{2}+1,i_1\le q-1,1\le i_0\le q-1\Big\}\\
			=&\Big\{\Big(\frac{q}{2},\frac{q}{2},\frac{q}{2}-1,0,0,0,q-1,i_2-\frac{q}{2}-1,\frac{q}{2}+i_1,\frac{q}{2}+1+i_0\Big)\ \Big|\\
			&\quad i_2\ge\frac{q}{2}+1,i_1\le q-1,1\le i_0\le q-1\Big\}.
		\end{align*}
		Hence, we have
		\begin{align*}
			\min R_4=&\min\{x\cdot q^{m-e-3}\mid x\in S_4,\ e=4\}\\
			=&\min\Big\{\Big(0,0,0,q-1,i_2-\frac{q}{2}-1,\frac{q}{2}+i_1,\frac{q}{2}+i_0,\frac{q}{2}-1,\frac{q}{2}-1,\frac{q}{2}\Big) \\ 
			&\quad i_2\ge\frac{q}{2}+1,i_1\le q-1,1\le i_0\le q-1 \Big\} \\
			=&\Big(0,0,0,q-1,0,\frac{q}{2},\frac{q}{2}+1,\frac{q}{2}-1,\frac{q}{2}-1,\frac{q}{2}\Big)_q>\delta_1.
		\end{align*}
		
		\item When $i_5=\frac{q}{2}-1,i_4=\frac{q}{2},i_3=\frac{q}{2}-1$ and $i_2\le\frac{q}{2}-2$, let
		\begin{equation*}
			a+i=\left(\frac{q}{2}-1,\frac{q}{2},\frac{q}{2},\frac{q}{2}-1+i_6,q-1,0,q-1,\frac{q}{2}-1+i_2,\frac{q}{2}-1+i_1,\frac{q}{2}+1+i_0\right)
		\end{equation*}
		and
		\begin{equation*}
			n-a=\left(\frac{q}{2},\frac{q}{2}-1,\frac{q}{2}-1,\frac{q}{2},\frac{q}{2},\frac{q}{2}-1,\frac{q}{2}-1,\frac{q}{2},\frac{q}{2},\frac{q}{2}\right)_q.
		\end{equation*}
		Thus, we have $\max\limits_{1\le j\le 5}L_j=2$. It then follows from Lemma~\ref{sec3 lem0} that $\min\limits_{i\in T_1}\CL(a+i)\ge q^7>\delta_1.$
		
		\item When $i_5=\frac{q}{2}-1,i_4=\frac{q}{2},i_3=\frac{q}{2}-1,i_2=\frac{q}{2},$ and
		\begin{equation*}
			\left\{\begin{aligned}
				&i_1\ge\frac{q}{2},\mathrm{\ or\ }\\
				&i_1=\frac{q}{2}-1,i_0\ge\frac{q}{2}+2,\mathrm{\ or\ }\\
				&i_6\le\frac{q}{2},i_1=\frac{q}{2}-1,i_0=\frac{q}{2}+1,\\
			\end{aligned}\right.
		\end{equation*} let
		\begin{equation*}
			a+i=\left(\frac{q}{2}-1,\frac{q}{2}-1,\frac{q}{2},\frac{q}{2}+i_6,q-1,0,0,0,\frac{q}{2}-1+i_1,\frac{q}{2}+i_0\right)
		\end{equation*}
		and
		\begin{equation*}
			n-a=\left(\frac{q}{2},\frac{q}{2},\frac{q}{2}-1,\frac{q}{2}-1,\frac{q}{2},\frac{q}{2},\frac{q}{2}-1,\frac{q}{2}-1,\frac{q}{2},\frac{q}{2}+1\right)_q.
		\end{equation*}
		Hence, we have $L_1=L_2=L_5=2$ and $L_3=L_4=3$. Hence $\max\limits_{1\le j\le 5}L_j=3$, and so $\min\limits_{i\in T_1}\CL(a+i)=\min\limits_{x\in X}\CL(x)=\min (R_3\cup R_4)$. Note that
		\begin{align*}
			S_3=&\Big\{\Big(\frac{q}{2}-1,\frac{q}{2}-1,\frac{q}{2},\frac{q}{2}+i_6,q-1,0,0,0,\frac{q}{2}-1+i_1,\frac{q}{2}+i_0\Big)\ \Big|\ i_6\le q-2,i_1=i_0=\frac{q}{2}\Big\}\\
			=&\Big\{\Big(\frac{q}{2}-1,\frac{q}{2}-1,\frac{q}{2},\frac{q}{2}+i_6,q-1,0,0,1,0,0\Big)\ \Big|\ i_6\le q-2\Big\},\ \mathrm{and}\\
			S_4=&\Big\{\Big(\frac{q}{2}-1,\frac{q}{2}-1,\frac{q}{2},\frac{q}{2}+i_6,q-1,0,0,0,\frac{q}{2}-1+i_1,\frac{q}{2}+i_0\Big)\ \Big|\\
			&\quad i_1=\frac{q}{2}-1,\ \mathrm{and\ either}\ i_0\ge\frac{q}{2}+2,\ \mathrm{or}\ i_6\le\frac{q}{2}\ \mathrm{and}\ i_0=\frac{q}{2}+1\Big\}\\
			=&\Big\{\Big(\frac{q}{2}-1,\frac{q}{2}-1,\frac{q}{2},\frac{q}{2}+i_6,q-1,0,0,0,q-1,i_0-\frac{q}{2}\Big)\ \Big|\\
			&\quad \mathrm{either}\ i_0\ge\frac{q}{2}+2,\ \mathrm{or}\ i_6\le\frac{q}{2}\ \mathrm{and}\ i_0=\frac{q}{2}+1\Big\}.
		\end{align*}
		Hence, we have
		\begin{align*}
			\min R_3=&\min\{x\cdot q^{m-2-3}\mid x\in S_3\}\\
			=&\min\Big\{\Big(0,0,0,q-1,q-1,\frac{q}{2},\frac{q}{2},\frac{q}{2}-1,\frac{q}{2}-1-i_6,1\Big)\ \Big|\ i_6\le q-2\Big\},\ \mathrm{and}\\
			\min R_4=&\min\{x\cdot q^{m-e-3}\mid x\in S_4,\ e=2\}\\
			=&\min\Big\{\Big(0,0,0,q-1,i_0-\frac{q}{2}-1,\frac{q}{2},\frac{q}{2},\frac{q}{2}-1,\frac{q}{2}-1-i_6,1\Big)\ \Big|\\
			&\quad \mathrm{either}\ i_0\ge\frac{q}{2}+2,\ \mathrm{or}\ i_6\le\frac{q}{2}\ \mathrm{and}\ i_0=\frac{q}{2}+1\Big\}\\
			=&\Big(0,0,0,q-1,0,\frac{q}{2},\frac{q}{2},\frac{q}{2}-2,q-1,1\Big)_q.
		\end{align*}
		Thus, $\min\limits_{i\in T_1}\CL(a+i)=\min R_4=\delta_1$.
		
		\item When $i_6\ge\frac{q}{2}+1,i_5=\frac{q}{2}-1,i_4=\frac{q}{2},i_3=\frac{q}{2}-1,i_2=\frac{q}{2},i_1=\frac{q}{2}-1,i_0=\frac{q}{2}+1$, let
		\begin{equation*}
			a+i=\left(\frac{q}{2},\frac{q}{2},\frac{q}{2},i_6-\frac{q}{2},0,0,q-1,0,0,2\right)
		\end{equation*}
		and
		\begin{equation*}
			n-a=\left(\frac{q}{2}-1,\frac{q}{2}-1,\frac{q}{2},\frac{q}{2},\frac{q}{2}-1,\frac{q}{2}-1,\frac{q}{2},\frac{q}{2},\frac{q}{2}-1,\frac{q}{2}\right)_q.
		\end{equation*}
		Note that $x_6\ne0$ for any $x\in X$. Hence, we have $\max\limits_{1\le j\le 5}L_j=2$. According to Lemma~\ref{sec3 lem0},  $\min\limits_{i\in T_1}\CL(a+i)\ge q^7>\delta_1$.
		
		\item When $i_5=\frac{q}{2}-1,i_4=\frac{q}{2},i_3=\frac{q}{2}-1,i_2=\frac{q}{2}$, and either $i_1=\frac{q}{2}-1,i_0\le\frac{q}{2},$ or $i_1\le\frac{q}{2}-2,$ let
		\begin{equation*}
			a+i=\left(\frac{q}{2}-1,\frac{q}{2},\frac{q}{2},\frac{q}{2}-1+i_6,q-1,0,q-1,q-1,\frac{q}{2}-1+i_1,\frac{q}{2}+1+i_0\right)
		\end{equation*}
		and
		\begin{equation*}
			n-a=\left(\frac{q}{2},\frac{q}{2}-1,\frac{q}{2}-1,\frac{q}{2},\frac{q}{2},\frac{q}{2}-1,\frac{q}{2}-1,\frac{q}{2},\frac{q}{2},\frac{q}{2}\right)_q.
		\end{equation*}
		Thus, we have $L_1=L_2=L_4=1,L_3=0$ and $L_5=3$. Hence $\max\limits_{1\le j\le 5}L_j=L_5=3$, and so $\min\limits_{i\in T_1}\CL(a+i)=\min\limits_{x\in X}\CL(x)=\min R_5$. Note that
		\begin{align*}
			S_5=&\Big\{\Big(\frac{q}{2}-1,\frac{q}{2},\frac{q}{2},\frac{q}{2}-1+i_6,q-1,0,q-1,q-1,\frac{q}{2}-1+i_1,\frac{q}{2}+1+i_0\Big)\ \Big|\\
			&\quad i_6\le q-2,i_1=\frac{q}{2}-1,i_0=\frac{q}{2}\Big\}\\
			=&\Big\{\Big(\frac{q}{2}-1,\frac{q}{2},\frac{q}{2},\frac{q}{2}-1+i_6,q-1,0,q-1,q-1,q-1,1\Big)\ \Big|\ i_6\le q-2\Big\}.
		\end{align*}
		Hence, we have
		\begin{align*}
			\min R_5=&\min\{x\cdot q^{2m-f-3}\mid x\in S_5,\ f=1\}\\
			=&\min\Big\{\Big(0,0,0,q-1,\frac{q}{2}-1,\frac{q}{2},\frac{q}{2},\frac{q}{2}-1+i_6,q-1,1\Big)\ \Big|\ i_6\le q-2\Big\}\\
			=&\Big(0,0,0,q-1,\frac{q}{2}-1,\frac{q}{2},\frac{q}{2},\frac{q}{2}-1,q-1,1\Big)_q>\delta_1.
		\end{align*}
		
		\item When $i_5=\frac{q}{2}-1,i_4\ne\frac{q}{2},i_3=\frac{q}{2}-1$ and $i_2\ne\frac{q}{2}-1,$ let
		\begin{equation*}
			a+i=\left(\frac{q}{2}-1,\frac{q}{2}-1,\frac{q}{2},\frac{q}{2}+i_6,q-2,\frac{q}{2}-1+i_4,q-1,\frac{q}{2}+i_2,\frac{q}{2}-1+i_1,\frac{q}{2}+i_0\right)
		\end{equation*}
		and
		\begin{equation*}
			n-a=\left(\frac{q}{2},\frac{q}{2},\frac{q}{2}-1,\frac{q}{2}-1,\frac{q}{2},\frac{q}{2},\frac{q}{2}-1,\frac{q}{2}-1,\frac{q}{2},\frac{q}{2}+1\right)_q.
		\end{equation*}
		Then we have $\max\limits_{1\le j\le 5}L_j=2$. It then follows from Lemma~\ref{sec3 lem0} that $\min\limits_{i\in T_1}\CL(a+i)\ge q^7>\delta_1.$
		
		\item When $i_5=\frac{q}{2}-1,i_3=\frac{q}{2}-1,i_2=\frac{q}{2}-1$, and either $i_4\ne\frac{q}{2}-1,$ or $i_4=\frac{q}{2}-1,i_1=\frac{q}{2}-1,$ let
		\begin{equation*}
			a+i=\left(\frac{q}{2}-1,\frac{q}{2},\frac{q}{2},\frac{q}{2}-1+i_6,q-2,\frac{q}{2}+i_4,q-1,q-2,\frac{q}{2}-1+i_1,\frac{q}{2}+1+i_0\right)
		\end{equation*}
		and
		\begin{equation*}
			n-a=\left(\frac{q}{2},\frac{q}{2}-1,\frac{q}{2}-1,\frac{q}{2},\frac{q}{2},\frac{q}{2}-1,\frac{q}{2}-1,\frac{q}{2},\frac{q}{2},\frac{q}{2}\right)_q.
		\end{equation*}
		Thus, we have $\max\limits_{1\le j\le 5}L_j=2$, and so $\min\limits_{i\in T_1}\CL(a+i)\ge q^7>\delta_1$ by Lemma~\ref{sec3 lem0}.
		
		\item When $i_5=\frac{q}{2}-1,i_4=\frac{q}{2}-1,i_3=\frac{q}{2}-1,i_2=\frac{q}{2}-1$ and $i_1\ne\frac{q}{2}-1,$ let
		\begin{equation*}
			a+i=\left(\frac{q}{2},\frac{q}{2}-1,\frac{q}{2}-1,\frac{q}{2}+i_6,q-1,q-2,q-2,q-1,\frac{q}{2}+i_1,\frac{q}{2}+i_0\right)
		\end{equation*}
		and
		\begin{equation*}
			n-a=\left(\frac{q}{2}-1,\frac{q}{2},\frac{q}{2},\frac{q}{2}-1,\frac{q}{2}-1,\frac{q}{2},\frac{q}{2},\frac{q}{2}-1,\frac{q}{2}-1,\frac{q}{2}+1\right)_q.
		\end{equation*}
		One can check that $\max\limits_{1\le j\le 5}L_j=2$. It follows from Lemma~\ref{sec3 lem0} that $\min\limits_{i\in T_1}\CL(a+i)\ge q^7>\delta_1.$ 
	\end{itemize}
	In summary, we have $a+i \in I_1$, and so $n+i=(a+i)+(n-a) \in I_1+I_2$ for any $i \in T_1$.
	
	\item $\boldsymbol{m=11:}$ we have
	\begin{equation*}
		i=(\underbrace{0,0,\ldots,0}_{m-r-1},i_r\ne0,i_{r-1},\ldots,i_1,i_0\ne0)_q,
	\end{equation*}
	where $0\leq r\leq 6$ as $$i<\delta_{1}=\frac{2q^8-q^5-2q^4+q}{2(q+1)}-1=\left(0,0,0,0,q-1,0,q-2,\frac{q}{2},\frac{q}{2}-1,\frac{q}{2},\frac{q}{2}-1\right)_q.$$ With the same notation as in Lemma~\ref{sec3 lem0}, take $X := a + T_1$. We only need to prove that there always exists an integer $a\in C_{\Delta_1}$ such that $\min\limits_{i\in T_1}\CL(a+i)\ge\delta_{1}$. Through calculation, the largest coset leader in $T_1$ is
	\begin{equation*}
		\left\{\begin{array}{ll}
			\left(0,0,0,0,q-1,0,q-2,\frac{q}{2},\frac{q}{2}-1,\frac{q}{2}-1,q-1\right)_q&\mathrm{if\ }q=4,\\
			\left(0,0,0,0,q-1,0,q-2,\frac{q}{2},\frac{q}{2}-1,\frac{q}{2},\frac{q}{2}-2\right)_q&\mathrm{if\ }q>4.
		\end{array}\right.
	\end{equation*}
	\begin{itemize}
		\item When $i_4\ne \frac{q}{2}-1$, and either $i_3\ne\frac{q}{2}$,  $i_2\ne\frac{q}{2}-1$, $i_1\ne\frac{q}{2}$, or $i_0\notin\{\frac{q}{2},\frac{q}{2}-1\}$, take $a:=n-\Delta_1$. Let
		\begin{equation*}
			a+i=n-\Delta_1+i=\left(\frac{q}{2},\frac{q}{2}-1,\frac{q}{2},\frac{q}{2}-1,\frac{q}{2}+i_6,\frac{q}{2}-1+i_5,\frac{q}{2}+i_4,\frac{q}{2}-1+i_3,\frac{q}{2}+i_2,\frac{q}{2}-1+i_1,\frac{q}{2}+1+i_0\right)
		\end{equation*}
		and
		\begin{equation*}
			y=\Delta_1=\left(\frac{q}{2}-1,\frac{q}{2},\frac{q}{2}-1,\frac{q}{2},\frac{q}{2}-1,\frac{q}{2},\frac{q}{2}-1,\frac{q}{2},\frac{q}{2}-1,\frac{q}{2},\frac{q}{2}\right)_q.
		\end{equation*}
		Since $x_4\ne\frac{q}{2}-1$, $(x_4,x_3)\notin \{(0,0),(q-1,q-1)\}$ for any $x\in X$. Then we have $(x_6,x_5,x_4,x_3)$, $(x_5,x_4,x_3,x_2)$, and $(x_4,x_3,x_2,x_1)\notin\{(0,0,0,0),(q-1,q-1,q-1,q-1)\}$, which implies that $L_4=L_5\le3$. Similarly, one can get that $L_1=L_2=L_3\le3$ if either $i_3\ne\frac{q}{2}$,  $i_2\ne\frac{q}{2}-1$, $i_1\ne\frac{q}{2}$, or $i_0\notin\{\frac{q}{2},\frac{q}{2}-1\}$. Since $i_5\le \frac{q}{2}-1$, $(x_5,x_4,x_3)\notin \{(0,0,0),(0,0,1)\}$. By Lemma~\ref{sec3 lem0}, $\min\limits_{i\in T_1}\CL(a+i)\ge q^7>\delta_1$.
		
		\item When $i_3=\frac{q}{2},i_2=\frac{q}{2}-1,i_1=\frac{q}{2}$ and $i_0\in\{\frac{q}{2},\frac{q}{2}-1\}$, take $a:=\Delta_1$. Let
		\begin{equation*}
			a+i=\Delta_1+i=\left(\frac{q}{2}-1,\frac{q}{2},\frac{q}{2}-1,\frac{q}{2},\frac{q}{2}-1+i_6,\frac{q}{2}+i_5,\frac{q}{2}+i_4,0,q-1,0,\frac{q}{2}+i_0\right)
		\end{equation*}
		and
		\begin{equation*}
			y=n-\Delta_1=\left(\frac{q}{2},\frac{q}{2}-1,\frac{q}{2},\frac{q}{2}-1,\frac{q}{2},\frac{q}{2}-1,\frac{q}{2},\frac{q}{2}-1,\frac{q}{2},\frac{q}{2}-1,\frac{q}{2}+1\right)_q.
		\end{equation*}
		Hence, we have $L_1=L_3=1,L_2=0,L_5=3$ and $L_4=4$. Hence $\max\limits_{1\le j\le 5}L_j=L_4=4$, and so $\min\limits_{i\in T_1}\CL(a+i)=\min\limits_{x\in X}\CL(x)=\min R_4$. Note that
		\begin{align*}
			S_4=&\Big\{\Big(\frac{q}{2}-1,\frac{q}{2},\frac{q}{2}-1,\frac{q}{2},\frac{q}{2}-1+i_6,\frac{q}{2}+i_5,\frac{q}{2}+i_4,0,q-1,0,\frac{q}{2}+i_0\Big)\ \Big|\\
			&\quad i_6=\frac{q}{2},i_5=\frac{q}{2}-1,i_4=\frac{q}{2},i_0\in\{\frac{q}{2},\frac{q}{2}-1\}\Big\}\\
			=&\Big\{\Big(\frac{q}{2}-1,\frac{q}{2},\frac{q}{2}-1,\frac{q}{2}+1,0,0,0,0,q-1,0,\frac{q}{2}+i_0\Big)\ \Big|\ i_0\in\{\frac{q}{2},\frac{q}{2}-1\}\Big\}.
		\end{align*}
		Hence, we have
		\begin{align*}
			\min R_4=&\min\{x\cdot q^{m-e-4}\mid x\in S_4,\ e=3\}\\
			=&\min\Big\{\Big(0,0,0,0,q-1,0,\frac{q}{2}-1+i_0,\frac{q}{2},\frac{q}{2}-1,\frac{q}{2},\frac{q}{2}-1\Big)\ \Big|\ i_0\in\{\frac{q}{2},\frac{q}{2}-1\}\Big\} \\
			=&\Big(0,0,0,0,q-1,0,q-2,\frac{q}{2},\frac{q}{2}-1,\frac{q}{2},\frac{q}{2}-1\Big)_q=\delta_1.
		\end{align*}
		
		\item When $i_4=\frac{q}{2}-1,i_3=\frac{q}{2}-1,i_2=\frac{q}{2},i_1=\frac{q}{2}-1$ and $i_0\in\{\frac{q}{2},\frac{q}{2}+1\}$, take $a:=n-\Delta_1$. Let
		\begin{equation*}
			a+i=n-\Delta_1+i=\left(\frac{q}{2},\frac{q}{2}-1,\frac{q}{2},\frac{q}{2}-1,\frac{q}{2}+i_6,\frac{q}{2}-1+i_5,q-1,q-1,0,q-1,i_0-\frac{q}{2}+1\right)
		\end{equation*}
		and
		\begin{equation*}
			y=\Delta_1=\left(\frac{q}{2}-1,\frac{q}{2},\frac{q}{2}-1,\frac{q}{2},\frac{q}{2}-1,\frac{q}{2},\frac{q}{2}-1,\frac{q}{2},\frac{q}{2}-1,\frac{q}{2},\frac{q}{2}\right)_q.
		\end{equation*}
		Then $L_1=L_2=1,L_3=0,L_4=2$ and $L_5=4$. Hence $\max\limits_{1\le j\le 5}L_j=L_5=4$, and so $\min\limits_{i\in T_1}\CL(a+i)=\min\limits_{x\in X}\CL(x)=\min R_5$. Note that
		\begin{align*}
			S_5=&\Big\{\Big(\frac{q}{2},\frac{q}{2}-1,\frac{q}{2},\frac{q}{2}-1,\frac{q}{2}+i_6,\frac{q}{2}-1+i_5,q-1,q-1,0,q-1,i_0-\frac{q}{2}+1\Big)\ \Big|\\
			&\quad i_6=\frac{q}{2}-1, i_5=\frac{q}{2},i_0\in\{\frac{q}{2},\frac{q}{2}+1\}\Big\}\\
			=&\Big\{\Big(\frac{q}{2},\frac{q}{2}-1,\frac{q}{2},\frac{q}{2}-1,q-1,q-1,q-1,q-1,0,q-1,i_0-\frac{q}{2}+1\Big)\ \Big|\ i_0\in\{\frac{q}{2},\frac{q}{2}+1\}\Big\}.
		\end{align*}
		Hence, we have
		\begin{align*}
			\min R_5=&\min\{x\cdot q^{2m-f-4}\mid x\in S_5,\ f=3\}\\
			=&\min\Big\{\Big(0,0,0,0,q-1,0,\frac{3q}{2}-1-i_0,\frac{q}{2},\frac{q}{2}-1,\frac{q}{2},\frac{q}{2}\Big)\ \Big|\ i_0\in\{\frac{q}{2},\frac{q}{2}+1\}\Big\} \\
			=&\Big(0,0,0,0,q-1,0,q-2,\frac{q}{2},\frac{q}{2}-1,\frac{q}{2},\frac{q}{2}\Big)_q>\delta_1.
		\end{align*}
		
		\item When $i_4=\frac{q}{2}-1$, and either $i_3\ne\frac{q}{2}-1,i_2\ne\frac{q}{2},i_1\ne\frac{q}{2}-1$, or $i_0\notin\{\frac{q}{2},\frac{q}{2}+1\}$, take $a := \Delta_{1}$. Let
		\begin{equation*}
			a+i=\Delta_1+i=\left(\frac{q}{2}-1,\frac{q}{2},\frac{q}{2}-1,\frac{q}{2},\frac{q}{2}-1+i_6,\frac{q}{2}+i_5,q-2,\frac{q}{2}+i_3,\frac{q}{2}-1+i_2,\frac{q}{2}+i_1,\frac{q}{2}+i_0\right)
		\end{equation*}
		and
		\begin{equation*}
			y=n-\Delta_1=\left(\frac{q}{2},\frac{q}{2}-1,\frac{q}{2},\frac{q}{2}-1,\frac{q}{2},\frac{q}{2}-1,\frac{q}{2},\frac{q}{2}-1,\frac{q}{2},\frac{q}{2}-1,\frac{q}{2}+1\right)_q.
		\end{equation*}
		One can easily obtain that $\max\limits_{1\le j\le 5}L_j=3$. It then follows from Lemma~\ref{sec3 lem0} that $\min\limits_{i\in T_1}\CL(a+i)=\min\limits_{x\in X}\CL(x)\ge q^7>\delta_1$.
	\end{itemize}
	In summary, we have $a+i \in I_1$, and so $n+i=(a+i)+(n-a) \in I_1+I_2$ for any $i \in T_1$. 
			\end{itemize}
		\end{itemize}
		Therefore, $\Z \subseteq I_1+I_2$.

		We then prove (\ref{sec3.2 thm1 equ2}). Suppose that $\delta_{1}=\Omega_{(q,m)}+1$ and $\delta_{2}=\Delta_2+1$, then $I_2=C_0\cup C_{\Delta_1}$.
		\begin{itemize}
				\item $\boldsymbol{m=2:}$ 
				Note that $\frac{q}{2}\in T\left(\delta_{1},\delta_{2}\right)$. Suppose on the contrary that $\frac{q}{2} \in I_1+I_2$, then there exist $x\in I_1$ and $y\in I_2$ such that
				\begin{equation*}
					x+y=\frac{q}{2}\ \text{or}\ x+y=\frac{q}{2}+n
				\end{equation*}
				since $0\leq x+y<2n$. If $y=0$, then $x=\frac{q}{2}\in T_1=\mathbb{Z}_n\setminus I_1$, a contradiction. Thus, we have $y\in C_{\Delta_1}$ and $x=n-y+\frac{q}{2}$. Note that $$\begin{aligned}
					\max\CL(x)=\max\limits_{y\in C_{\Delta_1}}\CL\Big(n-y+\frac{q}{2}\Big)=\Big(0,\frac{q}{2}+1\Big)_q<\frac{q}{2}+2=\delta_{1},
				\end{aligned}$$ which contradicts $x\in I_1$.
				\item $\boldsymbol{m=3:}$ 
				Note that $\frac{q^2-q}{2}\in T\left(\delta_{1},\delta_{2}\right)$. Suppose on the contrary that $\frac{q^2-q}{2} \in I_1+I_2$, then there exist $x\in I_1$ and $y\in I_2$ such that
				\begin{equation*}
					x+y=\frac{q^2-q}{2}\ \text{or}\ x+y=\frac{q^2-q}{2}+n
				\end{equation*}
				since $0\leq x+y<2n$. If $y=0$, then $x=\frac{q^2-q}{2}\in T_1=\mathbb{Z}_n\setminus I_1$, a contradiction. Thus, we have $y\in C_{\Delta_1}$ and $x=n-y+\frac{q^2-q}{2}$. Note that $$\max\CL(x)=\max\limits_{y\in C_{\Delta_1}}\CL\Big(n-y+\frac{q^2-q}{2}\Big)=\Big(0,\frac{q}{2}-1,1\Big)_q<\frac{q^2-q}{2}+1=\delta_{1},$$ which contradicts $x\in I_1$.
				\item $\boldsymbol{m=4:}$ 
				Note that $\frac{q^3-q^2-q}{2}\in T\left(\delta_{1},\delta_{2}\right)$. Suppose on the contrary that $\frac{q^3-q^2-q}{2} \in I_1+I_2$, then there exist $x\in I_1$ and $y\in I_2$ such that
				\begin{equation*}
					x+y=\frac{q^3-q^2-q}{2}\ \text{or}\ x+y=\frac{q^3-q^2-q}{2}+n
				\end{equation*}
				since $0\leq x+y<2n$. If $y=0$, then $x=\frac{q^3-q^2-q}{2}\in T_1=\mathbb{Z}_n\setminus I_1$, a contradiction. Thus, we have $y\in C_{\Delta_1}$ and $x=n-y+\frac{q^3-q^2-q}{2}$. Note that $$\begin{aligned}
					\max\CL(x)=\max\limits_{y\in C_{\Delta_1}}\CL\Big(n-y+\frac{q^3-q^2-q}{2}\Big)=\Big(0,\frac{q}{2}-1,1,1\Big)_q<\frac{q^3-q^2-q}{2}+1=\delta_{1},
				\end{aligned}$$ which contradicts $x\in I_1$.
				\item $\boldsymbol{m=5:}$ $\frac{q^3+q^2+q}{2}-1\in T\left(\delta_{1},\delta_{2}\right)$. Suppose on the contrary that $\frac{q^3+q^2+q}{2}-1 \in I_1+I_2$, then there exist $x\in I_1$ and $y\in I_2$ such that
				\begin{equation*}
					x+y=\frac{q^3+q^2+q}{2}-1\ \text{or}\ x+y=\frac{q^3+q^2+q}{2}-1+n
				\end{equation*}
				since $0\leq x+y<2n$. If $y=0$, then $x=\frac{q^3+q^2+q}{2}-1\in T_1=\mathbb{Z}_n\setminus I_1$, a contradiction. Thus, we have $y\in C_{\Delta_1}$ and $x=n-y+\frac{q^3+q^2+q}{2}-1$. Note that \begin{align*}
\max\CL(x)&=\max\limits_{y\in C_{\Delta_1}}\CL\Big(n-y+\frac{q^3+q^2+q}{2}-1\Big)\\
&=\Big(0,0,q-2,\frac{q}{2},\frac{q}{2}-1\Big)_q=\frac{2q^3-3q^2+q}{2}-1<\frac{2q^3-3q^2+q}{2}=\delta_{1},
\end{align*} which contradicts $x\in I_1$.
		\item $\boldsymbol{m=6:}$
		Note that $\frac{q^4-q^3+q^2-q}{2}\in T\left(\delta_{1},\delta_{2}\right)$. Suppose on the contrary that $\frac{q^4-q^3+q^2-q}{2} \in I_1+I_2$, then there exist $x\in I_1$ and $y\in I_2$ such that
		\begin{equation*}
			x+y=\frac{q^4-q^3+q^2-q}{2}\ \text{or}\ x+y=\frac{q^4-q^3+q^2-q}{2}+n
		\end{equation*}
		since $0\leq x+y<2n$. If $y=0$, then $x=\frac{q^4-q^3+q^2-q}{2}\in T_1=\mathbb{Z}_n\setminus I_1$, a contradiction. Thus, we have $y\in C_{\Delta_1}$ and $x=n-y+\frac{q^4-q^3+q^2-q}{2}$. Note that $$\begin{aligned}
			\max\CL(x)=\max\limits_{y\in C_{\Delta_1}}\CL\Big(n-y+\frac{q^4-q^3+q^2-q}{2}\Big)=\Big(0,0,q-2,q-1,\frac{q}{2},\frac{q}{2}\Big)_q<\frac{2q^4-2q^3-q^2+q}{2}+1=\delta_{1},
		\end{aligned}$$ which contradicts $x\in I_1$.
		\item $\boldsymbol{m=7:}$
		Note that $\frac{q^5-q^4-q^3+2q^2+2q}{2}\in T\left(\delta_{1},\delta_{2}\right)$. Suppose on the contrary that $\frac{q^5-q^4-q^3+2q^2+2q}{2} \in I_1+I_2$, then there exist $x\in I_1$ and $y\in I_2$ such that
		\begin{equation*}
			x+y=\frac{q^5-q^4-q^3+2q^2+2q}{2}\ \text{or}\ x+y=\frac{q^5-q^4-q^3+2q^2+2q}{2}+n
		\end{equation*}
		since $0\leq x+y<2n$. If $y=0$, then $x=\frac{q^5-q^4-q^3+2q^2+2q}{2}\in T_1=\mathbb{Z}_n\setminus I_1$, a contradiction. Thus, we have $y\in C_{\Delta_1}$ and $x=n-y+\frac{q^5-q^4-q^3+2q^2+2q}{2}$. Note that $$\begin{aligned}
			&\max\CL(x)=\max\limits_{y\in C_{\Delta_1}}\CL\Big(n-y+\frac{q^5-q^4-q^3+2q^2+2q}{2}\Big)\\
			=&\Big(0,0,\frac{q}{2}-1,\frac{q}{2},0,0,q-1\Big)_q=\frac{q^5-q^4+2q}{2}-1<\frac{q^5-q^4+2q}{2}=\delta_{1},
		\end{aligned}$$ which contradicts $x\in I_1$.
		\item $\boldsymbol{m=8:}$
		Note that $\frac{q^7-q}{2(q+1)}\in T\left(\delta_{1},\delta_{2}\right)$. Suppose on the contrary that $\frac{q^7-q}{2(q+1)} \in I_1+I_2$, then there exist $x\in I_1$ and $y\in I_2$ such that
		\begin{equation*}
			x+y=\frac{q^7-q}{2(q+1)}\ \text{or}\ x+y=\frac{q^7-q}{2(q+1)}+n
		\end{equation*}
		since $0\leq x+y<2n$. If $y=0$, then $x=\frac{q^7-q}{2(q+1)}\in T_1=\mathbb{Z}_n\setminus I_1$, a contradiction. Thus, we have $y\in C_{\Delta_1}$ and $x=n-y+\frac{q^7-q}{2(q+1)}$. Note that $$\begin{aligned}
			\max\CL(x)=\max\limits_{y\in C_{\Delta_1}}\CL\Big(n-y+\frac{q^7-q}{2(q+1)}\Big)=\Big(0,0,q-1,0,q-2,q-1,\frac{q}{2},\frac{q}{2}\Big)_q<\frac{2q^7-3q^3+q}{2(q+1)}+1=\delta_{1},
		\end{aligned}$$ which contradicts $x\in I_1$.
		\item $\boldsymbol{m=9:}$
		Note that $\frac{q^7-2q^3+q}{2(q+1)}\in T\left(\delta_{1},\delta_{2}\right)$. Suppose on the contrary that $\frac{q^7-2q^3+q}{2(q+1)} \in I_1+I_2$, then there exist $x\in I_1$ and $y\in I_2$ such that
		\begin{equation*}
			x+y=\frac{q^7-2q^3+q}{2(q+1)}\ \text{or}\ x+y=\frac{q^7-2q^3+q}{2(q+1)}+n
		\end{equation*}
		since $0\leq x+y<2n$. If $y=0$, then $x=\frac{q^7-2q^3+q}{2(q+1)}\in T_1=\mathbb{Z}_n\setminus I_1$, a contradiction. Thus, we have $y\in C_{\Delta_1}$ and $x=n-y+\frac{q^7-2q^3+q}{2(q+1)}$. Note that $$\begin{aligned}
			\max\CL(x)=\max\limits_{y\in C_{\Delta_1}}\CL\Big(n-y+\frac{q^7-2q^3+q}{2(q+1)}\Big)=\Big(0,0,0,\frac{q}{2}-1,\frac{q}{2},\frac{q}{2}-1,0,q-1,1\Big)_q<\frac{q^7-2q^3+q}{2(q+1)}=\delta_{1},
		\end{aligned}$$ which contradicts $x\in I_1$.
		\item $\boldsymbol{m=10:}$
		Note that $\frac{(q^7+q)(q+2)}{2(q+1)}\in T\left(\delta_{1},\delta_{2}\right)$. Suppose on the contrary that $\frac{(q^7+q)(q+2)}{2(q+1)} \in I_1+I_2$, then there exist $x\in I_1$ and $y\in I_2$ such that
		\begin{equation*}
			x+y=\frac{(q^7+q)(q+2)}{2(q+1)}\ \text{or}\ x+y=\frac{(q^7+q)(q+2)}{2(q+1)}+n
		\end{equation*}
		since $0\leq x+y<2n$. If $y=0$, then $x=\frac{(q^7+q)(q+2)}{2(q+1)}\in T_1=\mathbb{Z}_n\setminus I_1$, a contradiction. Thus, we have $y\in C_{\Delta_1}$ and $x=n-y+\frac{(q^7+q)(q+2)}{2(q+1)}$. Note that $$\begin{aligned}
			&\max\CL(x)=\max\limits_{y\in C_{\Delta_1}}\CL\Big(n-y+\frac{(q^7+q)(q+2)}{2(q+1)}\Big)\\
			=&\Big(0,0,0,q-1,0,\frac{q}{2},\frac{q}{2},\frac{q}{2}-2,q-1,1\Big)_q<\frac{2q^7-2q^6+q^5+q^4+q^3-2q^2-2q}{2}+2=\delta_{1},
		\end{aligned}$$ which contradicts $x\in I_1$.
		\item $\boldsymbol{m=11:}$
		Note that $\frac{q^8+2q^7+q^4}{2(q+1)}-1\in T\left(\delta_{1},\delta_{2}\right)$. Suppose on the contrary that $\frac{q^8+2q^7+q^4}{2(q+1)}-1 \in I_1+I_2$, then there exist $x\in I_1$ and $y\in I_2$ such that
		\begin{equation*}
			x+y=\frac{q^8+2q^7+q^4}{2(q+1)}-1\ \text{or}\ x+y=\frac{q^8+2q^7+q^4}{2(q+1)}-1+n
		\end{equation*}
		since $0\leq x+y<2n$. If $y=0$, then $x=\frac{q^8+2q^7+q^4}{2(q+1)}-1\in T_1=\mathbb{Z}_n\setminus I_1$, a contradiction. Thus, we have $y\in C_{\Delta_1}$ and $x=n-y+\frac{q^8+2q^7+q^4}{2(q+1)}-1$. Note that 
		$$\begin{aligned}
			\max\CL(x)=&\max\limits_{y\in C_{\Delta_1}}\CL\Big(n-y+\frac{q^8+2q^7+q^4}{2(q+1)}-1\Big)=\Big(0,0,0,0,q-1,0,q-2,\frac{q}{2},\frac{q}{2}-1,\frac{q}{2},\frac{q}{2}-1\Big)_q\\
			<&\frac{2q^8-q^5-2q^4+q}{2(q+1)}=\delta_{1},
		\end{aligned}$$ 
		which contradicts $x\in I_1$.
		Hence, we have $T\left(\Omega_{(q,m)}+1,\Delta_2+1\right)\ne\emptyset$.
		\end{itemize}
		The proof is completed.
\end{proof}

\section{\texorpdfstring{Proof of Theorem~\ref{sec3.2 thm2}}{Proof of Theorem 5}}\label{apdx b}

\begin{proof}
	For $5\le m\le 11$, we have $\Lambda_{(q,m)}\le\Omega_{(q,m)}$, where $\Omega_{(q,m)}$ is as in Table~\ref{sec3.2 thm1 table1}. Hence, we only need to prove the case of $m\ge13$ by Theorem~\ref{sec3.2 thm1}. Suppose that $\delta_{1}=\Lambda_{(q,m)}$ and $\delta_{2}=\Delta_1$, then $T_1=C_{1}\cup C_{2}\cup\cdots\cup C_{\Lambda_{(q,m)}-1}$ and $T_2=C_{1}\cup C_{2}\cup\cdots\cup C_{\Delta_1-1}$, and so $I_1=\mathbb{Z}_n\setminus T_1$ and $I_2=C_0\cup C_{\Delta_1}$. Similar to the proof of Theorem~\ref{sec3.2 thm1} (i), it suffices to show that $i\in I_{1}+I_{2}$ for any coset leader $i\in\mathbb{Z}_n$. We first consider the case where $\frac{m-1}{2}$ is odd. 
	\begin{itemize}
		\item If $i \in I_1$, then $i = i+0\in I_1+I_2$.
		\item If $i\in\mathbb{Z}_n\setminus I_1=T_1$, we have
		\begin{equation*}
			i=(\underbrace{0,0,\ldots,0}_{m-r-1},i_r\ne0,i_{r-1},\ldots,i_1,i_0\ne0)_q,
		\end{equation*}
		where $0\leq r\leq \frac{m+1}{2}$ as $$i<\delta_{1}=\Lambda_{(q,m)}=\Big(\underbrace{\vphantom{\frac{q-1}{2}}0,\dots,0}_{\frac{m-1}{2}},\frac{q}{2},\frac{q}{2},\frac{q}{2}-1,\dots,\frac{q}{2},\frac{q}{2}-1,\frac{q}{2}\Big)_q.$$ With the same notation as in Lemma~\ref{sec3 lem0}, take $X := a + T_1$. We only need to prove that there always exists an integer $a\in C_{\Delta_1}$ such that $\min\limits_{i\in T_1}\CL(a+i)\ge\delta_{1}$. Through calculation, the largest coset leader in $T_1$ is
		\begin{equation*}
			\Big(\underbrace{\vphantom{\frac{q-1}{2}}0,\dots,0}_{\frac{m-1}{2}},\frac{q}{2},\frac{q}{2},\frac{q}{2}-1,\dots,\frac{q}{2},\frac{q}{2}-1,\frac{q}{2}-1\Big)_q.
		\end{equation*}
		\begin{itemize}
			\item When $i_1\ne\frac{q}{2}$, take $a:=n-\Delta_1$. Let
			\begin{equation*}
				a+i=n-\Delta_1+i=\left(\frac{q}{2},\frac{q}{2}-1,\dots,\frac{q}{2},\frac{q}{2}-1+i_{\frac{m-1}{2}},\frac{q}{2}+i_{\frac{m-3}{2}},\dots,\frac{q}{2}+i_2,\frac{q}{2}-1+i_1,\frac{q}{2}+1+i_0\right)
			\end{equation*}
			and
			\begin{equation*}
				n-a=\Delta_{1}=\left(\frac{q}{2}-1,\frac{q}{2},\dots,\frac{q}{2}-1,\frac{q}{2},\frac{q}{2}\right)_q.
			\end{equation*}
			Since $i_1\ne\frac{q}{2}$, $(x_1,x_0)\notin\{(0,0),(0,1)\}$, which implies that $L_1=L_2=1$ and $L_3=L_4=L_5=\frac{m-1}{2}$. Hence $\max\limits_{1\le j\le 5}L_j=\frac{m-1}{2}$, and so $\min\limits_{i\in T_1}\CL(a+i)=\min\limits_{x\in X}\CL(x)=\min (R_3\cup R_4\cup R_5)$. Note that
			\begin{align*}
				S_3=&\Big\{\Big(\frac{q}{2},\frac{q}{2}-1,\dots,\frac{q}{2},\frac{q}{2}-1+i_{\frac{m-1}{2}},\frac{q}{2}+i_{\frac{m-3}{2}},\dots,\frac{q}{2}+i_2,\frac{q}{2}-1+i_1,\frac{q}{2}+1+i_0\Big)\ \Big|\\
				&\quad i_1=\frac{q}{2}+1,\ i_0=i_2=\dots=i_{\frac{m-3}{2}}=\frac{q}{2}-1,\ i_3=i_5=\dots=i_{\frac{m-1}{2}}=\frac{q}{2}\Big\}\\
				=&\Big\{\Big(\frac{q}{2},\frac{q}{2}-1,\dots,\frac{q}{2},\frac{q}{2}-1,\frac{q}{2}+1,\underbrace{0,\dots,0}_{\frac{m-3}{2}},1,0\Big)_q\Big\},\\
				S_4=&\Big\{\Big(\frac{q}{2},\frac{q}{2}-1,\dots,\frac{q}{2},\frac{q}{2}-1+i_{\frac{m-1}{2}},\frac{q}{2}+i_{\frac{m-3}{2}},\dots,\frac{q}{2}+i_2,\frac{q}{2}-1+i_1,\frac{q}{2}+1+i_0\Big)\ \Big|\\
				&\quad 1\le i_0\le\frac{q}{2}-2,\ i_1=\frac{q}{2}+1,\ i_2=\dots=i_{\frac{m-3}{2}}=\frac{q}{2}-1,\ i_3=i_5=\dots=i_{\frac{m-1}{2}}=\frac{q}{2}\Big\}\\
				=&\Big\{\Big(\frac{q}{2},\frac{q}{2}-1,\dots,\frac{q}{2},\frac{q}{2}-1,\frac{q}{2}+1,\underbrace{0,\dots,0}_{\frac{m-1}{2}},\frac{q}{2}+1+i_0\Big)\ \Big|\ 1\le i_0\le\frac{q}{2}-2\Big\},\ \mathrm{and}\\
				S_5=&\Big\{\Big(\frac{q}{2},\frac{q}{2}-1,\dots,\frac{q}{2},\frac{q}{2}-1+i_{\frac{m-1}{2}},\frac{q}{2}+i_{\frac{m-3}{2}},\dots,\frac{q}{2}+i_2,\frac{q}{2}-1+i_1,\frac{q}{2}+1+i_0\Big)\ \Big|\\
				&\quad \frac{q}{2}\le i_0\le q-1,\ i_1=\frac{q}{2}-1,\ i_2=\dots=i_{\frac{m-3}{2}}=\frac{q}{2}-1,\ i_3=i_5=\dots=i_{\frac{m-1}{2}}=\frac{q}{2}\Big\}\\
				=&\Big\{\Big(\frac{q}{2},\frac{q}{2}-1,\dots,\frac{q}{2},\frac{q}{2}-1,\frac{q}{2},\underbrace{q-1,\dots,q-1}_{\frac{m-1}{2}},i_0+1-\frac{q}{2}\Big)\ \Big|\ \frac{q}{2}\le i_0\le q-1\Big\}.
			\end{align*}
			Hence, we have
			\begin{align*}
				\min R_3=&\min\{x\cdot q^{m-1-\frac{m-1}{2}}\mid x\in S_3\}\\
				=&\Big(\underbrace{\vphantom{\frac{q-1}{2}}0,\dots,0}_{\frac{m-1}{2}},q-1,\frac{q}{2}-1,\frac{q}{2},\dots,\frac{q}{2}-1,\frac{q}{2},\frac{q}{2}-1\Big)_q>\delta_1,\\
				\min R_4=&\min\{x\cdot q^{m-e-\frac{m-1}{2}}\mid x\in S_4,\ e=1\}\\
				=&\min\Big\{\Big(\underbrace{\vphantom{\frac{q-1}{2}}0,\dots,0}_{\frac{m-1}{2}},\frac{q}{2}+i_0,\frac{q}{2}-1,\frac{q}{2},\dots,\frac{q}{2}-1,\frac{q}{2},\frac{q}{2}-1\Big)\ \Big|\ 1\le i_0\le\frac{q}{2}-2\Big\}\\
				=&\Big(\underbrace{\vphantom{\frac{q-1}{2}}0,\dots,0}_{\frac{m-1}{2}},\frac{q}{2}+1,\frac{q}{2}-1,\frac{q}{2},\dots,\frac{q}{2}-1,\frac{q}{2},\frac{q}{2}-1\Big)_q>\delta_1,\ \mathrm{and}\\
				\min R_5=&\min\{x\cdot q^{2m-f-\frac{m-1}{2}}\mid x\in S_5,\ f=1\}\\
				=&\min\Big\{\Big(\underbrace{\vphantom{\frac{q-1}{2}}0,\dots,0}_{\frac{m-1}{2}},\frac{3q}{2}-1-i_0,\frac{q}{2},\frac{q}{2}-1,\dots,\frac{q}{2},\frac{q}{2}-1,\frac{q}{2}\Big)\ \Big|\ \frac{q}{2}\le i_0\le q-1\Big\}\\
				=&\Big(\underbrace{\vphantom{\frac{q-1}{2}}0,\dots,0}_{\frac{m-1}{2}},\frac{q}{2},\frac{q}{2},\frac{q}{2}-1,\dots,\frac{q}{2},\frac{q}{2}-1,\frac{q}{2}+1\Big)_q>\delta_1.
			\end{align*}
			Hence, we have $\min\limits_{i\in T_1}\CL(a+i)>\delta_1$. 
			\item When $i_1=\frac{q}{2}$, take $a:=\Delta_1$. Let
			\begin{equation*}
				a+i=\Delta_1+i=\left(\frac{q}{2}-1,\frac{q}{2},\dots,\frac{q}{2}-1,\frac{q}{2}+i_{\frac{m-1}{2}},\frac{q}{2}-1+i_{\frac{m-3}{2}},\dots,\frac{q}{2}+i_2,0,\frac{q}{2}+i_0\right)
			\end{equation*}
			and
			\begin{equation*}
				n-a=n-\Delta_{1}=\left(\frac{q}{2},\frac{q}{2}-1,\dots,\frac{q}{2},\frac{q}{2}-1,\frac{q}{2}+1\right)_q.
			\end{equation*}
			Note that $(x_1,x_0)\notin\{(0,0),(0,1)\}$, which implies that $L_1=L_2=1$, $L_3=L_4=\frac{m-1}{2}$ and $L_5=\frac{m-3}{2}$. Hence $\max\limits_{1\le j\le 5}L_j=\frac{m-1}{2}$, and so $\min\limits_{i\in T_1}\CL(a+i)=\min\limits_{x\in X}\CL(x)=\min (R_3\cup R_4)$. Note that
			\begin{align*}
				S_3=&\Big\{\Big(\frac{q}{2}-1,\frac{q}{2},\dots,\frac{q}{2}-1,\frac{q}{2}+i_{\frac{m-1}{2}},\frac{q}{2}-1+i_{\frac{m-3}{2}},\dots,\frac{q}{2}+i_2,0,\frac{q}{2}+i_0\Big)\ \Big|\\
				&\quad i_0=i_2=\dots=i_{\frac{m-3}{2}}=\frac{q}{2},\ i_3=i_5=\dots=i_{\frac{m-1}{2}}=\frac{q}{2}-1\Big\}\\
				=&\Big\{\Big(\frac{q}{2}-1,\frac{q}{2},\dots,\frac{q}{2}-1,\frac{q}{2},\frac{q}{2},\underbrace{0,\dots,0}_{\frac{m-3}{2}},1,0\Big)_q\Big\},\\
				S_4=&\Big\{\Big(\frac{q}{2}-1,\frac{q}{2},\dots,\frac{q}{2}-1,\frac{q}{2}+i_{\frac{m-1}{2}},\frac{q}{2}-1+i_{\frac{m-3}{2}},\dots,\frac{q}{2}+i_2,0,\frac{q}{2}+i_0\Big)\ \Big|\\
				&\quad i_0\le\frac{q}{2}-1,\ i_2=\dots=i_{\frac{m-3}{2}}=\frac{q}{2},\ i_3=i_5=\dots=i_{\frac{m-1}{2}}=\frac{q}{2}-1\Big\}\\
				=&\Big\{\Big(\frac{q}{2}-1,\frac{q}{2},\dots,\frac{q}{2}-1,\frac{q}{2},\frac{q}{2},\underbrace{0,\dots,0}_{\frac{m-1}{2}},\frac{q}{2}+i_0\Big)\ \Big|\ 1\le i_0\le\frac{q}{2}-1\Big\}.
			\end{align*}
			Hence, we have
			\begin{align*}
				\min R_3=&\min\{x\cdot q^{m-1-\frac{m-1}{2}}\mid x\in S_3\}\\
				=&\Big(\underbrace{\vphantom{\frac{q-1}{2}}0,\dots,0}_{\frac{m-1}{2}},q-1,\frac{q}{2},\frac{q}{2}-1,\dots,\frac{q}{2},\frac{q}{2}-1,\frac{q}{2}\Big)_q>\delta_1,\ \mathrm{and}\\
				\min R_4=&\min\{x\cdot q^{m-e-\frac{m-1}{2}}\mid x\in S_4,\ e=1\}\\
				=&\min\Big\{\Big(\underbrace{\vphantom{\frac{q-1}{2}}0,\dots,0}_{\frac{m-1}{2}},\frac{q}{2}+i_0-1,\frac{q}{2},\frac{q}{2}-1,\dots,\frac{q}{2},\frac{q}{2}-1,\frac{q}{2}\Big)\ \Big|\ 1\le i_0\le\frac{q}{2}-1\Big\}\\
				=&\Big(\underbrace{\vphantom{\frac{q-1}{2}}0,\dots,0}_{\frac{m-1}{2}},\frac{q}{2},\frac{q}{2},\frac{q}{2}-1,\dots,\frac{q}{2},\frac{q}{2}-1,\frac{q}{2}\Big)_q=\delta_1.
			\end{align*}
			Hence, we have $\min\limits_{i\in T_1}\CL(a+i)=\delta_1$. 
		\end{itemize}
		Therefore, we have $a+i\in I_1$, and so $n+i=(a+i)+(n-a)$ and $i \in I_1+I_2$.
	\end{itemize}
	We now show the case where $\frac{m-1}{2}$ is even.
	\begin{itemize}
		\item If $i \in I_1$, then $i = i+0\in I_1+I_2$.
		\item If $i\in\mathbb{Z}_n\setminus I_1=T_1$, we have
		\begin{equation*}
			i=(\underbrace{0,0,\ldots,0}_{m-r-1},i_r\ne0,i_{r-1},\ldots,i_1,i_0\ne0)_q,
		\end{equation*}
		where $0\leq r\leq \frac{m+1}{2}$ as $$i<\delta_{1}=\Lambda_{(q,m)}=\Big(\underbrace{\vphantom{\frac{q-1}{2}}0,\dots,0}_{\frac{m-1}{2}},\frac{q}{2},\frac{q}{2},\frac{q}{2}-1,\dots,\frac{q}{2},\frac{q}{2}-1\Big)_q.$$ With the same notation as in Lemma~\ref{sec3 lem0}, take $X := a + T_1$. We only need to prove that there always exists an integer $a\in C_{\Delta_1}$ such that $\min\limits_{i\in T_1}\CL(a+i)\ge\delta_{1}$. Through calculation, the largest coset leader in $T_1$ is
		\begin{equation*}
			\left\{
			\begin{aligned}
				&\Big(\underbrace{\vphantom{\frac{q-1}{2}}0,\dots,0}_{\frac{m-1}{2}},\frac{q}{2},\frac{q}{2},\frac{q}{2}-1,\dots,\frac{q}{2},\frac{q}{2}-1,\frac{q}{2}-2\Big)_q && \mathrm{if}\ q>4;\\
				&\Big(\underbrace{\vphantom{\frac{q-1}{2}}0,\dots,0}_{\frac{m-1}{2}},\frac{q}{2},\frac{q}{2},\frac{q}{2}-1,\dots,\frac{q}{2},\frac{q}{2}-2,q-1\Big)_q && \mathrm{if}\ q=4.
			\end{aligned}\right.
		\end{equation*}
		\begin{itemize}
			\item When $i_1\ne\frac{q}{2}$, take $a:=n-\Delta_1$. Let
			\begin{equation*}
				a+i=n-\Delta_1+i=\left(\frac{q}{2},\frac{q}{2}-1,\dots,\frac{q}{2},\frac{q}{2}-1,\frac{q}{2}+i_{\frac{m-1}{2}},\frac{q}{2}-1+i_{\frac{m-3}{2}},\dots,\frac{q}{2}+i_2,\frac{q}{2}-1+i_1,\frac{q}{2}+1+i_0\right)
			\end{equation*}
			and
			\begin{equation*}
				n-a=\Delta_{1}=\left(\frac{q}{2}-1,\frac{q}{2},\dots,\frac{q}{2}-1,\frac{q}{2},\frac{q}{2}\right)_q.
			\end{equation*}
			Since $i_1\ne\frac{q}{2}$, $(x_1,x_0)\notin\{(0,0),(0,1)\}$, which implies that $L_1=L_2=1$ and $L_3=L_4=L_5=\frac{m-1}{2}$. Hence $\max\limits_{1\le j\le 5}L_j=\frac{m-1}{2}$, and so $\min\limits_{i\in T_1}\CL(a+i)=\min\limits_{x\in X}\CL(x)=\min (R_3\cup R_4\cup R_5)$. Note that
			\begin{align*}
				S_3=&\Big\{\Big(\frac{q}{2},\frac{q}{2}-1,\dots,\frac{q}{2},\frac{q}{2}-1,\frac{q}{2}+i_{\frac{m-1}{2}},\frac{q}{2}-1+i_{\frac{m-3}{2}},\dots,\frac{q}{2}+i_2,\frac{q}{2}-1+i_1,\frac{q}{2}+1+i_0\Big)\ \Big|\\
				&\quad i_1=\frac{q}{2}+1,\ i_0=i_2=\dots=i_{\frac{m-1}{2}}=\frac{q}{2}-1,\ i_3=i_5=\dots=i_{\frac{m-3}{2}}=\frac{q}{2}\Big\}\\
				=&\Big\{\Big(\frac{q}{2},\frac{q}{2}-1,\dots,\frac{q}{2},\frac{q}{2}-1,\frac{q}{2},\frac{q}{2},\underbrace{0,\dots,0}_{\frac{m-3}{2}},1,0\Big)_q\Big\},\\
				S_4=&\Big\{\Big(\frac{q}{2},\frac{q}{2}-1,\dots,\frac{q}{2},\frac{q}{2}-1,\frac{q}{2}+i_{\frac{m-1}{2}},\frac{q}{2}-1+i_{\frac{m-3}{2}},\dots,\frac{q}{2}+i_2,\frac{q}{2}-1+i_1,\frac{q}{2}+1+i_0\Big)\ \Big|\\
				&\quad i_0\le\frac{q}{2}-2,\ i_1=\frac{q}{2}+1,\ i_2=\dots=i_{\frac{m-1}{2}}=\frac{q}{2}-1,\ i_3=i_5=\dots=i_{\frac{m-3}{2}}=\frac{q}{2}\Big\}\\
				=&\Big\{\Big(\frac{q}{2},\frac{q}{2}-1,\dots,\frac{q}{2},\frac{q}{2}-1,\frac{q}{2},\frac{q}{2},\underbrace{0,\dots,0}_{\frac{m-1}{2}},\frac{q}{2}+1+i_0\Big)\ \Big|\ 1\le i_0\le\frac{q}{2}-2\Big\},\ \mathrm{and}\\
				S_5=&\Big\{\Big(\frac{q}{2},\frac{q}{2}-1,\dots,\frac{q}{2},\frac{q}{2}-1,\frac{q}{2}+i_{\frac{m-1}{2}},\frac{q}{2}-1+i_{\frac{m-3}{2}},\dots,\frac{q}{2}+i_2,\frac{q}{2}-1+i_1,\frac{q}{2}+1+i_0\Big)\ \Big|\\
				&\quad i_0\ge\frac{q}{2},\ i_1=\frac{q}{2}-1,\ i_2=\dots=i_{\frac{m-1}{2}}=\frac{q}{2}-1,\ i_3=i_5=\dots=i_{\frac{m-3}{2}}=\frac{q}{2}\Big\}\\
				=&\Big\{\Big(\frac{q}{2},\frac{q}{2}-1,\dots,\frac{q}{2},\frac{q}{2}-1,\underbrace{q-1,\dots,q-1}_{\frac{m-1}{2}},i_0+1-\frac{q}{2}\Big)\ \Big|\ \frac{q}{2}\le i_0\le q-1\Big\}.
			\end{align*}
			Hence, we have
			\begin{align*}
				\min R_3=&\min\{x\cdot q^{m-1-\frac{m-1}{2}}\mid x\in S_3\}=\Big(\underbrace{\vphantom{\frac{q-1}{2}}0,\dots,0}_{\frac{m-1}{2}},q-1,\frac{q}{2}-1,\frac{q}{2},\dots,\frac{q}{2}-1,\frac{q}{2}\Big)_q>\delta_1,\\
				\min R_4=&\min\{x\cdot q^{m-e-\frac{m-1}{2}}\mid x\in S_4,\ e=1\}\\
				=&\min\Big\{\Big(\underbrace{\vphantom{\frac{q-1}{2}}0,\dots,0}_{\frac{m-1}{2}},\frac{q}{2}+i_0,\frac{q}{2}-1,\frac{q}{2},\dots,\frac{q}{2}-1,\frac{q}{2}\Big)\ \Big|\ 1\le i_0\le\frac{q}{2}-2\Big\}\\
				=&\Big(\underbrace{\vphantom{\frac{q-1}{2}}0,\dots,0}_{\frac{m-1}{2}},\frac{q}{2}+1,\frac{q}{2}-1,\frac{q}{2},\dots,\frac{q}{2}-1,\frac{q}{2}\Big)_q>\delta_1,\ \mathrm{and}\\
				\min R_5=&\min\{x\cdot q^{2m-f-\frac{m-1}{2}}\mid x\in S_5,\ f=1\}\\
				=&\min\Big\{\Big(\underbrace{\vphantom{\frac{q-1}{2}}0,\dots,0}_{\frac{m-1}{2}},\frac{3q}{2}-1-i_0,\frac{q}{2},\frac{q}{2}-1,\dots,\frac{q}{2},\frac{q}{2}\Big)\ \Big|\ \frac{q}{2}\le i_0\le q-1\Big\}\\
				=&\Big(\underbrace{\vphantom{\frac{q-1}{2}}0,\dots,0}_{\frac{m-1}{2}},\frac{q}{2},\frac{q}{2},\frac{q}{2}-1,\dots,\frac{q}{2},\frac{q}{2}\Big)_q>\delta_1.
			\end{align*}
			Hence, we have $\min\limits_{i\in T_1}\CL(a+i)>\delta_1$.
			\item When $i_1=\frac{q}{2}$, take $a:=\Delta_1$. For even $\frac{m-1}{2}$, let
			\begin{equation*}
				a+i=\Delta_1+i=\left(\frac{q}{2}-1,\frac{q}{2},\dots,\frac{q}{2}-1,\frac{q}{2},\frac{q}{2}-1+i_{\frac{m-1}{2}},\frac{q}{2}+i_{\frac{m-3}{2}},\dots,\frac{q}{2}+i_2,0,\frac{q}{2}+i_0\right)
			\end{equation*}
			and
			\begin{equation*}
				n-a=n-\Delta_{1}=\left(\frac{q}{2},\frac{q}{2}-1,\dots,\frac{q}{2},\frac{q}{2}-1,\frac{q}{2}+1\right)_q.
			\end{equation*}
			Note that $(x_1,x_0)\notin\{(0,0),(0,1)\}$, which implies that $L_1=L_2=1$, $L_3=L_4=\frac{m-1}{2}$ and $L_5=\frac{m-3}{2}$. Hence $\max\limits_{1\le j\le 5}L_j=\frac{m-1}{2}$, and so $\min\limits_{i\in T_1}\CL(a+i)=\min\limits_{x\in X}\CL(x)=\min (R_3\cup R_4)$. Note that
			\begin{align*}
				S_3=&\Big\{\Big(\frac{q}{2}-1,\frac{q}{2},\dots,\frac{q}{2}-1,\frac{q}{2},\frac{q}{2}-1+i_{\frac{m-1}{2}},\frac{q}{2}+i_{\frac{m-3}{2}},\dots,\frac{q}{2}+i_2,0,\frac{q}{2}+i_0\Big)\ \Big|\\
				&\quad i_0=i_2=\dots=i_{\frac{m-1}{2}}=\frac{q}{2},\ i_3=i_5=\dots=i_{\frac{m-3}{2}}=\frac{q}{2}-1\Big\}\\
				=&\Big\{\Big(\frac{q}{2}-1,\frac{q}{2},\dots,\frac{q}{2}-1,\frac{q}{2},\frac{q}{2}-1,\frac{q}{2}+1,\underbrace{0,\dots,0}_{\frac{m-3}{2}},1,0\Big)_q\Big\},\ \mathrm{and}\\
				S_4=&\Big\{\Big(\frac{q}{2}-1,\frac{q}{2},\dots,\frac{q}{2}-1,\frac{q}{2},\frac{q}{2}-1+i_{\frac{m-1}{2}},\frac{q}{2}+i_{\frac{m-3}{2}},\dots,\frac{q}{2}+i_2,0,\frac{q}{2}+i_0\Big)\ \Big|\\
				&\quad i_0\le\frac{q}{2}-1,\ i_2=\dots=i_{\frac{m-1}{2}}=\frac{q}{2},\ i_3=i_5=\dots=i_{\frac{m-3}{2}}=\frac{q}{2}-1\Big\}\\
				=&\Big\{\Big(\frac{q}{2}-1,\frac{q}{2},\dots,\frac{q}{2}-1,\frac{q}{2},\frac{q}{2}-1,\frac{q}{2}+1,\underbrace{0,\dots,0}_{\frac{m-1}{2}},\frac{q}{2}+i_0\Big)\ \Big|\ 1\le i_0\le\frac{q}{2}-1\Big\}.
			\end{align*}
			Hence, we have
			\begin{align*}
				\min R_3=&\min\{x\cdot q^{m-1-\frac{m-1}{2}}\mid x\in S_3\}=\Big(\underbrace{\vphantom{\frac{q-1}{2}}0,\dots,0}_{\frac{m-1}{2}},q-1,\frac{q}{2},\frac{q}{2}-1,\dots,\frac{q}{2},\frac{q}{2}-1\Big)_q>\delta_1,\ \mathrm{and}\\
				\min R_4=&\min\{x\cdot q^{m-e-\frac{m-1}{2}}\mid x\in S_4,\ e=1\}\\
				=&\min\Big\{\Big(\underbrace{\vphantom{\frac{q-1}{2}}0,\dots,0}_{\frac{m-1}{2}},\frac{q}{2}+i_0-1,\frac{q}{2},\frac{q}{2}-1,\dots,\frac{q}{2},\frac{q}{2}-1\Big)\ \Big|\ 1\le i_0\le\frac{q}{2}-1\Big\}\\
				=&\Big(\underbrace{\vphantom{\frac{q-1}{2}}0,\dots,0}_{\frac{m-1}{2}},\frac{q}{2},\frac{q}{2},\frac{q}{2}-1,\dots,\frac{q}{2},\frac{q}{2}-1\Big)_q=\delta_1.
			\end{align*}
			Hence, we have $\min\limits_{i\in T_1}\CL(a+i)=\delta_1$.
		\end{itemize}
	\end{itemize}
	Therefore, we have $a+i\in I_1$, and so $n+i=(a+i)+(n-a)$ and $i \in I_1+I_2$. The proof is completed.
\end{proof}

\section{\texorpdfstring{Proof of Theorem~\ref{sec3.2 thm3}}{Proof of Theorem 6}}\label{apdx c}

\begin{proof}
	For $m=6$ or $10$, we have $\Lambda_{(q,m)}\le\Omega_{(q,m)}$, where $\Omega_{(q,m)}$ is as in Table~\ref{sec3.2 thm1 table1}. Hence, we only need to prove the case of $m\ge14$ by Theorem~\ref{sec3.2 thm1}. Suppose that $\delta_{1}=\Lambda_{(q,m)}$ and $\delta_{2}=\Delta_1$, then $T_1=C_{1}\cup C_{2}\cup\cdots\cup C_{\Lambda_{(q,m)}-1}$ and $T_2=C_{1}\cup C_{2}\cup\cdots\cup C_{\Delta_1-1}$, and so $I_1=\mathbb{Z}_n\setminus T_1$ and $I_2=C_0\cup C_{\Delta_1}$. Similar to the proof of Theorem~\ref{sec3.2 thm1} (i), it suffices to show that $i\in I_{1}+I_{2}$ for any coset leader $i\in\mathbb{Z}_n$. We first consider the case where $\frac{m}{2}\equiv1\bmod 4$. 
	\begin{itemize}
		\item If $i \in I_1$, then $i = i+0\in I_1+I_2$.
		\item If $i\in\mathbb{Z}_n\setminus I_1=T_1$, we have
		\begin{equation*}
			i=(\underbrace{0,0,\ldots,0}_{m-r-1},i_r\ne0,i_{r-1},\ldots,i_1,i_0\ne0)_q,
		\end{equation*}
		where $0\leq r\leq \frac{m}{2}$ as $$i<\delta_{1}=\Lambda_{(q,m)}=\Big(\underbrace{\vphantom{\frac{q-1}{2}}0,\dots,0}_{\frac{m}{2}-1},\frac{q}{2}-1,\frac{q}{2},\frac{q}{2},\frac{q}{2},\frac{q}{2}-1,\frac{q}{2}-1,\dots\frac{q}{2},\frac{q}{2},\frac{q}{2}-1,\frac{q}{2}-1,\frac{q}{2},\frac{q}{2},\frac{q}{2}-1,\frac{q}{2}\Big)_q.$$ With the same notation as in Lemma~\ref{sec3 lem0}, take $X := a + T_1$. We only need to prove that there always exists an integer $a\in C_{\Delta_1}$ such that $\min\limits_{i\in T_1}\CL(a+i)\ge\delta_{1}$. Through calculation, the largest coset leader in $T_1$ is
		\begin{equation*}
			\Big(\underbrace{\vphantom{\frac{q-1}{2}}0,\dots,0}_{\frac{m}{2}-1},\frac{q}{2}-1,\frac{q}{2},\frac{q}{2},\frac{q}{2},\frac{q}{2}-1,\frac{q}{2}-1,\dots\frac{q}{2},\frac{q}{2},\frac{q}{2}-1,\frac{q}{2}-1,\frac{q}{2},\frac{q}{2},\frac{q}{2}-1,\frac{q}{2}-1\Big)_q.
		\end{equation*}
		\begin{itemize}
			\item When $i_2\ne\frac{q}{2}-1$ and $i_1\ne\frac{q}{2}-1$, let
			\begin{align*}
					a+i=\Big(&\frac{q}{2},\frac{q}{2}-1,\frac{q}{2}-1,\frac{q}{2},\dots,\frac{q}{2}+i_{\frac{m}{2}},\frac{q}{2}-1+i_{\frac{m}{2}-1},\frac{q}{2}-1+i_{\frac{m}{2}-2},\frac{q}{2}+i_{\frac{m}{2}-3},\\
					&\dots,\frac{q}{2}+i_{5},\frac{q}{2}-1+i_{4},\frac{q}{2}-1+i_{3},\frac{q}{2}+i_2,\frac{q}{2}+i_1,\frac{q}{2}+i_0\Big)
				\end{align*}
			and
			\begin{equation*}
				n-a=\left(\frac{q}{2}-1,\frac{q}{2},\frac{q}{2},\frac{q}{2}-1,\dots,\frac{q}{2}-1,\frac{q}{2},\frac{q}{2},\frac{q}{2}-1,\frac{q}{2}-1,\frac{q}{2}+1\right)_q.
			\end{equation*}
			Since $i_2\ne\frac{q}{2}-1$ and $i_1\ne\frac{q}{2}-1$, $(x_1,x_0)\notin\{(0,0),(0,1)\}$ and $(x_2,x_1)\notin\{(0,0),(0,1),(q-1,q-1)\}$, which implies that $L_1=L_2=L_3=1$, and $L_4=L_5=\frac{m}{2}-1$. Hence $\max\limits_{1\le j\le 5}L_j=\frac{m}{2}-1$, and so $\min\limits_{i\in T_1}\CL(a+i)=\min\limits_{x\in X}\CL(x)=\min (R_4\cup R_5)$. Note that
			\begin{align*}
				S_4=&\Big\{\Big(\frac{q}{2},\frac{q}{2}-1,\frac{q}{2}-1,\frac{q}{2},\dots,\frac{q}{2}+i_{\frac{m}{2}},\frac{q}{2}-1+i_{\frac{m}{2}-1},\frac{q}{2}-1+i_{\frac{m}{2}-2},\frac{q}{2}+i_{\frac{m}{2}-3},\\
				&\quad\dots,\frac{q}{2}+i_{5},\frac{q}{2}-1+i_{4},\frac{q}{2}-1+i_{3},\frac{q}{2}+i_2,\frac{q}{2}+i_1,\frac{q}{2}+i_0\Big)\ \Big|\\
				&\quad \Big| i_2=\frac{q}{2},i_{\frac{m}{2}}=\frac{q}{2}-1,i_3=i_4=i_7=i_8=\dots=i_{\frac{m}{2}-2}=i_{\frac{m}{2}-1}=\frac{q}{2},\\
				&\quad i_5=i_6=i_9=i_{10}=\dots=i_{\frac{m}{2}-4}=i_{\frac{m}{2}-3}=\frac{q}{2}-1,i_1\le q-1,1\le i_0\le q-1\Big\}\\
				=&\Big\{\Big(\frac{q}{2},\frac{q}{2}-1,\frac{q}{2}-1,\frac{q}{2},\dots,\frac{q}{2},\frac{q}{2}-1,\frac{q}{2}-1,\frac{q}{2}+1,\underbrace{0,\dots,0}_{\frac{m}{2}-1},\frac{q}{2}+i_1,\frac{q}{2}+i_0\Big)\ \Big|\\
				&\quad i_1\le q-1,1\le i_0\le q-1\Big\},\ \mathrm{and}\\
				S_5=&\Big\{\Big(\frac{q}{2},\frac{q}{2}-1,\frac{q}{2}-1,\frac{q}{2},\dots,\frac{q}{2}+i_{\frac{m}{2}},\frac{q}{2}-1+i_{\frac{m}{2}-1},\frac{q}{2}-1+i_{\frac{m}{2}-2},\frac{q}{2}+i_{\frac{m}{2}-3},\\
				&\quad\dots,\frac{q}{2}+i_{5},\frac{q}{2}-1+i_{4},\frac{q}{2}-1+i_{3},\frac{q}{2}+i_2,\frac{q}{2}+i_1,\frac{q}{2}+i_0\Big)\ \Big|\\
				&\quad i_1\ge\frac{q}{2},i_2=\frac{q}{2}-2,i_3=i_4=i_7=i_8=\dots=i_{\frac{m}{2}-2}=i_{\frac{m}{2}-1}=\frac{q}{2}, \\
				&\quad i_5=i_6=i_9=i_{10}=\dots=i_{\frac{m}{2}-4}=i_{\frac{m}{2}-3}=i_{\frac{m}{2}}=\frac{q}{2}-1, 1\le i_0\le q-1\Big\}\\
				=&\Big\{\Big(\frac{q}{2},\frac{q}{2}-1,\frac{q}{2}-1,\frac{q}{2},\dots,\frac{q}{2},\frac{q}{2}-1,\frac{q}{2}-1,\frac{q}{2},\underbrace{q-1,\dots,q-1}_{\frac{m}{2}-1},i_1-\frac{q}{2},\frac{q}{2}+i_0\Big)\ \Big|\\
				&\quad i_1\ge\frac{q}{2},\ 1\le i_0\le q-1\Big\}.
			\end{align*}
			Hence, we have
			\begin{align*}
				\min R_4=&\min\{x\cdot q^{m-e-(\frac{m}{2}-1)}\mid x\in S_4,\ e=2\}\\
				=&\min\Big\{\Big(\underbrace{\vphantom{\frac{q-1}{2}}0,\dots,0}_{\frac{m}{2}-1},\frac{q}{2}+i_1,\frac{q}{2}-1+i_0,\frac{q}{2}-1,\frac{q}{2},\frac{q}{2},\frac{q}{2}-1,\dots,\frac{q}{2}-1,\frac{q}{2},\frac{q}{2},\frac{q}{2}-1\Big)_q\\
				&\quad\Big|\ i_1\le q-1,\ 1\le i_0\le q-1\Big\}\\
				=&\Big(\underbrace{\vphantom{\frac{q-1}{2}}0,\dots,0}_{\frac{m}{2}-1},\frac{q}{2},\frac{q}{2},\frac{q}{2}-1,\frac{q}{2},\frac{q}{2},\frac{q}{2}-1,\dots, \frac{q}{2}-1,\frac{q}{2},\frac{q}{2},\frac{q}{2}-1\Big)_q>\delta_1,\ \mathrm{and}\\
				\min R_5=&\min\{x\cdot q^{2m-f-(\frac{m}{2}-1)}\mid x\in S_5,\ f=2\}\\
				=&\min\Big\{\Big(\underbrace{\vphantom{\frac{q-1}{2}}0,\dots,0}_{\frac{m}{2}-1},\frac{3q}{2}-1-i_1,\frac{q}{2}-i_0,\frac{q}{2},\frac{q}{2}-1,\frac{q}{2}-1,\frac{q}{2},\dots,\frac{q}{2},\frac{q}{2}-1,\frac{q}{2}-1,\frac{q}{2}\\
				&\quad \frac{q}{2},\frac{q}{2}-1,\frac{q}{2}-1,\frac{q}{2}+1\Big)_q\Big|\ \frac{q}{2}\le i_1\le q-1,\ 1\le i_0\le q-1\Big\}\\
				=&\Big(\underbrace{\vphantom{\frac{q-1}{2}}0,\dots,0}_{\frac{m}{2}-1},\frac{q}{2}-1,\frac{q}{2}+1,\frac{q}{2},\frac{q}{2}-1,\frac{q}{2}-1,\frac{q}{2},\dots,\\
				&\quad \frac{q}{2},\frac{q}{2}-1,\frac{q}{2}-1,\frac{q}{2},\frac{q}{2},\frac{q}{2}-1,\frac{q}{2}-1,\frac{q}{2}+1\Big)_q>\delta_1.
			\end{align*}
			Hence, we have $\min\limits_{i\in T_1}\CL(a+i)>\delta_1$. 
			\item When $i_2\ne\frac{q}{2}-1$ and $i_1=\frac{q}{2}-1$, let
			\begin{align*}
					a+i=\Big(&\frac{q}{2}-1,\frac{q}{2}-1,\frac{q}{2},\frac{q}{2},\dots,\frac{q}{2}-1+i_{\frac{m}{2}},\frac{q}{2}-1+i_{\frac{m}{2}-1},\frac{q}{2}+i_{\frac{m}{2}-2},\frac{q}{2}+i_{\frac{m}{2}-3},\\
					&\dots,\frac{q}{2}-1+i_{5},\frac{q}{2}-1+i_{4},\frac{q}{2}+i_{3},\frac{q}{2}+i_2,q-2,\frac{q}{2}+i_0\Big)
				\end{align*}
			and
			\begin{equation*}
				n-a=\left(\frac{q}{2},\frac{q}{2},\frac{q}{2}-1,\frac{q}{2}-1,\dots,\frac{q}{2},\frac{q}{2},\frac{q}{2}-1,\frac{q}{2}-1,\frac{q}{2},\frac{q}{2}+1\right)_q.
			\end{equation*}
			Then we have $(x_2,x_1)\notin\{(0,0),(q-1,q-1)\}$ and $(x_1,x_0)\notin\{(0,0),(0,1)\}$, which implies that $L_1=L_2=1$, $L_3=0$, $L_4=\frac{m}{2}-1$ and $L_5=\frac{m}{2}-2$. Hence $\max\limits_{1\le j\le 5}L_j=L_4=\frac{m}{2}-1$, and so $\min\limits_{i\in T_1}\CL(a+i)=\min\limits_{x\in X}\CL(x)=\min R_4$. Note that
			\begin{align*}
				S_4=&\Big\{\Big(\frac{q}{2}-1,\frac{q}{2}-1,\frac{q}{2},\frac{q}{2},\dots,\frac{q}{2}-1+i_{\frac{m}{2}},\frac{q}{2}-1+i_{\frac{m}{2}-1},\frac{q}{2}+i_{\frac{m}{2}-2},\frac{q}{2}+i_{\frac{m}{2}-3},\\
				&\quad \dots,\frac{q}{2}-1+i_{5},\frac{q}{2}-1+i_{4},\frac{q}{2}+i_{3},\frac{q}{2}+i_2,q-2,\frac{q}{2}+i_0\Big)\ \Big|\\
				&\quad i_2=\frac{q}{2},\ i_3=\frac{q}{2}-1,\ i_4=i_5=i_8=i_9=\dots=i_{\frac{m}{2}-1}=i_{\frac{m}{2}}=\frac{q}{2},\\
				&\qquad i_6=i_7=i_{10}=i_{11}=\dots=i_{\frac{m}{2}-3}=i_{\frac{m}{2}-2}=\frac{q}{2}-1,\ 1\le i_0\le q-1\Big\}\\
				=&\Big\{\Big(\frac{q}{2}-1,\frac{q}{2}-1,\frac{q}{2},\frac{q}{2},\dots,\frac{q}{2}-1,\frac{q}{2}-1,\frac{q}{2},\frac{q}{2}+1,\underbrace{0,\dots,0}_{\frac{m}{2}-1},q-2,\frac{q}{2}+i_0\Big)\ \Big|\ 1\le i_0\le q-1\Big\}.
			\end{align*}
			Hence, we have
			\begin{align*}
				\min R_4=&\min\{x\cdot q^{m-e-\frac{m}{2}+1}\mid x\in S_4,\ e=2\}\\
				=&\min\Big\{\Big(\underbrace{\vphantom{\frac{q-1}{2}}0,\dots,0}_{\frac{m}{2}-1},q-2,\frac{q}{2}-1+i_0,\frac{q}{2},\frac{q}{2},\frac{q}{2}-1,\frac{q}{2}-1,\dots,\frac{q}{2},\frac{q}{2},\frac{q}{2}-1,\frac{q}{2}-1\Big)\ \Big|\\
				&\quad 1\le i_0\le q-1\Big\}\\
				=&\Big(\underbrace{\vphantom{\frac{q-1}{2}}0,\dots,0}_{\frac{m}{2}-1},q-2,\frac{q}{2},\frac{q}{2},\frac{q}{2},\frac{q}{2}-1,\frac{q}{2}-1,\dots,\frac{q}{2},\frac{q}{2},\frac{q}{2}-1,\frac{q}{2}-1\Big)_q>\delta_1.
			\end{align*}
			\item When $i_2=\frac{q}{2}-1$ and $i_1=\frac{q}{2}-1$, let
			\begin{align*}
					a+i=\Big(&\frac{q}{2}-1,\frac{q}{2},\frac{q}{2},\frac{q}{2}-1,\dots,\frac{q}{2}-1+i_{\frac{m}{2}},\frac{q}{2}+i_{\frac{m}{2}-1},\frac{q}{2}+i_{\frac{m}{2}-2},\frac{q}{2}-1+i_{\frac{m}{2}-3},\\
					&\dots,\frac{q}{2}-1+i_{5},\frac{q}{2}+i_{4},\frac{q}{2}+i_{3},q-2,q-2,\frac{q}{2}+1+i_0\Big)
				\end{align*}
			and
			\begin{equation*}
				n-a=\left(\frac{q}{2},\frac{q}{2}-1,\frac{q}{2}-1,\frac{q}{2},\dots,\frac{q}{2},\frac{q}{2}-1,\frac{q}{2}-1,\frac{q}{2},\frac{q}{2},\frac{q}{2}\right)_q.
			\end{equation*}
			Then we have $L_1=L_2=1$, $L_3=0$ and $L_4=L_5=\frac{m}{2}-2$. By Lemma~\ref{sec3 lem0}, we have $\min\limits_{i\in T_1}\CL(a+i)\ge q^{\frac{m}{2}+1}>\delta_1$.
			\item When $i_2=\frac{q}{2}-1$ and $i_1\ne\frac{q}{2}-1$, let
			\begin{align*}
					a+i=\Big(&\frac{q}{2},\frac{q}{2},\frac{q}{2}-1,\frac{q}{2}-1,\dots,\frac{q}{2}+i_{\frac{m}{2}},\frac{q}{2}+i_{\frac{m}{2}-1},\frac{q}{2}-1+i_{\frac{m}{2}-2},\frac{q}{2}-1+i_{\frac{m}{2}-3},\\
					&\dots,\frac{q}{2}+i_{5},\frac{q}{2}+i_{4},\frac{q}{2}-1+i_{3},q-2,\frac{q}{2}+i_1,\frac{q}{2}+1+i_0\Big)
				\end{align*}
			and
			\begin{equation*}
				n-a=\left(\frac{q}{2}-1,\frac{q}{2}-1,\frac{q}{2},\frac{q}{2},\dots,\frac{q}{2}-1,\frac{q}{2}-1,\frac{q}{2},\frac{q}{2},\frac{q}{2}-1,\frac{q}{2}\right)_q.
			\end{equation*}
			Then we have $L_1=L_2=L_3=1$, $L_4=\frac{m}{2}-2$ and $L_5=\frac{m}{2}-1$. Hence $\max\limits_{1\le j\le 5}L_j=L_5=\frac{m}{2}-1$, and so $\min\limits_{i\in T_1}\CL(a+i)=\min\limits_{x\in X}\CL(x)=\min R_5$. Note that
			\begin{align*}
				S_5=&\Big\{\Big(\frac{q}{2},\frac{q}{2},\frac{q}{2}-1,\frac{q}{2}-1,\dots,\frac{q}{2}+i_{\frac{m}{2}},\frac{q}{2}+i_{\frac{m}{2}-1},\frac{q}{2}-1+i_{\frac{m}{2}-2},\frac{q}{2}-1+i_{\frac{m}{2}-3},\\
				&\quad \dots,\frac{q}{2}+i_{5},\frac{q}{2}+i_{4},\frac{q}{2}-1+i_{3},q-2,\frac{q}{2}+i_1,\frac{q}{2}+1+i_0\Big)\ \Big|\\
				&\quad i_1\ge\frac{q}{2},i_3=\frac{q}{2},i_4=i_5=i_8=i_9=\dots=i_{\frac{m}{2}-1}=i_{\frac{m}{2}}=\frac{q}{2}-1,\\
				&\quad i_6=i_7=i_{10}=i_{11}=\dots=i_{\frac{m}{2}-3}=i_{\frac{m}{2}-2}=\frac{q}{2},1\le i_0\le q-1\Big\}\\
				=&\Big\{\Big(\frac{q}{2},\frac{q}{2},\frac{q}{2}-1,\frac{q}{2}-1,\dots,\frac{q}{2},\frac{q}{2},\frac{q}{2}-1,\frac{q}{2}-1,\underbrace{q-1,\dots,q-1}_{\frac{m}{2}-1},i_1-\frac{q}{2},\frac{q}{2}+1+i_0\Big)\ \Big|\\
				&\quad i_1\ge\frac{q}{2},1\le i_0\le q-1\Big\}.
			\end{align*}
			Hence, we have
			\begin{align*}
				\min R_5=&\min\{x\cdot q^{2m-f-\frac{m}{2}+1}\mid x\in S_5,\ f=2\}\\
				=&\min\Big\{\Big(\underbrace{\vphantom{\frac{q-1}{2}}0,\dots,0}_{\frac{m}{2}-1},\frac{3q}{2}-1-i_1,\frac{q}{2}-1-i_0,\frac{q}{2},\frac{q}{2},\frac{q}{2}-1,\frac{q}{2}-1,\\
				&\qquad \dots,\frac{q}{2},\frac{q}{2},\frac{q}{2}-1,\frac{q}{2}-1,\frac{q}{2},\frac{q}{2},\frac{q}{2}-1,\frac{q}{2}\Big)_q\\
				&\qquad \Big|\ \frac{q}{2}\le i_1\le q-1,\ 1\le i_0\le q-1\Big\} \\
				=&\Big(\underbrace{\vphantom{\frac{q-1}{2}}0,\dots,0}_{\frac{m}{2}-1},\frac{q}{2}-1,\frac{q}{2},\frac{q}{2},\frac{q}{2},\frac{q}{2}-1,\frac{q}{2}-1,\dots,\frac{q}{2},\frac{q}{2},\frac{q}{2}-1,\frac{q}{2}-1,\frac{q}{2},\frac{q}{2},\frac{q}{2}-1,\frac{q}{2}\Big)_q=\delta_1.
			\end{align*}
		\end{itemize}
		Therefore, we have $a+i\in I_1$, and so $n+i=(a+i)+(n-a)$ and $i \in I_1+I_2$.
	\end{itemize}
	We now prove the case where $\frac{m}{2}\equiv3\bmod 4$.
	\begin{itemize}
		\item If $i \in I_1$, then $i = i+0\in I_1+I_2$.
		\item If $i\in\mathbb{Z}_n\setminus I_1=T_1$, we have
		\begin{equation*}
			i=(\underbrace{0,0,\ldots,0}_{m-r-1},i_r\ne0,i_{r-1},\ldots,i_1,i_0\ne0)_q,
		\end{equation*}
		where $0\leq r\leq \frac{m}{2}$ as $$i<\delta_{1}=\Lambda_{(q,m)}=\Big(\underbrace{\vphantom{\frac{q-1}{2}}0,\dots,0}_{\frac{m}{2}-1},\frac{q}{2}-1,\frac{q}{2},\frac{q}{2},\frac{q}{2},\frac{q}{2}-1,\frac{q}{2}-1,\dots\frac{q}{2},\frac{q}{2},\frac{q}{2}-1,\frac{q}{2}-1,\frac{q}{2},\frac{q}{2}\Big)_q.$$ With the same notation as in Lemma~\ref{sec3 lem0}, take $X := a + T_1$. We only need to prove that there always exists an integer $a\in C_{\Delta_1}$ such that $\min\limits_{i\in T_1}\CL(a+i)\ge\delta_{1}$. Through calculation, the largest coset leader in $T_1$ is
		\begin{equation}\label{a3 equ1}
			\Big(\underbrace{\vphantom{\frac{q-1}{2}}0,\dots,0}_{\frac{m}{2}-1},\frac{q}{2}-1,\frac{q}{2},\frac{q}{2},\frac{q}{2},\frac{q}{2}-1,\frac{q}{2}-1,\dots\frac{q}{2},\frac{q}{2},\frac{q}{2}-1,\frac{q}{2}-1,\frac{q}{2},\frac{q}{2}-1\Big)_q.
		\end{equation}
		\begin{itemize}
			\item When $i_2\ne\frac{q}{2}-1$ and $i_1\ne\frac{q}{2}-1$, let
			\begin{align*}
					a+i=\Big(&\frac{q}{2},\frac{q}{2}-1,\frac{q}{2}-1,\frac{q}{2},\dots,\frac{q}{2},\frac{q}{2}-1,\frac{q}{2}-1+i_{\frac{m}{2}},\frac{q}{2}+i_{\frac{m}{2}-1},\\
					&\dots,\frac{q}{2}+i_{5},\frac{q}{2}-1+i_{4},\frac{q}{2}-1+i_{3},\frac{q}{2}+i_2,\frac{q}{2}+i_1,\frac{q}{2}+i_0\Big)
				\end{align*}
			and
			\begin{equation*}
				n-a=\left(\frac{q}{2}-1,\frac{q}{2},\frac{q}{2},\frac{q}{2}-1,\dots,\frac{q}{2}-1,\frac{q}{2},\frac{q}{2},\frac{q}{2}-1,\frac{q}{2}-1,\frac{q}{2}+1\right)_q.
			\end{equation*}
			Since $i_2\ne\frac{q}{2}-1$ and $i_1\ne\frac{q}{2}-1$, $(x_1,x_0)\notin\{(0,0),(0,1)\}$ and $(x_2,x_1)\notin\{(0,0),(0,1),(q-1,q-1)\}$, which implies that $L_1=L_2=L_3=1$, By \eqref{a3 equ1}, we have $i_{\frac{m}{2}}\le\frac{q}{2}-1$, and so $L_4=L_5=\frac{m}{2}-2$. By Lemma~\ref{sec3 lem0}, we have $\min\limits_{i\in T_1}\CL(a+i)\ge q^{\frac{m}{2}+1}>\delta_1$. 
			\item When $i_2\ne\frac{q}{2}-1$ and $i_1=\frac{q}{2}-1$, let
			\begin{align*}
					a+i=\Big(&\frac{q}{2}-1,\frac{q}{2}-1,\frac{q}{2},\frac{q}{2},\dots,\frac{q}{2}-1,\frac{q}{2}-1,\frac{q}{2}+i_{\frac{m}{2}},\frac{q}{2}+i_{\frac{m}{2}-1},\\
					&\dots,\frac{q}{2}-1+i_{5},\frac{q}{2}-1+i_{4},\frac{q}{2}+i_{3},\frac{q}{2}+i_2,q-2,\frac{q}{2}+i_0\Big)
				\end{align*}
			and
			\begin{equation*}
				n-a=\left(\frac{q}{2},\frac{q}{2},\frac{q}{2}-1,\frac{q}{2}-1,\dots,\frac{q}{2},\frac{q}{2},\frac{q}{2}-1,\frac{q}{2}-1,\frac{q}{2},\frac{q}{2}+1\right)_q.
			\end{equation*}
			Then we have $(x_2,x_1)\notin\{(0,0),(q-1,q-1)\}$ and $(x_1,x_0)\notin\{(0,0),(0,1)\}$, which implies that $L_1=L_2=1$, $L_3=0$, $L_4=\frac{m}{2}-1$ and $L_5=\frac{m}{2}-2$. Hence $\max\limits_{1\le j\le 5}L_j=L_4=\frac{m}{2}-1$, and so $\min\limits_{i\in T_1}\CL(a+i)=\min\limits_{x\in X}\CL(x)=\min R_4$. Note that
			\begin{align*}
				S_4=&\Big\{\Big(\frac{q}{2}-1,\frac{q}{2}-1,\frac{q}{2},\frac{q}{2},\dots,\frac{q}{2}-1,\frac{q}{2}-1,\frac{q}{2}+i_{\frac{m}{2}},\frac{q}{2}+i_{\frac{m}{2}-1},\dots,\\
				&\quad \frac{q}{2}-1+i_{5},\frac{q}{2}-1+i_{4},\frac{q}{2}+i_{3},\frac{q}{2}+i_2,q-2,\frac{q}{2}+i_0\Big)\ \Big|\ i_2=\frac{q}{2},i_3=\frac{q}{2}-1,1\le i_0\le q-1\\  &i_4=i_5=i_8=i_9=\dots=i_{\frac{m}{2}-4}=i_{\frac{m}{2}-3}=\frac{q}{2},i_6=i_7=i_{10}=i_{11}=\dots=i_{\frac{m}{2}-1}=i_{\frac{m}{2}}=\frac{q}{2}-1\Big\}\\
				=&\Big\{\Big(\frac{q}{2}-1,\frac{q}{2}-1,\frac{q}{2},\frac{q}{2},\dots,\frac{q}{2}-1,\frac{q}{2}-1,\frac{q}{2},\frac{q}{2},\frac{q}{2}-1,\frac{q}{2},\underbrace{0,\dots,0}_{\frac{m}{2}-1},q-2,\frac{q}{2}+i_0\Big)\ \Big|\\
				&\quad 1\le i_0\le q-1\Big\}.
			\end{align*}
			Hence, we have
			\begin{align*}
				\min R_4=&\min\{x\cdot q^{m-e-\frac{m}{2}+1}\mid x\in S_4,\ e=2\}\\
				=&\min\Big\{\Big(\underbrace{\vphantom{\frac{q-1}{2}}0,\dots,0}_{\frac{m}{2}-1},q-2,\frac{q}{2}-1+i_0,\frac{q}{2},\frac{q}{2},\frac{q}{2}-1,\frac{q}{2}-1,\dots,\\
				&\quad \frac{q}{2},\frac{q}{2},\frac{q}{2}-1,\frac{q}{2}-1,\frac{q}{2},\frac{q}{2}\Big)\ \Big|\ 1\le i_0\le q-1\Big\}>\delta_1.
			\end{align*}
			\item When $i_2=\frac{q}{2}-1$ and $i_1=\frac{q}{2}-1$, let
			\begin{align*}
					a+i=\Big(&\frac{q}{2}-1,\frac{q}{2},\frac{q}{2},\frac{q}{2}-1,\dots,\frac{q}{2}-1,\frac{q}{2},\frac{q}{2}+i_{\frac{m}{2}},\frac{q}{2}-1+i_{\frac{m}{2}-1},\\
					&\dots,\frac{q}{2}-1+i_{5},\frac{q}{2}+i_{4},\frac{q}{2}+i_{3},q-2,q-2,\frac{q}{2}+1+i_0\Big)
				\end{align*}
			and
			\begin{equation*}
				n-a=\left(\frac{q}{2},\frac{q}{2}-1,\frac{q}{2}-1,\frac{q}{2},\dots,\frac{q}{2},\frac{q}{2}-1,\frac{q}{2}-1,\frac{q}{2},\frac{q}{2},\frac{q}{2}\right)_q.
			\end{equation*}
			Then we have $L_1=L_2=1$, $L_3=0$ and $L_4=L_5=\frac{m}{2}-2$. By Lemma~\ref{sec3 lem0}, we have $\min\limits_{i\in T_1}\CL(a+i)\ge q^{\frac{m}{2}+1}>\delta_1$.
			\item When $i_2=\frac{q}{2}-1$ and $i_1\ne\frac{q}{2}-1$, let
			\begin{align*}
					a+i=\Big(&\frac{q}{2},\frac{q}{2},\frac{q}{2}-1,\frac{q}{2}-1,\dots,\frac{q}{2},\frac{q}{2},\frac{q}{2}-1+i_{\frac{m}{2}},\frac{q}{2}-1+i_{\frac{m}{2}-1},\\
					&\dots,\frac{q}{2}+i_{5},\frac{q}{2}+i_{4},\frac{q}{2}-1+i_{3},q-2,\frac{q}{2}+i_1,\frac{q}{2}+1+i_0\Big)
				\end{align*}
			and
			\begin{equation*}
				n-a=\left(\frac{q}{2}-1,\frac{q}{2}-1,\frac{q}{2},\frac{q}{2},\dots,\frac{q}{2}-1,\frac{q}{2}-1,\frac{q}{2},\frac{q}{2},\frac{q}{2}-1,\frac{q}{2}\right)_q.
			\end{equation*}
			Then we have $L_1=L_2=L_3=1$ and $L_4=\frac{m}{2}-2$. By \eqref{a3 equ1}, $i_{\frac{m}{2}}\le\frac{q}{2}-1$, and hence $L_5=\frac{m}{2}-2$. According to Lemma~\ref{sec3 lem0}, we have $\min\limits_{i\in T_1}\CL(a+i)\ge q^{\frac{m}{2}+1}>\delta_1$. 
		\end{itemize}
		Therefore, we have $a+i\in I_1$, and so $n+i=(a+i)+(n-a)$ and $i \in I_1+I_2$.
	\end{itemize}
	 The proof is completed.
\end{proof}



\end{document}